\documentclass[%
 reprint,
 amsmath,amssymb,
 aps,
]{revtex4-2}

\usepackage{graphicx}
\usepackage{dcolumn}
\usepackage{bm}
\usepackage{physics}

\begin{document}

\preprint{APS/123-QED}

\title{Model-free characterization of topological edge and corner states in mechanical networks}

\author{Marcelo Guzman}
 \altaffiliation[Also at ]{Department of Physics and Astronomy, University of Pennsylvania.}
 \email{mguzmanj@sas.upenn.edu}
\author{David Carpentier}%
\author{Denis Bartolo}%
\affiliation{%
 Univ. Lyon, ENS de Lyon, Univ. Claude Bernard, CNRS, Laboratoire de Physique, F-69342, Lyon, France
}%

\author{Xiaofei Guo}
\author{Corentin Coulais}
\affiliation{%
 Institute of Physics, Universiteit van Amsterdam, 1098 XH Amsterdam, The Netherlands
}%

\date{\today}

\begin{abstract}
Topological materials can host edge and corner states  that are protected from disorder and material imperfections.
In particular, the topological edge states of mechanical structures present unmatched opportunities for achieving robust responses in wave guiding, sensing, computation, and filtering. 
However, determining whether a mechanical structure is topologically nontrivial and features topologically-protected modes has hitherto relied on theoretical models.
This strong requirement has limited  the experimental and practical significance of topological mechanics to laboratory demonstrations. 
Here, we introduce and validate an experimental method to detect the topologically protected zero modes of mechanical structures without resorting to any modeling step. 
Our practical method is based on a simple electrostatic analogy: topological zero modes are akin to electric charges. To detect them, we identify elementary mechanical molecules and measure their chiral polarization, a recently introduced  
marker of topology in chiral phases. 
Topological zero modes are then identified as singularities  of the polarization field.
Our method readily applies to any mechanical structure and effectively detects the edge and corner  states of regular and higher-order topological insulators.  
Our findings  extend the reach of chiral topological phases beyond  designer materials, and allow their direct  experimental investigation.  
\end{abstract}

\title{Model-free characterization of topological edge and corner states in mechanical networks}



\maketitle

"To measure is to know''. This century old tenet traced back to Lord Kelvin~\cite{kelvin1891popular} has been foundational 
 in the discovery of material properties. 
The ancient greeks discovered magnetism  long before the  concepts of magnetic moment and spin were introduced~\cite{Hartmann1999Magnetic}, Hooke discovered the laws of elasticity before the introduction of tensorial mechanics~\cite{Hooke}, 
and the quantized conductance of the
Quantum Hall effect was discovered prior to the concepts of topological phases~\cite{Klitzing,thouless1998topological}.
In this article we follow Kelvin's wisdom, and  
we  experimentally reveal  the topological phases of mechanical metamaterials, and locate their soft and weak spots without resorting to any {\em a priori} modeling. 

This insight on structural mechanics~\cite{Kadic_Review2013,Bertoldi2017Flexible} is atypical from the perspective of metamaterial science. 
Metamaterials are designed~\cite{Kadic_Review2013,Bertoldi2017Flexible}. 
It is only after a necessary modeling step that they can be engineered, and eventually achieve properties out of reach of conventional matter. 
In particular,  the field of topological mechanics has bloomed since the identification of a formal correspondence between the linear dynamics of isostatic lattices of beads and springs, and tight binding models of  topological insulators in quantum matter~\cite{Kane2014Topological,Mao2018,Prodan2009}.  
This correspondence then led to a new stream of design strategies for mechanical structures featuring polar mechanical response~\cite{Rocklin2017,Bilal2017,Coulais_Nature2017}, or directional phononic wave channels~\cite{Nash2015,Peano2015}. 
Today virtually all types of electronic topological phases have found a mechanical counterpart, from the topological insulators and superconductors of the tenfold 
classification~\cite{Susstrunk2016Classification}, to  non-Hermitian~\cite{Ghatak2020,Coulais_review2021}, quantum spin-Hall~\cite{Susstrunk2015} and even higher order~\cite{Serra2018,Saremi2018,Xue2019,Ni2019} topological phases. 

\begin{figure*}
        \centering
        \includegraphics[width=0.9\textwidth]{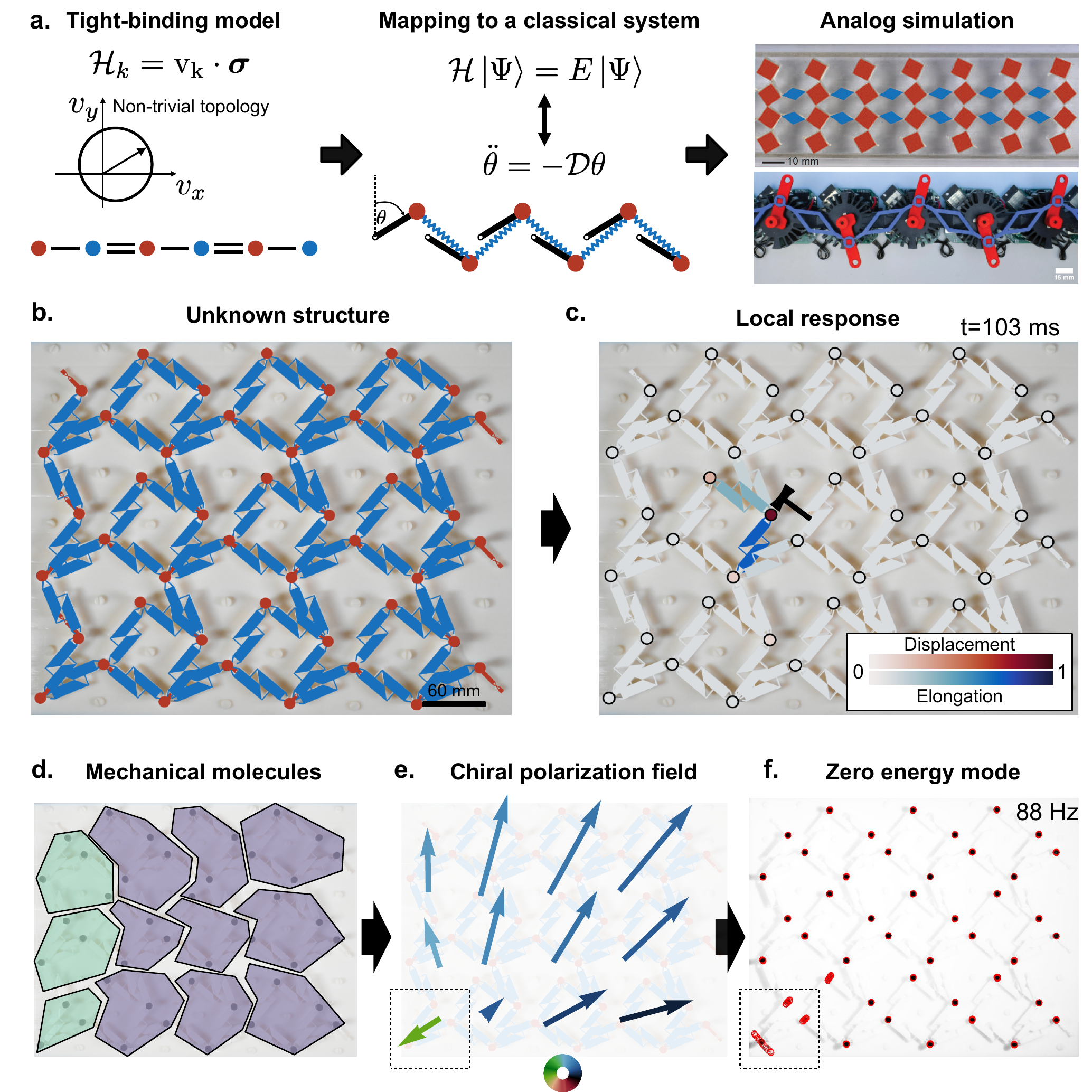}
        \caption{{\bf  Two complementary perspectives on topological Mechanics.} {\bf a.} Topological metamaterial design. The starting point is the tight binding Hamiltonian of a quantum electronic system.  
        We give the example of a 1D chiral Hamiltonian $\mathcal H_{k}=\mathbf v_k^i\cdot\bm \sigma^i$, where the  $\sigma_i$ are  two  Pauli matrices and $\mathbf{v}$ a real 2D vector indexed by the wave vector $k$ . The Hamiltonian is characterized by a topological index.
        Here, given the unit cell sketched in the figure,  $\mathcal H$ is associated to a nonvanishing winding number, the topological index of chiral Hamiltonians, 
        given by the winding of $\mathbf v_k$ around the origin as $\mathbf{k}$ varies.  
        The Hamiltonian dynamics is then mapped on the classical mechanics of mechanically coupled structures such as a network of beads and springs. The bead and spring network can then then be approximately realized with a variety of materials. The upper right panel shows a classical realization of the celebrated SSH Hamiltonian~\cite{Coulais_Nature2017}, while the bottom right picture shows a realization of its non-Hermitian counterparts~\cite{Ghatak2020}~ 
        {\bf b.} Model-free detection of topological zero modes in unknown mechanical structures. The starting point is now an {\em  a priori} unknown mechanical structure made of passive units such as connected beads, springs and beams. 
        {\bf c.} We identify both the displacement and elongation degrees of freedom and measure their dynamic response to a point force. 
        The blue and red colormaps indicate the instantaneous magnitude of the displacements and elongations in response to a local point force, SI and Figure 2 for more details. 
        {\bf d.} and {\bf e.} The weighted average of the displacements and elongations allows us to define mechanical molecules and their associated  local chiral polarization $\bm \Pi(\bm r)$, which generalizes the concept of topological polarization introduced in~\cite{Kane2014Topological}.The big bulk molecules are highlighted in violet while the smallest, belonging to the left edge, in green.
        {\bf f.} Assemblies of  mechanical molecules that share a chiral polarization pointing along the same direction feature the same spectral topology. 
        Conversely, singularities (i.e discontinuities of the order of the lattice spacing) in the polarization field reveal the existence of localized zero energy modes:  floppy modes and states of self-stress. Here the discontinuity is localized at the bottom left corner (dashed square) and the presence of a corner floppy mode is verified by shaking the whoel sample at the lowest eigenfrequency: $88$Hz.
        The red circles track the positions of the beads over time. 
               }
        \label{fig.reasoning}
\end{figure*}

These diverse realizations all stem from the same design protocol, which relies on rather abstract concepts. 
We summarize it in Fig.~\ref{fig.reasoning}a.   The starting point is a tight binding Hamiltonian $\mathcal H$ known to feature a so-called nontrivial topological index~\cite{Hasan_Review}. 
$\mathcal H$ typically describes the quantum dynamics of electrons in a solid.  
 The next step consists in using a formal mapping from the quantum dynamics ruled by $\mathcal H$ to a classical Newtonian   system~\cite{Kane2014Topological,Susstrunk2016Classification}. 
 Finally, a mechanical structure is designed and engineered to perform an analog simulation of the classical dynamics. 
 The key property of the resulting structures is the existence of  edge states protected by the non-trivial topology of the bulk-vibration modes, 
a consequence of the  bulk-boundary correspondence principle.
 
In this article, we take an alternative perspective on topological mechanics. 
We place measurements as a primary tool to inquire about the properties of mechanical structures. 
We aim at answering a basic question: given a rigid structure (isostatic or hyperstatic) assembled from bead, springs or beams, can one predict the existence and location of its topologically protected floppy, or self-stress, modes without resorting to any theoretical model?  
The motivation is clear as these two  modes can either offer functional capabilities, or limit the range of applications of a mechanical structure. 
In mechanical insulators, floppy modes are localized soft spots which respond nonlinearly to vanishingly small forces~\cite{Chen2014,Mao_Review,Paulose2015,Bilal2017,Ma2018}, while self stresses are weak spots precursor of failure upon external loads, and cannot be detected from deformations fields~\cite{Calladine1978,Paulose2015,
Zhang2018,
Widstrand2022}.%

To predict and locate the soft and weak spots in natural or man-made structures, we introduce a generic method inspired by classical electrostatics~\cite{Bertoldi2017Flexible}. 
We illustrate it in Fig.~\ref{fig.reasoning}b. 
Starting from an unknown structure, we poke it locally and measure the resulting displacements and deformations. 
From the spatial correlations between the deformation and displacement fields we then identify mechanical ``molecules''. 
They are defined as the most strongly coupled displacement and deformation degrees of freedom. 
From the measured responses, we can then define a polarization $\bm \Pi$ for each molecule. 
This so-called chiral polarization encodes both the local rigidity, and spectral topology of the metamaterial, and was theoretically defined in~\cite{guzman2020geometry}.
$\bm \Pi(\mathbf r)$ is a local material property that 
 generalizes earlier macroscopic markers of topological phase models: the topological polarization introduced by Kane and Lubensky~\cite{Kane2014Topological}, the topological contribution to the polarization of 
 electronic insulators~\cite{Resta2007} and the mean chiral displacement of 
 photonic metamaterials~\cite{Cardano2017,Roberts2022}. 
We explain how to measure $\bm \Pi$, and show that floppy  and self-stress modes correspond to net mechanical charges  signaled by topological defects in the chiral polarization field. 
Our minimal method predicts the location of the zero-energy modes  without resorting to any theoretical model, or abstract topological concepts. 
Combining experiments, simulations and theory we establish the robustness of our predictions to uncontrolled dissipation processes, nonlinearities and material imperfections.  

\section{Method: Basic concepts and tutorial example}
\subsubsection{Poking the mechanical SSH chain}
\begin{figure*}
        \centering
        \includegraphics[width=0.85\textwidth]{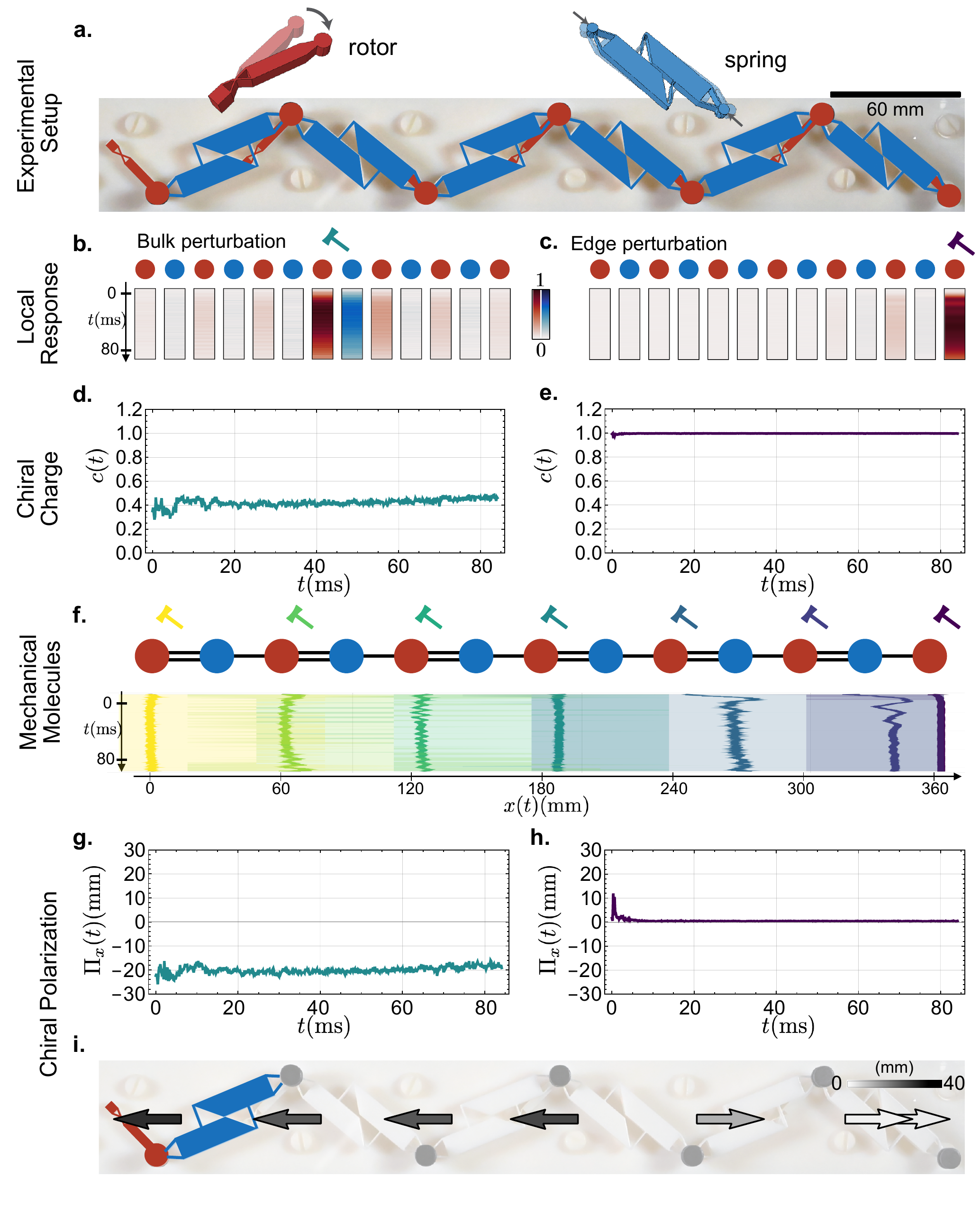}
        \caption{{\bf Mechanical molecules and chiral polarization in a 1D mechanical insulator.}\\
        {\bf a.} A 3D printed mechanical chain assembled from seven rotors (red) connected by springs (blue). The system hosts a floppy mode (FM) in the rightmost node. See also SI.
        {\bf b. } Chiral representation of beads (red circles) and springs (blue circles). 
        We locally excite the fourth node of the sample by poking the bead with a hammer. The response of the elongation and displacement degrees of freedom propagate in time and space (experimental measurements).
        {\bf c.} In contrast, poking the last node results in a localized response  where almost no elongation is observed.
        {\bf d.} To account for the nature of the response, whether it is displacement or elongation-dominated, we trace its chiral charge. The bulk response is characterized by an oscillating charge aroung $c=0.4$.
        {\bf e.} In contrast, the rightmost perturbation is characterized by $c=1$ typical of a floppy mode. 
        {\bf f.} The response, as seen in panels {\bf b} and {\bf c} remains centered around  the excited node. 
        Its spatial extent is captured, for example, by the confidence interval of the probability distribution defined by $\Psi$  here set to 95\%.
        The mechanical molecules are confined regions in space of size comparable with the length-scale of the repeating motif (bottom).
        {\bf g,h,i.}  We assign the chiral polarization  $\bm \Pi(t)$ to each mechanical molecule. It is a vector which characterizes the (chiral) topological  phase of the system.
        The source-like discontinuity of the field signals the floppy mode. In practice a discontinuity of the chiral polarization corresponds to a change of one of its coordinates of the order of the lattice spacing $a=60\text{mm}$.
        }
        \label{fig.responseCharacterization1}
\end{figure*}
For the sake of clarity, we first explain and benchmark our method using a tutorial example.
We consider the structure shown in Fig.~\ref{fig.responseCharacterization1}a. 
It is a compliant realization of the Kane-Lubensky chain first introduced in~\cite{Kane2014Topological}.
This 3D printed structure is made of repeated units hosting one angular degree of freedom (the bead position showed in red), and one spring (showed in blue). 
When the rest angle $\theta_0$ is non-zero, the isostatic metamaterial is a mechanical insulator~\cite{Kane2014Topological,Paulose2015Topologica}. 
Figs.~\ref{fig.responseCharacterization1}b and c show that when we
locally poke one bead with a hammer, the response remains localized in space, see also SI where we provide a detailed description of the experiments. 
This seemingly mundane operation reveals two essential features: Firstly, when hitting a bead, the neighboring springs evolve their elongation, but do not respond symmetrically: the bulk response of the metamaterial is intrinsically polar, Fig.~\ref{fig.responseCharacterization1}b. 
Secondly, when poking the rightmost bead, the elongation of the neighbouring springs remains unchanged, they  therefore come at no elastic energy cost and define a floppy mode, Fig.~\ref{fig.responseCharacterization1}c. 
We now make our observations more quantitative. Our goal is to: (i) experimentally identify  the topological character of the metamaterial spectrum based on our measurements only, and (ii) predict the nature and location of the topologically protected zero-energy modes with a minimal sampling protocol. 
This protocol is of particular relevance for self-stress modes which, unlike floppy modes, have never been revealed by {\em any} non-destructive measurement~\cite{Paulose2015,Widstrand2022}. 
As a matter of fact, a self-stress is a mechanical state where  finite stresses results in no forces and hence in no bead dynamics. 

Our method fundamentally relies on the concepts of chiral polarization, 
and more broadly on  chiral symmetry, a generic feature  of mechanical networks.  
 These two concepts were both thoroughly discussed in Ref.~\cite{guzman2020geometry} from a sole theoretical perspective.
In the remainder of the manuscript, we build on these concepts to introduce a model-free characterization of topological edge and corner modes.
\subsubsection{Chiral symmetry} 
All mechanical structures can be described as assemblies of beads connected by springs. 
This elementary observation translates in a fundamental symmetry: chiral symmetry. 
We now show the practical relevance of this seemingly formal observation. 
In short, we note $\bm u(t)$ the vector formed by the displacements of all beads, and $\bm e(t)$ the displacements of all springs.
The two quantities are linearly related via the connectivity of the mechanical network: $\bm e= \mathcal Q^{\rm T}\bm u$,
where $\mathcal Q^{\rm T}$ is the so-called compatibility matrix~\cite{Mao_Review}. 
When the spring response is linear, the elastic energy modes are identical to the spectrum of the matrix 
\begin{equation}
\mathcal H=\begin{pmatrix}
0&\mathcal Q\\\mathcal Q^{\text{T}}&0
\end{pmatrix} \ .
\end{equation}
From a condensed matter perspective $\mathcal H$ would be the Hamiltonian a free quantum particle evolving on a bipartite lattice  defined by the bead and spring positions respectively, red and blue sites in  Fig.~\ref{fig.responseCharacterization1}b.
Crucially, the lattice connects only blue sites to red sites and vice versa. 
There is no connection between sites of identical nature. 

This simple geometrical property formally translates to chiral symmetry: $\mathcal H$ anticommutes with the diagonal chiral operator: $ \mathcal H \mathcal C = - \mathcal C \mathcal H $, 
where $\mathcal C=\begin{pmatrix}
1&0\\0&-1
\end{pmatrix}$. 
To see how this symmetry rules the dynamics of the network, we recall that the Schr\"odinger equation defined by $\mathcal H$  describes the  dynamics of the wave function $\ket{\Psi(t)}=\left( \mathbf u(t),\mathbf e(t)\right)$ which encodes the full information about both the displacements and elongations in the network~\cite{Mao_Review} (Units are chosen so that the bead mass and spring stiffness are both equal to one). 
We explicitly recall in the SI how to map the Newtonian dynamics of the assembly of beads and springs on the above Shr\"odinger picture.

\subsubsection{Chiral Charge} 
The structure of $\mathcal C$ suggests an intuitive electrostatic analogy. We can assign a $+1$ charge to the displacement degrees of freedom, and a $-1$ charge to the spring deformations. 
This analogy is useful to quantify the nature of the response to a mechanical perturbation. 
When poked, a mechanical structure undergoes vibrations, where the beads move and the springs deform. 
The presence of asymmetric lever arms---always at stake in topological Maxwell lattices---implies an imbalance between deformations and displacements. 
To quantify this imbalance on the response to a perturbation  applied to the $i^{\rm th}$ bead, we define the chiral charge $c_i(t)$ as
\begin{align}
    c_i(t)=\frac{\bm u^\text{T}\bm u-\bm e^\text{T}\bm e}{\bm u^\text{T}\bm u+\bm e^\text{T}\bm e} = \frac{\expval{\mathcal C}{\Psi}}{\braket{\Psi}} \ .
\end{align}
Positive (resp. negative) chiral charges indicate the preponderance of displacements (resp. stresses), the extreme cases being floppy modes ($c=1$) and states of self stress ($c=-1$) as exemplified 
by our experiments  in Fig.~\ref{fig.responseCharacterization1}d. 
In electrostatics, charges are bound to molecules. To pursue the analogy further, we therefore need to define ``mechanical molecules''.
\subsubsection{Mechanical molecules}
We now define the ``mechanical  molecules'' hosting the chiral charges: the ensemble of beads and springs that are the more strongly coupled and separated by less than one lattice spacing.
To do so, we hammer all the rotors of the chain one at a time.  
We then  track  the bead displacements $\bm u(t)$ and  spring elongations $\bm e(t)$. 
Our experiments show that the resulting wave function $\Psi(t)$ is: (i) localized in space and (ii) asymmetrically distributed around the bead that has been perturbed. 
In the experiment showed in Fig.~\ref{fig.responseCharacterization1}b, the shape of $\Psi(t)$ gives a visual indication of which beads and springs are the most strongly coupled, thereby defining what we dub a ''mechanical molecule''.
This definition can be made more systematic. 
In all that follows, we experimentally define a molecule by the ensemble of beads and springs within the region of space that corresponds to the $95\%$ confidence interval associated to the distribution $\Psi(t)$, Fig.~\ref{fig.responseCharacterization1}f, see also SI.
 
Intuitively, the pairs of beads and springs that form the molecules are those that are coupled by the largest lever arm: in our experiments, when a bead moves, it  primarily deforms the spring sitting on its right hand side. 
In the language of topological insulators the mechanical, or chiral, molecules 
define the units cells compatible with the atomic-limit 
of the linearized Hamiltonian $\mathcal H$~\cite{vanderbilt2018berry,guzman2020geometry}.
In mechanics the atomic limit is reached when the lever arm between two molecules is vanishingly small, i.e. when the rotor and the spring are colinear.

We finally introduce the center of the mechanical molecules as the time average of the instantaneous weighted position $\bm r_i(t)$ defined by 
\begin{equation}
    \bm r_i(t)=\frac{\bm u^\text{T}\mathcal R_u\bm u+\bm e^\text{T}\mathcal R_e\bm e}{\bm u^\text{T}\bm u+\bm e^\text{T}\bm e}=\frac{\expval{\mathcal R}{\Psi}}{\braket{\Psi}}.
    \label{Eq.position1D}
\end{equation}
$\mathcal R_u$ (resp. $\mathcal R_e$) is a diagonal matrix whose entries are the  equilibrium coordinates of the beads (resp.  springs), and $\mathcal R$ is the combined position operator. 

\subsection{Chiral polarization}
In the Kane-Lubensky chain, as in all Maxwell lattices, the molecules include as many beads as springs. 
The molecule is therefore ``geometrically neutral''. 
However, the structure showed in Fig.~\ref{fig.responseCharacterization1}f is not tesselated by an integer number of neutral molecules.
This observation suggests that the rightmost edge of the structure might host a net $+1$ chiral charge, {\it i.e.}  a floppy mode. 
It corresponds to a bead whose displacement is not coupled to spring elongations.

Our electrostatic analogy then begs for defining another material property:  a mechanical polarization. 
Just as one would define an electrostatic dipole, we can define locally a  polarization vector $\bm \Pi(t)$ as the response function to a perturbation applied to the $i^{\rm th}$ bead:
\begin{align}
    \bm \Pi_i(t)&=
    2\left(\frac{\bm u^\text{T}\mathcal R_u\bm u-\bm e^\text{T}\mathcal R_e\bm e}{\bm u^\text{T}\bm u+\bm e^\text{T}\bm e}-
    c_i(t)\bm r_i(t)\right)
    \label{eq.chiralPol}\\
    &=
    2\left(\frac{\expval{\mathcal C\mathcal R}{\Psi(t)}}{\braket{\Psi(t)}}
    -c_i(t)\bm r_i(t)\right).\nonumber
    \label{eq.chiralPol}
\end{align}
This quantity is nothing else but the chiral polarization introduced in~\cite{guzman2020geometry}.
In Fig.~\ref{fig.responseCharacterization1}g, we plot the $x$-component of the time-averaged chiral polarization measured in our experiments. 
Away from the floppy mode $\langle \bm \Pi(t)\rangle_t$ is uniform and points towards the left hand side. 
In simple terms, the polarization $\bm \Pi_i$ always points from the poked bead to the spring whose deformation is the most prominent due to the lever effect.
This is the reason why chains with right-leaning and left-leaning rotors have opposite polarizations, as we will see  in the next section.

However, close to the right edge of the sample,  the chiral polarization field is singular and changes sign. 
This singularity coincides with the location of the floppy mode, Fig.~\ref{fig.responseCharacterization1}c. 
We show below that this observation has a fundamental origin. 
It is deeply rooted in the bulk boundary correspondence of topological insulators. 
To explain it we first need to relate the measure of $\bm \Pi_i$ to the topological index of a chiral topological phases.

In~\cite{guzman2020geometry}, 
we showed that, in the bulk of a periodic Maxwell lattice whose dynamics is described by a Bloch Hamiltonian $\mathcal H(\bm k)$,
\begin{equation}
\bm \Pi_i=p-a w, 
\label{Eq.windingH}
\end{equation}
where $w$ is the winding number of $\mathcal H(\bm k)$, and $p$ is the geometrical polarization of the unit cell of size $a$ used to define the Bloch Hamiltonian.
Strictly speaking this formula applies when the initial perturbation excites a so-called Wannier state of $\mathcal H$~\cite{vanderbilt2018berry}, a condition which is in principle never met in our experiments. 
However, we numerically show in SI that \eqref{Eq.windingH} and \eqref{eq.chiralPol} lead to the same measure of $\bm \Pi_i$, which hardly depends on the specifics of the poking protocol (provided that it is localized in space).
At long times $\bm \Pi_i(t)$ weakly fluctuates around a well defined average value for $\langle \bm \Pi(t)\rangle_t$. 
This numerical result implies that the measurement of a net chiral polarization field in an experiment unambiguously signals a nontrivial bulk topology. 
We can draw three conclusions from this central result:

(i) In crystalline lattices, away from the edge, translation invariance implies that the chiral polarization field is homogeneous. 
From a {\em single} poking experiment, we can  then measure the bulk chiral polarization, and therefore deduce the topological nature of the metamaterial without resorting to any theoretical model.   

(ii) In our experiments the 3D printed structures dissipate  energy. 
In addition, by hammering them,  we excite a variety of structural modes which are not captured by the naive bead and spring model we use to predict the value of $\bm \Pi$.
 Nonetheless, all our results are in excellent agreement with the  prediction of the chiral polarization based on the Kane-Lubensky model with no free fitting parameter. 
 We measure $\Pi_x=21.5\pm 1.2$ mm and our linear theory predicts $\Pi_x=30$ mm. 
 This agreement echoes the strong robustness of our method to geometrical, material and protocol imperfections.

(iii) \eqref{eq.chiralPol} does not rely on any underlying periodic structure, or any assumption about the isostaticity of the mechanical structure. 
In the remainder of the article, we demonstrate that this method further applies to heterogeneous, and hyperstatic structures.  

\subsection{Bulk boundary correspondence and optimal detection of zero energy modes}
In principle, the detection of floppy modes and states of self stress  in an unknown sample requires measuring the displacement and stress response everywhere in space, and at all frequencies. 
Here we show how to circumvent this costly procedure. 

To do so, we build on the relation between $\bm \Pi_i(t)$ and the winding number of a chiral Hamiltonian to take advantage of the bulk-boundary-correspondence principle. 
This principle relates the bulk topology of a material to the existence of edge states~\cite{Mao_Review,guzman2020geometry,Jezequel2022}. 
Topologically protected zero modes exist when~\cite{guzman2020geometry}:
(i) the chiral polarization in the bulk does not vanish, and 
(ii) the metamaterial is not composed of an integer number of mechanical molecules. 
This last condition equivalently means that the  lattice  
cannot 
be tesselated with units cells compatible with the atomic limit of $\mathcal H$~\cite{guzman2020geometry}. 
As conjectured above, this situation is analogous to a molecular chain hosting dielectric dipoles. When a chain includes an integer number of molecules the overall charge vanishes. 
Conversely when a fraction of a molecule is added at its end, the chain hosts an additional uncompensated charge. 
The sign of the charge is then determined by the orientation of the electric dipole.  

We can use this analogy to locate mechanical zero modes in the structure showed in  Fig.~\ref{fig.responseCharacterization1}a. 
Firstly, we note that it cannot be fully tesselated with mechanical molecules, in Fig.~\ref{fig.responseCharacterization1}e, we see that an extra bead must be added on the right hand side to complete the chain. Secondly, $\langle \bm\Pi(t)\rangle_t$ is finite and points towards the left hand side in the bulk. 
 Altogether these observations tell us that a floppy mode (positive chiral charge) must exist at the rightmost end of the chain, as confirmed by our observations and measurements, Fig.~\ref{fig.responseCharacterization1}c and e.
 We stress that our method requires poking a {\em single} bead in the bulk to predict and locate a floppy boundary mode.  

\begin{figure*}
        \centering
        \includegraphics[width=0.7\textwidth]{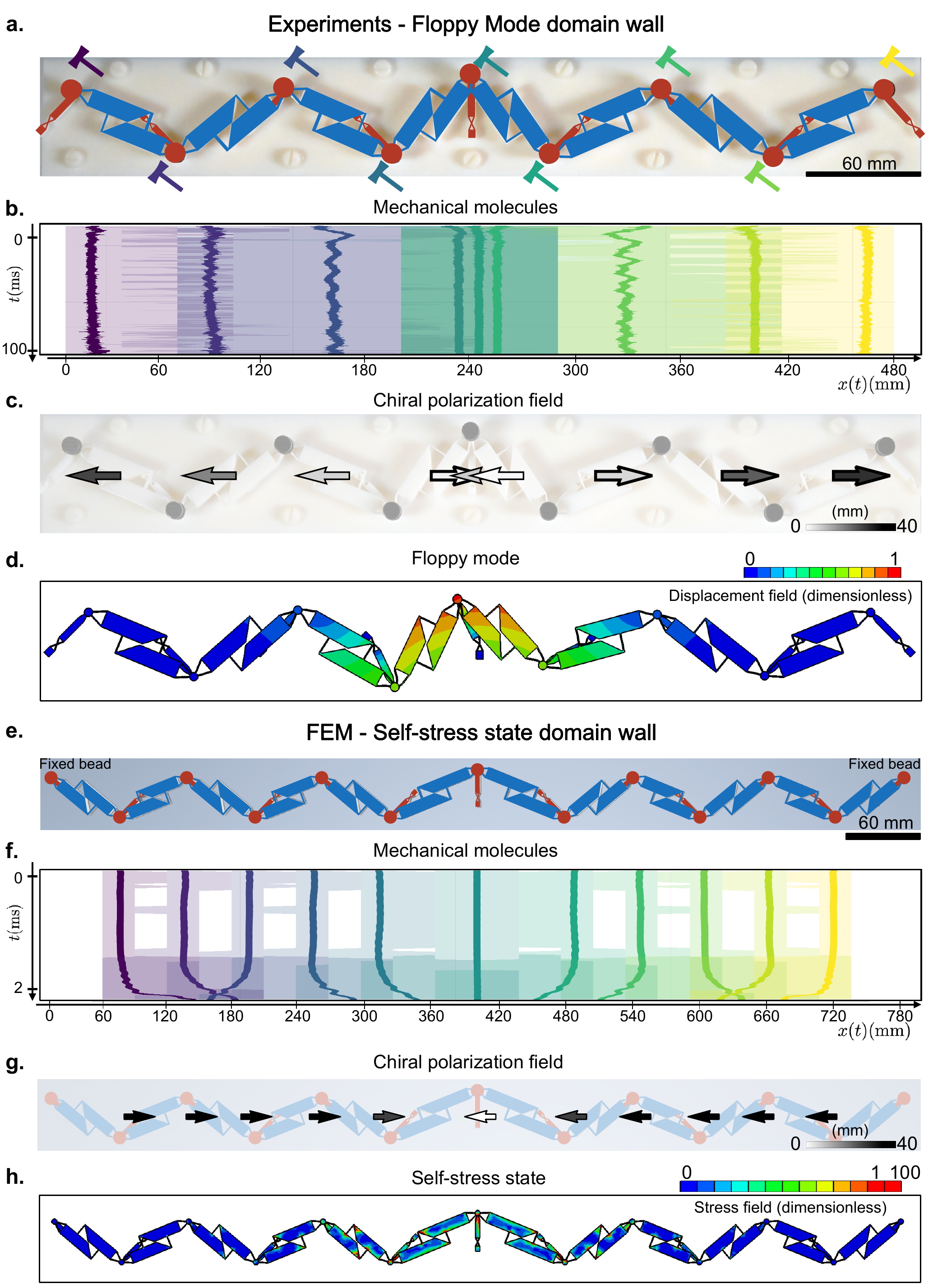} 
        \caption{{\bf Detecting topologically protected zero modes from poking experiments.} {\bf a.} 3D printed mechanical chain with an angular discontinuity at the central rotor. We locally perturb each node by poking them with a hammer. Each perturbation corresponds to a different color.
        {\bf b.} To characterize 
        the response to the nine poking experiments, we track the average positions $\mathbf r_i(t)$ (solid line), and the  extent of the associated chiral molecules defined by the $95\%$ confidence interval associated to $\Psi$ (shaded regions).
        Away from the structural discontinuity, the molecules includes one bead and one spring. 
        At the boundary between the left and right domains, the chiral molecules overlap.
        {\bf c.} Average chiral polarization field, located at the average centers of perturbations. The arrows are shown with fixed length of $30$(mm) and the magnitude is depicted with a grayscale. From left to right, the polarization switches sign from negative to positive. The midpoint is characterized by a discontinuity in the direction of the field. The source of the discontinuity indicates the presence of the floppy mode.
        {\bf d.} The prediction is confirmed by FEM simulations. Here we show the lowest energy mode of the system. It corresponds to a localized displacement mode, i.e. a floppy mode localized at the boundary between the two topologically distinct Kane-Lubensky chains.
        {\bf e.} Using FEM simulations we analyse the complementary version of the system shown in {\bf a}, where left and right sides are swapped. Moreover, we fix both ending beads in order to avoid the edge floppy modes.
        {\bf f.} From each perturbation, implemented by imposing local displacements on the nodes at $t=0$, we track the center and the extent of the mechanical molecules (confidence interval of 95\%). 
        Due to the absence of damping in the simulations, the perturbations do not fade away. They instead  bounce back and forth in the finite system.
        We therefore consider the evolution of the molecules in a small time window of $2$ms.
        {\bf g.} The resulting chiral polarization field displays a sink-like discontinuity at the boundary between the two  distinct domains. This observation is consistent with the existence of a topological state of self stress, {\bf h} (FEM simulations).
        }
        \label{fig.BulkBoundary}
\end{figure*}

To further confirm the predictive power of our method, we performed  additional experiments and finite element method (FEM) simulations. 
We first consider the heterogeneous metamaterials showed in Fig.~\ref{fig.BulkBoundary}a. 
It is composed of two Kane-Lubensky chains leaning in opposite directions.
We  use the same procedure as above to experimentally identify the mechanical molecules~Fig.~\ref{fig.BulkBoundary}b, and their chiral polarization, Fig.~\ref{fig.BulkBoundary}c.  
We find again that we cannot tesselate the full metamaterial with a single type of mechanical molecule, and that the chiral polarizations of the two domains leaning in opposite directions have opposite signs, Fig.~\ref{fig.BulkBoundary}c.
The $\langle \bm \Pi(\bm r, t)\rangle_t$ field has a positive singular divergence at the boundary between the two domains. 
More specifically, the jump in the $x$ component of $\langle \bm \Pi(t)\rangle_t$ is of the order of one half of a unit cell $a$: $|\Pi_x^{\text{left}}-\Pi_x^{\text{right}}|\approx 0.77a$. 
These observations imply that a topologically protected floppy mode must exist at the junction between the two domains. 
The first eigenmode of the FEM simulation unambiguously confirms this prediction Fig.~\ref{fig.BulkBoundary}d. 

Similarly,  we perform a FEM analysis of the second heterogeneous structure showed in Fig.~\ref{fig.BulkBoundary}e.
In this case the $\bm \Pi$ field has a negative singular divergence.
($|\Pi_x^{\text{left}}-\Pi_x^{\text{right}}|\approx 0.79 a$), Fig.~\ref{fig.BulkBoundary}g.
We therefore expect the localization of a state of self-stress at the boundary between the two domains, as demonstrated in~\cite{Kane2014Topological}.
It is worth noting that self-stress states cannot be directly detected from linear displacements modes. 
However, here we can numerically track them thanks to the sub-unit resolution of the FEM simulations for the lowest energy mode, Fig.~\ref{fig.BulkBoundary}h, confirming our prediction.

To conclude, 
we stress on the efficiency of our method.
When a metamaterial is made of a polycrystalline structure, only one poking experiment per domain is enough to predict the existence and the nature of localized zero modes.
One poking experiment in the bulk of each domain gives an information about the orientation of the chiral polarization of the corresponding crystal. 
Any discontinuity in the polarization of the order of, or larger than, one lattice spacing between two domains  implies the existence of topologically protected zero modes at the boundary.
This procedure is undoubtedly simpler, and much faster, than any modeling process which would rely on the computation of topological invariants.
Our method is also more efficient than any protocol based on a full spectral characterization, which  requires an extensive number of independent measurements, see also SI.

\begin{figure*}
        \centering
        \includegraphics[width=\textwidth]{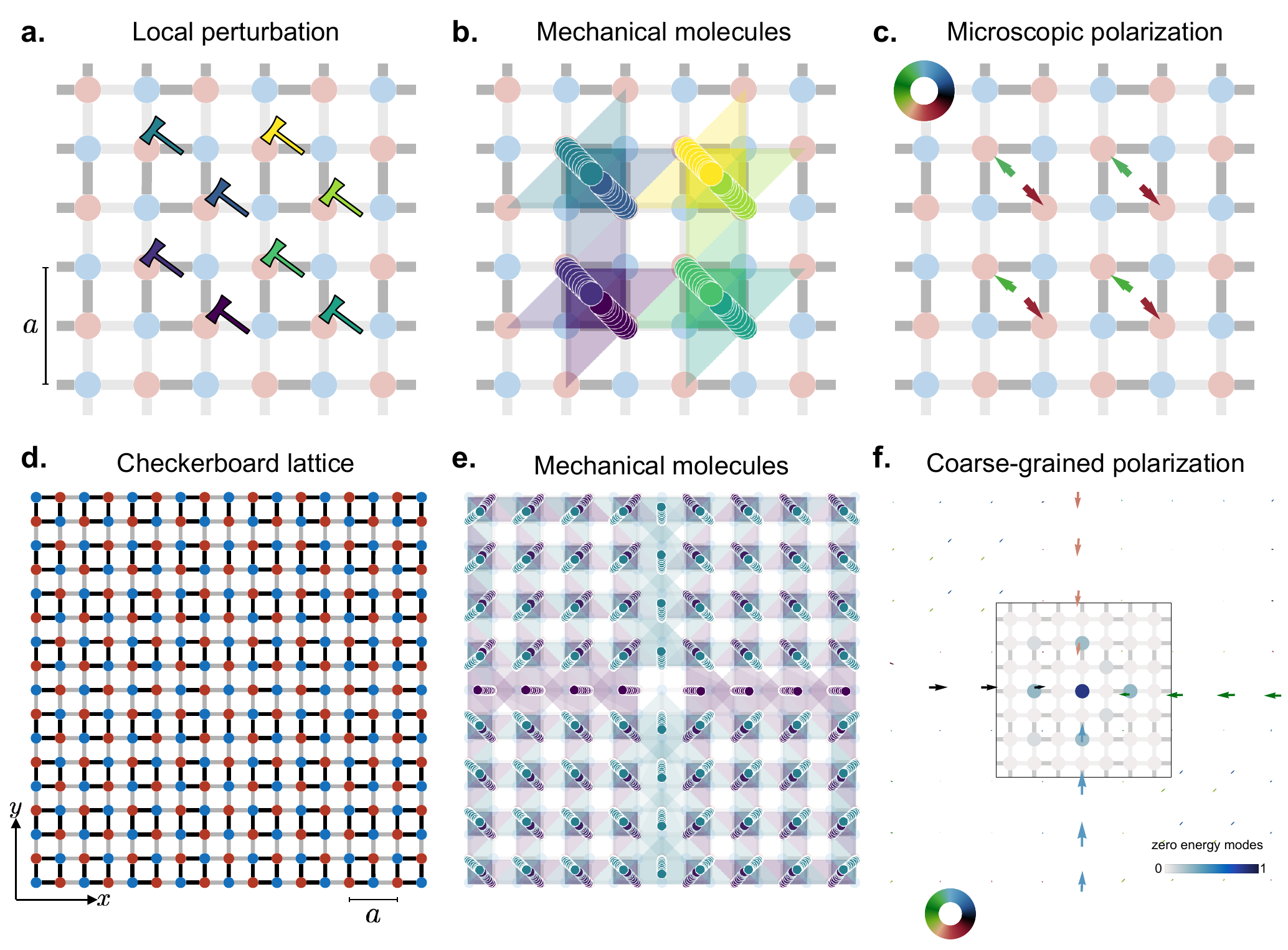}
        \caption{{\bf Polyatomic mechanical molecules} {\bf a.} Model-free characterization of the checkerboard lattice through local perturbations (hammers). The chiral network connects degrees of freedom (red disks) to constraints (blue disks). The unit-cell length is $a$.
        {\bf b.} The centers of the perturbations $\bm r(t)$ jiggle around confined regions with periodicity $~a$. 
        Unlike the 1D example, here two centers revolve around the same area, they both define the mechanical molecule.
        This is further confirmed by the spatial spreading of the perturbations (colored polygons). 
        it is defined through the Mahalanobis distance computed from $\Psi(t)$ (see SI).
        Here we superimpose the polygons for each time step.
        The opaque regions where the perturbations overlap include the two beads and two springs that form the mechanical molecule.
        {\bf c.} Each perturbation leads to a time-average polarization, illustrated by the arrows. The polarization field is highly discontinuous at the microscopic scale. However, it vanishes when coarse graining over the mechanical molecules. 
        {\bf d.} Finite size checkerboard lattice of 17x17 nodes.
        {\bf e.} We simulate the evolution of local perturbations over all the degrees of freedom, as in {\bf a}. To reveal the spatial arrangement of the mechanical molecules, we superimpose the colored polygons as in {\bf b}. The outcome clearly distinguishes two kinds of molecules: complex molecules of two degrees of freedom and two constraints, and simple molecules of one degree of freedom and one constraint.
        {\bf f.} Coarse graining the field unveils four sectors of null polarization connected by finite domains of finite polarization. Furthermore the field has a sink-like discontinuity at the center; the system hosts a localized self-stress state.
        }
        \label{fig.checkerboard}
\end{figure*}

\section{Zero modes and topological phases of higher dimensional metamaterials.}
\label{sec.complexmol}
Our model-free characterization of topologically polarized metamaterials has so far been limited to a canonical case where each mechanical molecule is made from a single pair of atoms. 
However, most mechanical structures have multiple degrees of freedom per unit cell. 
They are made of ``polyatomic mechanical molecules'', or in the 
language of band theory, each unit cell includes several Wannier centers.
The strategy to detect topologically protected zero modes  remains the same: (i) we poke the system to identify the mechanical molecules, (ii) we define their chiral polarization, (iii) if the material cannot be tesselated with an integer number of mechanical molecules, we look for discontinuities in the polarization field and hence resolve the location of states of self-stress and floppy modes.

To clearly demonstrate this generalization, we first build on the  well known theoretical model sketched in Fig.~\ref{fig.checkerboard}a, which we characterize numerically. 
This model is a paradigmatic example of a higher order topological insulator, it consists of a chiral tight-binding Hamiltonian defined on the checkerboard lattice~\cite{Benalcazar2017Quantized}, and maps on the dynamics of the mechanical structure discussed in~\cite{serra2018observation}.

\subsection{Polyatomic mechanical molecules}
The checkerboard lattice includes two degrees of freedom, and therefore two Wannier centers, per unit cell.  
The identifications of the beads and springs that define the mechanical molecules hence requires an extra step. 
Without the {\em a priori} knowledge of the bond strength, we need to study the response to more than two local perturbations to define our mechanical molecules, Fig.~\ref{fig.checkerboard}a. 
 Fig.~\ref{fig.checkerboard}b shows the time evolution of the trajectories of the centers $\mathbf r_n(t)$, see \eqref{Eq.position1D}, and the spreading of $\Psi(t)$  in response to localized perturbations.
We can then readily generalize the method introduced in the previous section, and  define the mechanical molecule as shown in Fig. \ref{fig.checkerboard}b.
\subsection{Chiral polarization texture}
We can now define the chiral polarization associated to the two beads poked within the mechanical molecules. 
We  use the same definition as in~\eqref{eq.chiralPol}.
Fig.~\ref{fig.checkerboard}c reveals that each mechanical molecule features an inner polarization texture where the two beads are associated to opposite chiral polarizations. 
In all that follows we disregard this microscopic structure reminiscent of the quadrupolar order discussed at large in~\cite{Benalcazar2017Quantized}.
We instead define the total chiral polarization by computing
\begin{align}
    \bm \Pi=\sum_{n=1}^N\langle \bm \Pi_n(t)\rangle_t,
    \label{eq.chiralPolmean}
\end{align}
where $N$ is the number of degrees of freedom per mechanical molecule.

\subsection{Detection of topologically protected corner states}
In Fig.~\ref{fig.checkerboard}d we show a structure assembled from four different checkerboard lattices. 
We find that  the whole structure cannot be tesselated by a single mechanical molecule. It is instead cut into four quadrants by two perpendicular lines of ``half-molecules'', Fig. \ref{fig.checkerboard}e. 

Fig.~\ref{fig.checkerboard}f shows the full chiral polarization field.
This pattern allows us to make two central predictions. 
Firstly, the structure is topologically heterogeneous. 
The four quadrants are topologically trivial as far as the chiral polarization is concerned (the structure is isostatic and $\bm \Pi$ is vanishingly small~\cite{guzman2020geometry}). 
There is no discontinuity of the chiral polarization in the directions normal to the edges of the sample, or to the domain walls. This observation implies that the metamaterial does not feature any topological edge state.
However, the cross-shaped domain wall form 1D topological insulators with incompatible chiral polarizations. 
We therefore predict the existence a zero energy mode at their intersection. 
As the polarization field points towards  the center of cross, it must host a negative chiral charge, i.e. a state of self stress.
This prediction is in perfect agreement with the distribution of the zero energy eigenmode found by diagonalizing $\mathcal H$ and plotted in Fig.~\ref{fig.checkerboard}f.
It is worth noting that we have detected the topologically protected zero mode at the intersection between four higher order topological phases without resorting to any higher order topological invariant, and without exploiting the fourfold  symmetry of the lattice~\cite{Benalcazar2017Quantized,serra2018observation}.
Our prediction solely stems from chiral symmetry, or more accurately from the incommensurability of the system with the mechanical molecules, and from the existence of a non-vanishing chiral polarization along the edges. 

This last step concludes the presentation of our method. We are now ready to apply it to actual metamaterials which are not mere analog simulations of  well established  models.

\section{Results}
\label{sec.Results}
We apply our method to predict the existence and locate the topologically protected zero  modes of the 3D printed structure shown in our first figure, Fig.~\ref{fig.reasoning}b.
This metamaterial is a piece of a crystal assembled from repeated units including 6 springs (blue) and 4 beads (red). 
Three beads are located at the end of rotors and are therefore associated to a single degree of freedom.
We know from the Maxwell-Calladine count that the structure is hyperstatic as the number of degrees of freedoms (5) is smaller than the number of springs (6). 
Each unit cell therefore features one state of self-stress.
Our goal is now to locate  possible additional localized zero modes.

We follow our method step by step. 
We hammer all the beads, each rotor once and each bead, associated to two  degrees of freedom twice, in two different directions.
We then compile our results to form the functions $\ket{\psi_i}$, where $i$ here indexes the different degrees of freedom.
In the SI, we show the extent of each perturbation defining the mechanical molecules shown in Fig.~\ref{fig.reasoning}d.
Now that we have identified the bulk mechanical molecule (viz. the atomic-limit unit cell), we  notice that it does not fully tesselate our metamaterial. 
Therefore zero energy modes can in principle be found at the edges, and, or, the corners of the sample.
To check whether they exist or not, we need to compute the average chiral polarization of each molecule both in the bulk and along the edges. 
To do so we apply \eqref{eq.chiralPolmean}
to each  mechanical molecule.
We find that the chiral polarization  differs in the bulk and in the edge regions, but  the amplitude of the jump in $\bm \Pi$ 
is smaller than one lattice spacing. This local change in the chiral polarization
does not correspond to a singularity in the continuum limit and therefore does not translate into any edge state.
In stark contrast, we observe  that the bottom and left edges feature two incompatible chiral polarization fields which feature a positive-divergence singularity at the bottom left corner. 
This observation implies that this corner must host a localized floppy mode.
In order to check our prediction, we shake our metamaterial at increasing frequencies and find that the first vibrational mode is indeed localized at the bottom left corner as illustrated by the superimposed image sequence of Fig. ~\ref{fig.reasoning}e.
The finite frequency of the vibration mode is due to the finite stiffness of the rotors' design.
In order to further confirm our predictions we perform FEM simulations, compute the vibration spectrum of the full structure  and indeed find that the lowest energy mode is a floppy mode located at the bottom left corner, see SI. 
We note that we have been extra cautious when measuring the chiral polarization of all mechanical molecules. An even faster protocol would have consisted in probing the chiral polarization of five molecules only: one in the bulk and one per edges.
The experimental measurements reported in this section  confirm the predictive power and the robustness of our method. 
We can detect experimentally all the  topologically-protected  zero  modes which live at the edge and corners of  unknown mechanical networks without resorting to any theoretical model.

\section{Discussion}
\label{sec.discussion}

To close this article we comment on the fundamental and practical implications of our findings.
From a fundamental perspective,
We note that the model studied in Fig.~\ref{fig.checkerboard} is the paradigmatic example of a Higher order Topological Insulator (HOTI)~\cite{Benalcazar2017Quantized}.
HOTIs are usually characterized by elaborated invariants of tight-binding Hamiltonians such as the nested Wilson loops~\cite{Benalcazar2017Quantized} or mirror-graded winding numbers~\cite{Neupert2018Topological}, which rely on crystalline symmetries. 
However, in Fig.~\ref{fig.checkerboard}d, we show that even though the spatial average of $\bm \Pi$ is not a topological marker of HOTIs, its local measurement 
provides a very efficient tool to experimentally predict and locate their associated corner states. 
In mechanics, higher order topological metamaterials are fundamentally chiral, therefore, as in electrostatics, the two only  ingredients needed to detect localized zero-energy states (i.e a local excess of chiral charges) are (i) the incommensurability of the mechanical  molecules with the global structure and (ii) a local divergence of the chiral polarization field.

From a more practical perspective, 
reviving Kelvin's wisdom in the context of topological mechanics has allowed us to provide a simple and practical method to identify chiral topological phases and detect their zero-energy modes. 
    We stress in particular the relevance of our approach to probe the existence of states of self-stresses that would escape any form of inspection based on (linear) spectral measurements. 
At this stage, it is also worth recalling that our method is not specific to periodic lattices. 
The very concepts of mechanical molecules, and chiral polarization exist in disordered networks as well, and can be be measured following the exact same procedures and formulas.

We conclude with two final remarks. 
We don't believe that Kelvin's tenet should be opposed to metamaterial design. 
On the contrary, we expect that our experimental method could be effectively used as an effective optimization tool. 
For instance we can think of genetic evolution strategies to evolve the structure of mechanical networks according to the measurement of their chiral polarization, either to avoid, or promote,  states of self-stress as local precursors of non-linear response and failure~\cite{Paulose2015Selectivebuckling}.

Finally, beyond the specifics of mechanics, our method readily applies to any form of chiral matter: 
from photonic metamaterials where the response to light impulses has already been used to characterize Hamiltonian models~\cite{cardano2017detection,maffei2018topological,St-Jean2017Lasing-in-}, to acoustic~\cite{Ni2019Observatio,serra2018observation,Weiner2020ThirdOrderAcoustic}, microwave~\cite{Bellec2014Manipulation}, and  electrical circuits~\cite{DiVentra2022CustodialChiral,Imhof2018Topolectri,kotwal2021active}.

\begin{acknowledgments}
This work was supported by ANR grant WTF, Idex ToRe project and by the European Research Council under grant agreement 852587.
\end{acknowledgments}

\appendix
\section*{Supplementary Information}

In the following we provide the theoretical basis for the electrostatic analogy in mechanics, we work out in detail the one dimensional case, and we further discuss the methodology proposed in the main text.
In the supporting information (SI) we present a detailed validation for the one and two dimensional systems, using experiments, FEM simulations, and linear simulations.
We also expand on the technical details regarding the data analysis, FEM simulations, and sample fabrication. 

\section{From bead-and-spring networks to chiral Hamiltonians}
\label{sec.chiralHamiltonian}
In this section, we provide a brief introduction to the correspondence between the linear dynamics of bead-and-spring networks, and the  quantum dynamics of a particle ruled by a chiral Hamiltonian. 
More specifically, we first define the linear chiral Hamiltonian $\mathcal H$. 
Our derivation is alternative to the  original correspondence introduced by Kane and Lubensky in Ref.~\cite{Kane2014Topological}.

Let us consider a collection of $N$ beads of mass $m$ and $N_c$ springs of stiffness $k$ in $d$ dimensions.
We denote $\bm u = (\bm u_1, \bm u_2, ..., \bm u_N)$ the vector of displacements, and $\bm e=(e_1,e_2,...,e_{N_c})$ the vector of elongations.
$\bm u$ and $\bm e$ are geometrically related by the linear relation
\begin{equation}
   \bm e= \mathcal Q^{\text{T}}\bm u, 
\end{equation}
 where $\mathcal Q^{\text{T}}$ is the so called compatibility matrix. For a linear chain $\mathcal Q^{\text{T}}$ is nothing else but the discrete difference operator. 
Its transpose $\mathcal Q$ is the equilibrium matrix. It relates forces acting on the beads $\bm f = (\bm f_1, \bm f_2, ..., \bm f_N)$ to spring tensions $\bm t=(t_1,t_2,...,t_{N_c})$ as 
\begin{equation}
\bm f=\mathcal Q \bm t.    
\end{equation}

Newton's equations, together with Hooke's law $\bm t=k\bm e$, define the dynamics of the beads in the linear response regime:
\begin{equation}
    \ddot{\bm u} = -\frac{k}{m}\mathcal Q\mathcal Q^{\text{T}}\bm u.
    \label{eq.harmonicOscillatorU}
\end{equation}
In the simplest case of a linear collection of beads connected by springs 
$\mathcal Q\mathcal Q^{\text{T}}$ reduces to the discrete Laplace operator. It is worth noting that 
although less standard,  the dynamics of the mechanical structure can be expressed as a function of the elongation variables simply by using the compatibility relation:
\begin{equation}
    \ddot{\bm e}=-\frac{k}{m}\mathcal Q^{\text{T}}\mathcal Q\bm e.
\end{equation}
These two equations can be combined into a single differential equation in terms of the joint vector of displacements and elongations $\ket{\Psi}=(\bm u, \bm e)$:
\begin{equation}
    \left(\partial_t^2-\mathcal H^2\right)\ket{\Psi}=0,
\label{eq.dynamical2}
\end{equation}
where $\mathcal H=\begin{pmatrix}
0&\mathcal Q\\\mathcal Q^{\text{T}}&0
\end{pmatrix}$, and units are chosen such that $k/m=1$ for simplicity. 
The operator on the left hand side can be recast as the product of two commuting operators $\left(\partial_t^2-\mathcal H^2\right)=-\left(i\partial_t-\mathcal H\right)\left(i\partial_t+\mathcal H\right)$, thus \eqref{eq.dynamical2} is equivalent to the Schrödinger equation
\begin{equation}
    i\partial_t\ket{\Psi}=\mathcal H\ket{\Psi}.
    \label{eq.schrodinger}
\end{equation}
\subsection{Chiral Symmetry and electrostatic analogy} 
\label{sec.EManalogy}
At this stage it is crucial to note that $\mathcal H$ is a chiral Hamiltonian. 
Chiral symmetry is defined by the anticommutation relation 
$\left\{\mathcal C, \mathcal H\right\}=0$, where the unitary chiral operator $\mathcal{C}$ is represented in the present case 
by the diagonal matrix  $\mathcal C_{ij}=\delta_{ij}$ for $i\leq dN$ and $\mathcal C_{ij}=-\delta_{ij}$  otherwise. 

The chiral symmetry of the bead-and-spring dynamics is central to our analysis.
We can readily state some straightforward consequences of this fundamental symmetry. 
Firstly,  chiral symmetry implies that for each eigenstate $\ket{\Psi_E}$ of finite energy $E$,
there exists a chiral partner $\ket{\Psi_{-E}}=\mathcal C \ket{\Psi_E}$ of opposite energy $-E$. 
The spectrum of $\mathcal H$ is thus symmetric.
Taking the opposite sign of $\mathcal H$ in \eqref{eq.schrodinger} therefore describes the same physics.
However, zero energy modes don't come by pairs of eigenstates, and are eigenstates of 
$\mathcal C$ with eigenvalues $+1$ (resp. $-1$) for floppy modes (resp. self-stress states).  
Secondly, by virtue of the Maxwell-Calladine index theorem, isostatic systems correspond to those having as many constraints as displacement degrees of freedom, {\it i.e.} a null total chiral charge 
defined as 
$\tr(\mathcal C)=0$~\cite{guzman2020geometry}.

Finally, the structure of $\mathcal C$ begs for an electrostatic analogy where the displacement 
degrees of freedom  plays the role of positive charges while the elongations play the role of negative charges, and positive charges only interact with negative charges. 
At this stage the analogy is merely superficial, we give it some substance in the following sections.

\section{Theoretical basis for the model-free characterization of mechanical structures}
\subsection{Wannier states and mechanical, molecules}
We first provide a theoretical benchmark to validate the model-free measurements discussed in the main text. 
To do so, we recall the definition of some quantities and concepts thoroughly discussed in Ref.~\cite{guzman2020geometry}.
Our starting point is the linear chiral Hamiltonian $\mathcal H$ defined in section~\ref{sec.chiralHamiltonian}, and 
an ensemble of Wanniers states associated to the negative spectrum.
While Wannier states are commonly defined as localized states built out of eigenstates of a single energy band, their definition can be 
extended to a set of bands. 
In the present case we focus on multiband Wannier states associated to the negative energy spectrum. 
They form a spatially localized basis of $\mathcal H$.
This set of Wannier functions is not unique. 
In fact, there are infinitely many possible choices depending on the degree of localization or their complex phases. 
However, in the context of electrons in crystals, 
maximally localized Wannier functions provide a good representation of atomic orbitals~\cite{vanderbilt2018berry}. 
By extension, in our experiments 
we expect the Wannier functions of $\mathcal H$ to define the support of ''mechanical molecules'', 
{\it i.e.} the pair of degrees of deformation and constraints that are dominantly coupled, similarly to an atomic orbital for electrons.

\subsection{How to compute the Wannier functions}
For 2D metamaterials we define the Wannier functions  as the set of functions spanning the negative energy space and  minimizing the  spreading functional
\begin{equation}
    \Omega[\left\{W_i\right\}]=\frac{1}{N}\sum_{j}^{N}\left(\expval{\mathcal R^2}{W_j}-\left|\expval{\mathcal R}{W_j}\right|^2\right),
\end{equation}
where $\mathcal R$ is the position operator.
Alternatives methods such as the matrix pencil method leads to sets of comparable localization~\cite{guzman2020geometry}.

For one-dimensional systems, a maximally localized basis of Wannier functions corresponds to the eigenstates of the projected position operator $\mathcal P\mathcal R_x\mathcal P$, where $\mathcal P=\sum_{E<0}\ket{\Psi_E}\bra{\Psi_E}$. 
Obviously, in a periodic lattice distinct Wannier functions are related by unit-cell translations, $\braket{x}{W_i}=\braket{x-ja}{W_{i-j}}$.
Note that by definition the Wannier functions are normalized: $\braket{W_i}=1$.

\subsection{Charge, location and polarization of mechanical molecules}
We use the electrostatic analogy introduced in Sec.~\ref{sec.EManalogy} to define the charge, the location and the polarization of mechanical molecules.

\subsubsection*{Chiral charge}
We introduce the concept of chiral charge as 
\begin{equation}
    c_{W_i}=\expval{\mathcal C}{W_i}.
\label{eq.chargeW}
\end{equation}
It represents the difference between the average number of deformation degrees of freedoms  ($+1$ charges) and constraints ($-1$ charges) within a mechanical molecule defined by the support of a Wannier function.
\subsubsection*{Chiral molecules}
Similarly we naturally define the  center of mass of the Wannier function as 
\begin{equation}
{\bm r_{W_i}}=\expval{\mathcal R}{W_i}.
 \label{eq.posW}
\end{equation}
In a periodic lattice, the $\bm r_{W_i}$ are deduced from one another by unit-cell translations. 
\subsubsection*{Chiral polarization}
Finally, as in conventional charged systems, we can define the chiral polarization of our mechanical molecules as
\begin{align}
\Pi_{W_i}=2\left(\expval{\mathcal C \mathcal R}{W_i}-c_{W_i}\bm r_{W_i}\right).
    \label{eq.polW}
\end{align}

\subsection{A tutorial example: Theoretical chiral polarization field in the periodic mechanical SSH chain}
\begin{figure}
        \centering
        \includegraphics[width=\columnwidth]{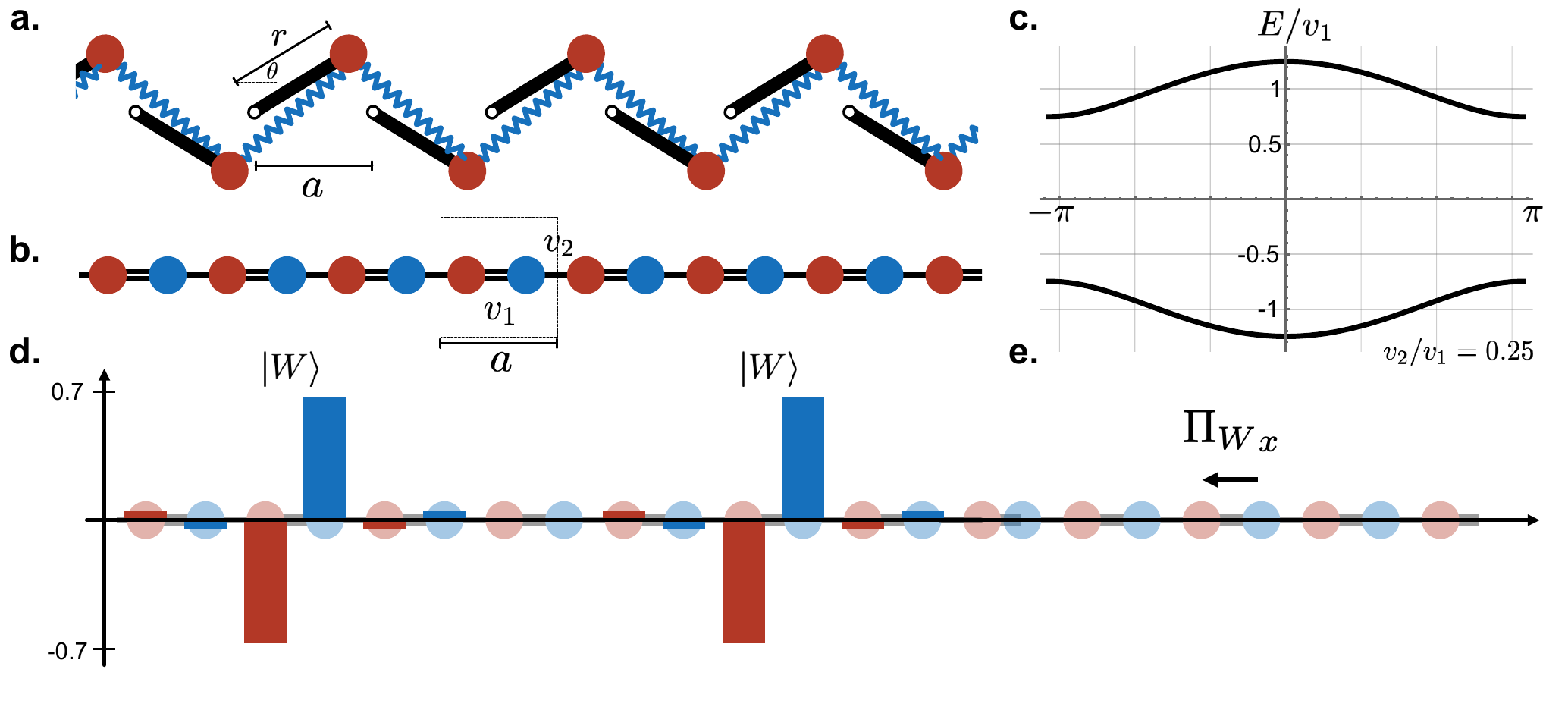}
        \caption{{\bf Periodic SSH model of the mechanical chain.} {\bf a.} The equilibrium configuration is described by the length of the rotor $r$, the unit-cell spacing $a$, and the equilibrium angle $\theta$. {\bf b.} Chiral one-dimensional representation of the mechanical chain. Beads and springs are  symbolized by red and blue circles, respectively. Small perturbations around the equilibrium configuration are described by the linear Hamiltonian $\mathcal H$. It corresponds to a tight-binding model known as the SSH model. In the picture, $v_2<v_1$.
        {\bf c.} Typical spectrum of a periodic chain. The values used are the same as in the experimental setup: $r=25.4$mm, $a=60$mm, and $\theta=\pi/4$.
        {\bf d.} Wannier functions are highly localized on the strongly linked sites (in this case $v_1$). Moreover, Wannier functions are translations of one another.
        {\bf e.} The chiral polarization is translationally invariant and connects the strongly connected sites. Here we represent only the horizontal component.
        }
        \label{fig.SI.theoryMechanicalChain}
\end{figure}
The mechanical chain of rotors under periodic boundary conditions is described by the SSH model~\cite{Kane2014Topological}, see fig.~\ref{fig.SI.theoryMechanicalChain}a,b. 
Exploiting the periodicity of the system, the Bloch Hamiltonian is described in momentum space as:
\begin{equation}
    \mathcal H_{\text{SSH}}(k)=\begin{pmatrix}0& v_1+v_2e^{ika}\\ v_1+v_2e^{-ika}&0\end{pmatrix},
\end{equation}
where the parameters $v_1$ and $v_2$ depends on the specific geometry of the metamaterial. 
They depend on the length of the rotor $r$, the equilibrium angle $\theta$, and the unit-cell length $a$:
\begin{align}
    &v_1=r\sin\theta\frac{a+2r\sin\theta}{\sqrt{a^2+4r^2\sin^2\theta}},\\
    &v_2=r\sin\theta\frac{a-2r\sin\theta}{\sqrt{a^2+4r^2\sin^2\theta}}.
\end{align}
The equilibrium position operator $\mathcal R$ is a diagonal matrix whose entries are the vectors of equilibrium positions of beads and springs:
$\mathcal R = \text{diag}\left(\bm r_{\text{bead}_1},\bm r_{\text{bead}_2},...,\bm r_{\text{bead}_N}, \bm r_{\text{spring}_1},...,\bm r_{\text{spring}_{N_c}}\right)$,
where the equilibrium positions are given by 
\begin{align}
    \bm r_{\text{bead}_j}&=(j a+r\cos\theta,(-1)^jr\sin\theta) ,\\
    \bm r_{\text{spring}_j}&=((j+1/2)a+r\cos\theta,0).
\end{align}
Since the problem in question is essentially one-dimensional, we are only interested in the horizontal component of the position operator. Its Fourier transform takes the simple form:
\begin{equation}
    \mathcal R_x(k)=\begin{pmatrix}r\cos\theta+i\partial_k&0\\0&a/2+r\cos\theta+i\partial_k\end{pmatrix}.
\end{equation}

In an infinite or a periodic chain, the chiral charge linked to a single Wannier function also corresponds to the average chiral charge $\tr(\mathcal C)$ of the whole system:
\begin{equation}
    c_{W_i}=\expval{\mathcal C}{W_i}=\frac{\tr\left(\mathcal C\mathcal P\right)}{\tr(\mathcal P)}=\frac{\tr(\mathcal C)}{2\tr(\mathcal P)}=0,
\end{equation}
where the last equality follows from the isostaticity of the network, the number of degrees of freedom is equal to the number of springs($N=N_c$).
The center of charges in a periodic chain, the $\bm r_{W_i}$, are deduced by one another by a unit-cell translation, Fig.~\ref{fig.SI.theoryMechanicalChain}d.

Finally we can define the chiral polarization of our mechanical molecules using \eqref{eq.polW}. 
Noting that the chiral charge is zero in the isostatic chain, we find that $\expval{\mathcal C \mathcal R}{W_i}$ is origin independent. 
In addition, as the Wannier functions are translations of one another, we can simplify \eqref{eq.polW} as
\begin{equation}
\left({\Pi}_{W_i}\right)_x=2 \expval{\mathcal C \mathcal R_x}{W_i}
=2\frac{\tr(\mathcal C\mathcal R_x\mathcal P)}{\tr(\mathcal P)}=\text{sign}(v_2-v_1)\frac{a}{2}.
    \label{eq.polWPeriodic}
\end{equation}
The above expression can be recast into the compact form 
$\Pi_x=wa-a/2$, with $w=\frac{1}{2}[\text{sign}(v_2-v_1)+1]$. 
In Ref.~\cite{guzman2020geometry} we showed that the integer $w$ is nothing else but the conventional topological index of chiral topological phases, {\it i.e.} the winding number of $\mathcal H$.
Distinct topological phases are therefore characterized by opposite chiral polarizations, Fig.~\ref{fig.SI.theoryMechanicalChain}e. 
\begin{figure}
        \centering
        \includegraphics[width=\columnwidth]{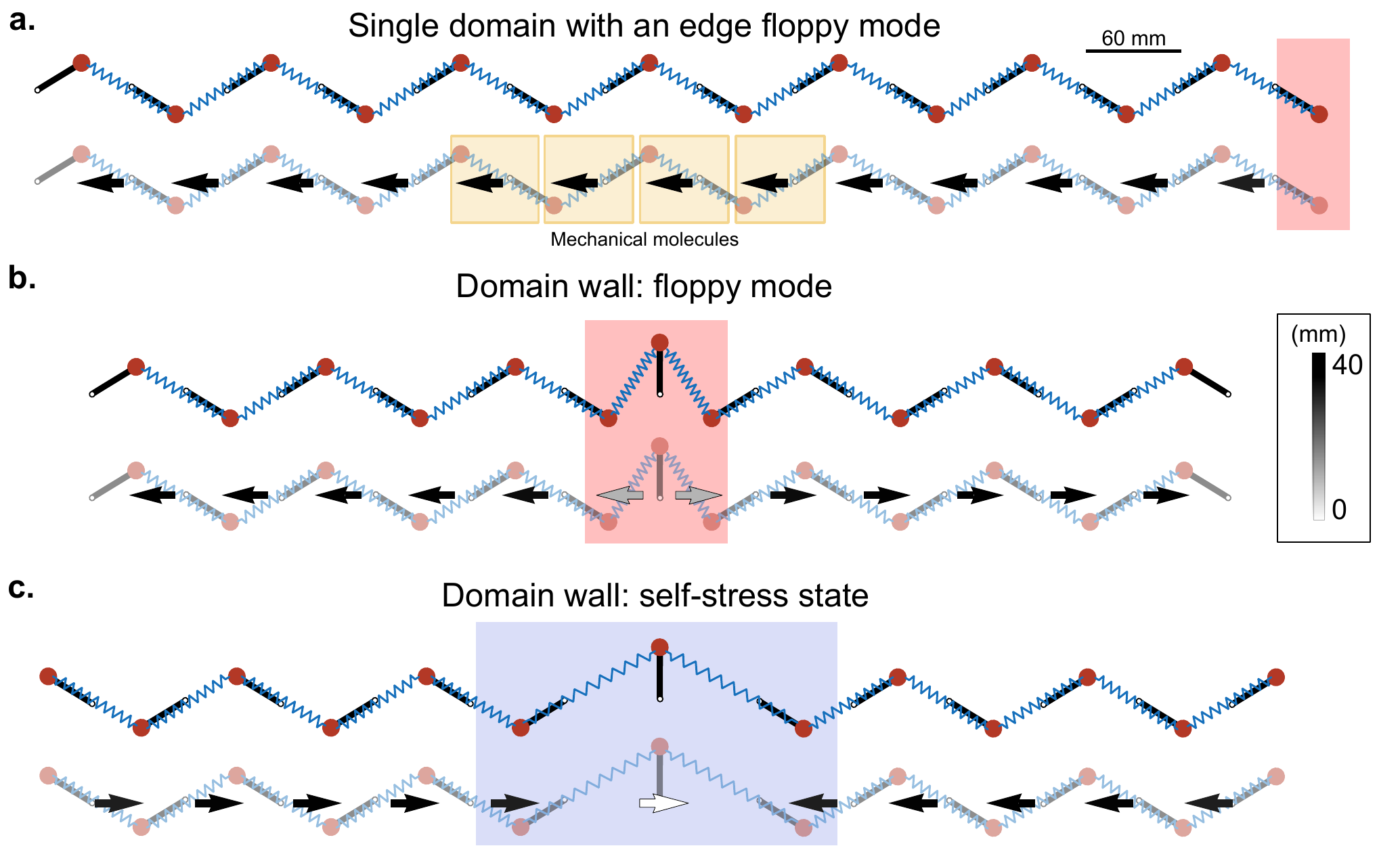}
        \caption{
        Theoretical chiral polarization field for the three examples shown in the main text: a single domain {\bf a}, two domains connected by a floppy mode {\bf b}, and two domains connected by a self-stress state {\bf c}.
        Per each case we show the finite chain (top) and chiral polarization field (bottom).
        The vector field is represented by constant-length arrows with magnitude indicated by the black-and-white colorbar.
        In all three cases we used the same physical parameters as in the experiment: rotor length $25.4$mm, unit-cell length $60$mm, and equilibrium angle $|\theta|=\pi/4$.
        }
        \label{fig.SI.theoryMechanicalChainfinite}
\end{figure}

\subsection{Chiral polarization field from Wannier functions in  finite mechanical chains}

Figure~\ref{fig.SI.theoryMechanicalChainfinite} illustrates the theoretical chiral polarization computed from the numerical values of the Wannier functions in three types of mechanical chains.
Figure~\ref{fig.SI.theoryMechanicalChainfinite}a shows a homogeneous system hosting a floppy mode at one end, representing the experimental chain of Fig.2a. We illustrate the support of the chiral molecules, their center, the localized floppy mode at the right edge and the chiral polarization field.
In Figs.~\ref{fig.SI.theoryMechanicalChainfinite}b and c, we show two heterogeneous chains assembled from two distinct topological phases characterized by opposite chiral polarizations, representing the systems of Figs.3a and e.
We emphasize that, as in electrostatics, a discontinuity of the chiral polarization field implies the existence of an isolated chiral charge at the junction between the two incompatible topological phases. 
This observation is known as the bulk-boundary correspondence and was thoroughly discussed in~\cite{guzman2020geometry}. 
More precisely, a discontinuity of the chiral polarization field of the order of one unit-cell length: $|\Pi_{x}^{\text{left}}-\Pi_x^{\text{right}}|\geq a$ distinguishes two distinct topological phases. 
Floppy modes and self-stress states are identified by source and sink-like discontinuities respectively~\cite{guzman2020geometry}, see 
Figs.~\ref{fig.SI.theoryMechanicalChainfinite}b and c.

\section{Methodology}

From the last section, we know that the whole information about the chiral polarization field relies on the ensemble of Wannier functions. 
They uniquely partition the system into molecules and determine their polarization. 
A natural question then is how to experimentally determine them. As we show below, this is practically impossible. 
However, we can circumvent the need of computing the Wannier functions by exploiting the dynamics of localized mechanical perturbations.

\subsection{The shortcomings of Wannier functions}

In a nutshell, knowing the low-energy eigenstates we can compute the local functions defining the chiral molecules and, ultimately, the chiral polarization field. 
In principle, the low-energy eigenstates could be measured experimentally  by shaking the sample at increasing frequencies. The eigenstates are then identified as the vibrational modes where the magnitude of the fluctuations is locally maximal.
While appealing in theory, this approach is doomed to fail. 
Although the eigenenergies are easily accessible, see SI of ref.~\cite{serra2018observation}, determining a set of low-energy chiral eigenstates is out of reach of any realistic experiment.
Whether due to the measurement accuracy, the intrinsic experimental noise, or the inevitable consequences of mechanical nonlinearities not captured by  $\mathcal H$,
the experimental eigenmodes  always differ from the theoretical modes.
An ensemble of eigenmodes measured in an experiment is generically not chiral: it 
includes a nonvanishing  projection on both 
$\mathcal P = \mathcal P_{E<0}$
and 
$ \mathcal P_{E>0} = \mathbb I-\mathcal P $. 
This inevitable feature makes it impossible to define chiral molecules and their polarization from experimentally measured modes. 
In the next sections, we show how to circumvent this fundamental limitation, and more precisely, how to use local mechanical excitations as effective proxies for Wannier functions. 

\subsection{Dynamics of the Wannier function in the mechanical chain}

To gain some insight, we first study the dynamics of the Wannier functions themselves, and show how their unitary evolution affects the chiral charges and polarizations.
The unitary evolution operator is given by $\mathcal U_t=e^{-it \mathcal H}$. Starting from a Wannier state $\ket{W_i}$, the wave function at time $t$ is given by $\ket{W_i(t)}=\mathcal U_t\ket{W_i}$

Using \eqref{eq.chargeW} we find
\begin{align}
    c_{W_i(t)}&=\expval{\mathcal U_{-t}\mathcal C \mathcal U_t}{W_i}
    =\frac{\tr(\mathcal U_{-t}\mathcal C\mathcal U_t\mathcal P)}{\tr(\mathcal P)}\nonumber\\
    &=\frac{\tr(\mathcal U_{-t}\mathcal C\mathcal P\mathcal U_t)}{\tr(\mathcal P)}
    =\frac{\tr(\mathcal C\mathcal P)}{\tr(\mathcal P)}=0,
\end{align}
where we use the periodicty of $\ket{W_i}$ to simplify the second equality. The third equality simplifies thanks to the commutation relation $[\mathcal P,\mathcal U^t]=0$, and the  fourth equality follows from the cyclicity of the trace operation. 
Ultimately we find that the chiral charge is time-independent and equal to zero in the isostatic SSH chain.

The same reasoning applies to the chiral polarization itself. The chiral polarization is time independent: $({\Pi}_{W_i})_x(t)=\text{sign}(v_2-v_1)\frac{a}{2}$.

\subsection{Numerical validation in the finite mechanical chain}

We now show that the long-time average of the chiral charge, position, and polarization do not fundamentally require initializing the dynamics with a pure Wannier state. 
We numerically show that the dynamic response of any localized mechanical 
perturbation makes it possible to determine the support of the Wannier states (the mechanical molecules), their center, and their chiral polarization. 

\begin{figure}
        \centering
        \includegraphics[width=\columnwidth]{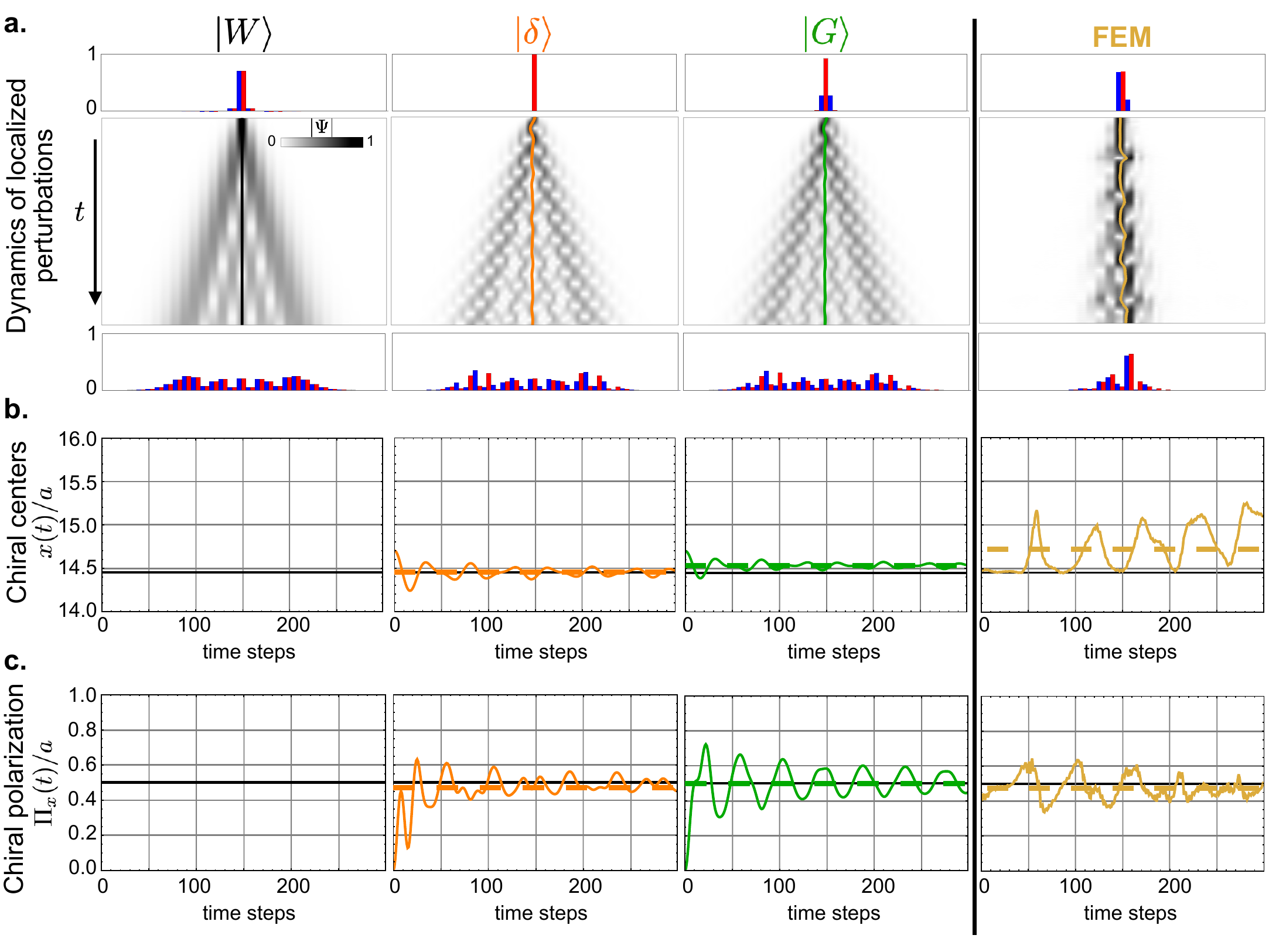}
        \caption{{\bf Local dynamics as proxies of the Wannier functions: Numerical validation in the one-dimensional model.} 
        {\bf a.} Evolution of four distinct localized perturbations in a one-dimensional mechanical chain of 31 rotors. The initial conditions are centered in the middle of the finite system. From left to right: Linear evolution of a Wannier function $\ket{W}$, a fully localized function on one degree of freedom $\ket{\delta}$, and a Gaussian function $\ket{G}$, and a finite element simulation of a single-node displacement. In all three linear cases the initial state spreads with at a rate fixed by the stiffness and the mass of the beads $~\sqrt{k/m}=1$.
        At each time step we compute the chiral center {\bf b} and the chiral polarization {\bf c}, both of them normalized by the unit-cell length $a$.
        For the Wannier case only the time signal is constant. Otherwise, both the position and the polarization oscillate around the expected value.
        The chiral center and polarization are thus defined as the time average of the signal (dashed line).
        Beyond the linear Hamiltonian, FEM simulations display the same behavior. 
        In all four cases we display 300 time steps. For the first three linear cases the time between two consecutive time steps is $dt=0.1$ and $a=1$ (dimensionless units). For the FEM simulation $dt=0.03$ms and $a=60$mm.
        }
        \label{fig.SI.validationSSH}
\end{figure}

As a benchmark comparison, we first detail the one-dimensional mechanical chain of rotors.
In Fig.~\ref{fig.SI.validationSSH} we compare the evolution of Wannier functions, and two types of localized wave functions: a function localized on a single degree of freedom, and a Gaussian function of a width comparable with the unit-cell length $a$.
Regardless of the initial state, all perturbations spread at the same speed, set by the ratio $v_1/v_2$. 
The difference between the two coupling constants $v_1$ and $v_2$ (the dimerization of the chain in the electronic language) results in a biased time evolution, the wave function takes more time to spread over the links associated with the weakest interactions. 
Figure \ref{fig.SI.validationSSH}d shows that all three functions lead to the same value of the time average center of the mechanical molecule  $\langle x(t)\rangle_t $.
Remarkably, this observation also holds for the chiral polarization as seen in Fig.~\ref{fig.SI.validationSSH}e.
Regardless of the initial condition, the chiral polarization oscillates around the theoretical  value defined by the Wannier functions.

We finally compare our linear model to the full FEM resolution of the mechanical problem (rightmost panels of Fig.~\ref{fig.SI.validationSSH}). We study the dynamics of a structure identical to our 3D printed material in response to the displacement of a single rotor at $t=0$. 
We find that both the position of the mechanical molecule and its chiral polarization merely fluctuate around the theoretical value predicted from our simple linear bead and spring model. 
This last example  confirms the robustness of our method to measure the centers of the chiral molecules and their 
chiral polarization using  a local poking experiment.

\subsection{Discussion: Linear numerics versus experiments}
In the previous section we show how the linear dynamics provides an excellent proxy for the Wannier functions in mechanical networks.
In reality, however, the dynamics due to local perturbations are subject to two major effects non-captured by the previous models: damping and non-linear effects.

Damping breaks the chiral formulation presented in section~\ref{sec.chiralHamiltonian}. 
The addition of velocity dependent forces forbids the chiral Hamiltonian decomposition of \eqref{eq.schrodinger}.
Nonetheless, the intrinsic difference between the beads and springs does not depend on any possible dissipation process.
This is where mechanics differs from other chiral platforms, such as electronics or photonics: the sublattice symmetry is intrinsic and not limited to any linear theory.
The simple and natural distinction between beads and springs is enough to accurately compute the chiral polarization field.

From an experimental point of view, the inclusion of damping is even desirable as it filters out the weakest interactions, giving even clearer signals of the strongly coupled degrees of freedom and constraints.

Finally, it is worth pointing out that in all the model-based descriptions, an a priori knowledge of the degrees of freedom is required. For example, the degree of freedom of a rotor is the angle variable $\theta$ and not the displacement vector $\bm u$.
The model-free characterization shown in the main text does not rely on this information.
All the data is treated in the most generic form: displacements and elongations.
In the next section we detail the data reconstruction from experiments and Finite Element (FE) simulations,  without resorting to any linear modeling.

\section{Validation: Two-dimensional mechanical system}
\subsection{Chiral polarization field from Wannier functions in 2D mechanical metamaterial}

Here we provide a benchmark comparison for the experimental measurements of the chiral polarization field shown in Fig.1 of the main text.

\begin{figure}
        \centering
        \includegraphics[width=\columnwidth]{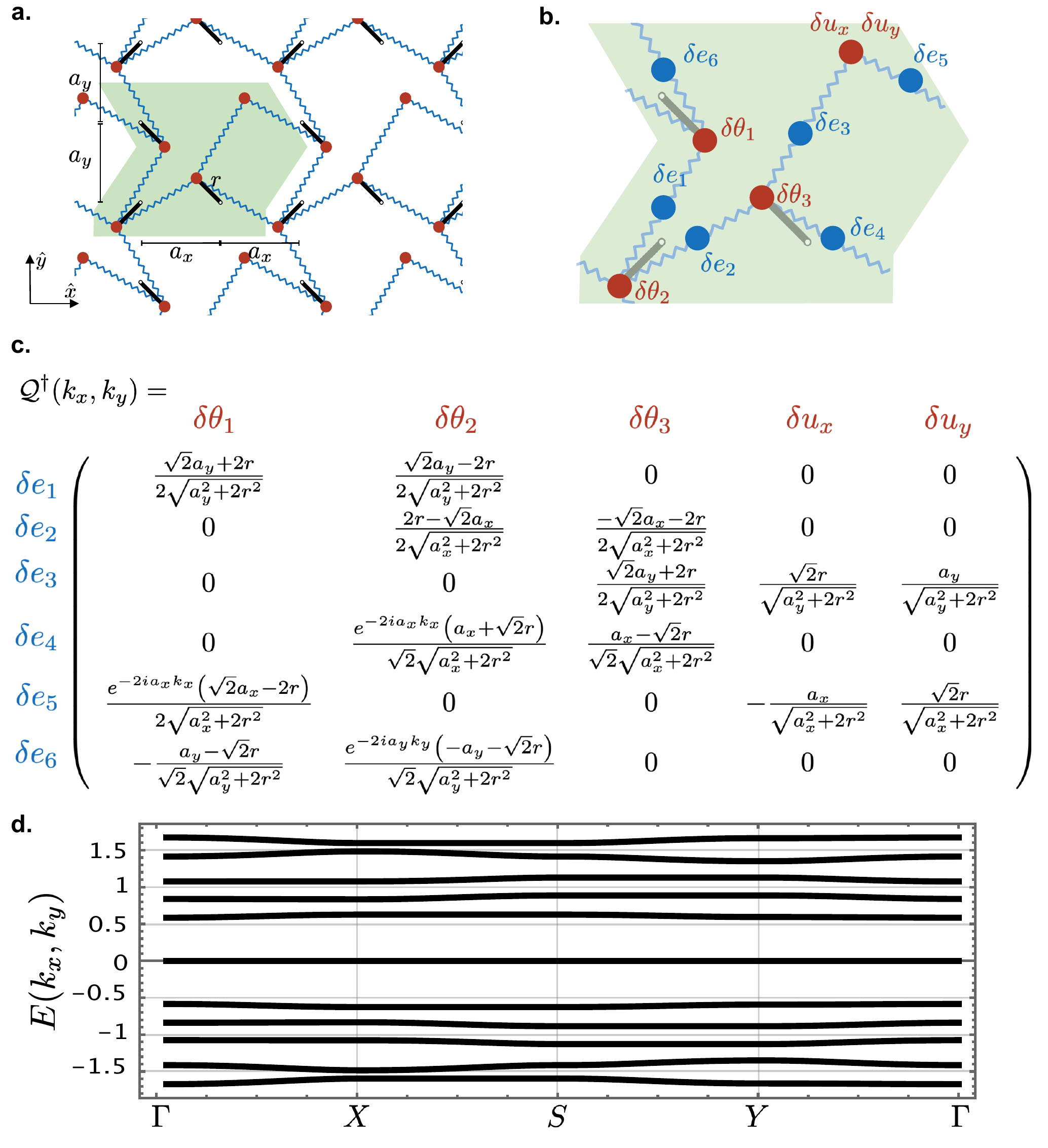}
        \caption{{\bf Linear model of the 2D mechanical metamaterial.}
        {\bf a.} Periodic collection of bead-and-springs modelling the 2D metamaterial shown in the main text. The unit-cell is highlighted in green.
        All the equilibrium angles are assumed to be $\pm\pi/4$ with respect to the $x$ axis.
        {\bf b.} Displacements and elongations in the chiral representation inside the unit cell.
        {\bf c.} Compatibility matrix in momentum space, according to the variables indicated in {\bf b}.
        {\bf d.} Eigenvalues of the Bloch Hamiltonian along the high-symmetry points of the square Brillouin zone.  
        The spectrum is symmetric due to the chiral symmetry. The zero-energy flat band reflects the hyperstaticity of the system.
        }
        \label{fig.SI.linearMechanical2D}
\end{figure}

Figure~\ref{fig.SI.linearMechanical2D} depicts the linear modelization of the system as a collection of beads and springs. 
The unit cell contains four beads and six springs: the metamaterial is hyperstatic.
From the local connectivity we find the compatibility matrix and its Fourier transform, 
Fig.~\ref{fig.SI.linearMechanical2D}c, from which we construct the Hamiltonian $\mathcal H$.

In momentum space, the compatibility matrix is rectangular as  there is one non-compensated constraint in each unit cell. The corresponding Hamiltonian, therefore, enjoys a zero-energy flat band, Fig.~\ref{fig.SI.linearMechanical2D}d.
The system is  hyperstatic, and the number of self-stress states is extensive in the bulk.

\begin{figure}
        \centering
        \includegraphics[width=\columnwidth]{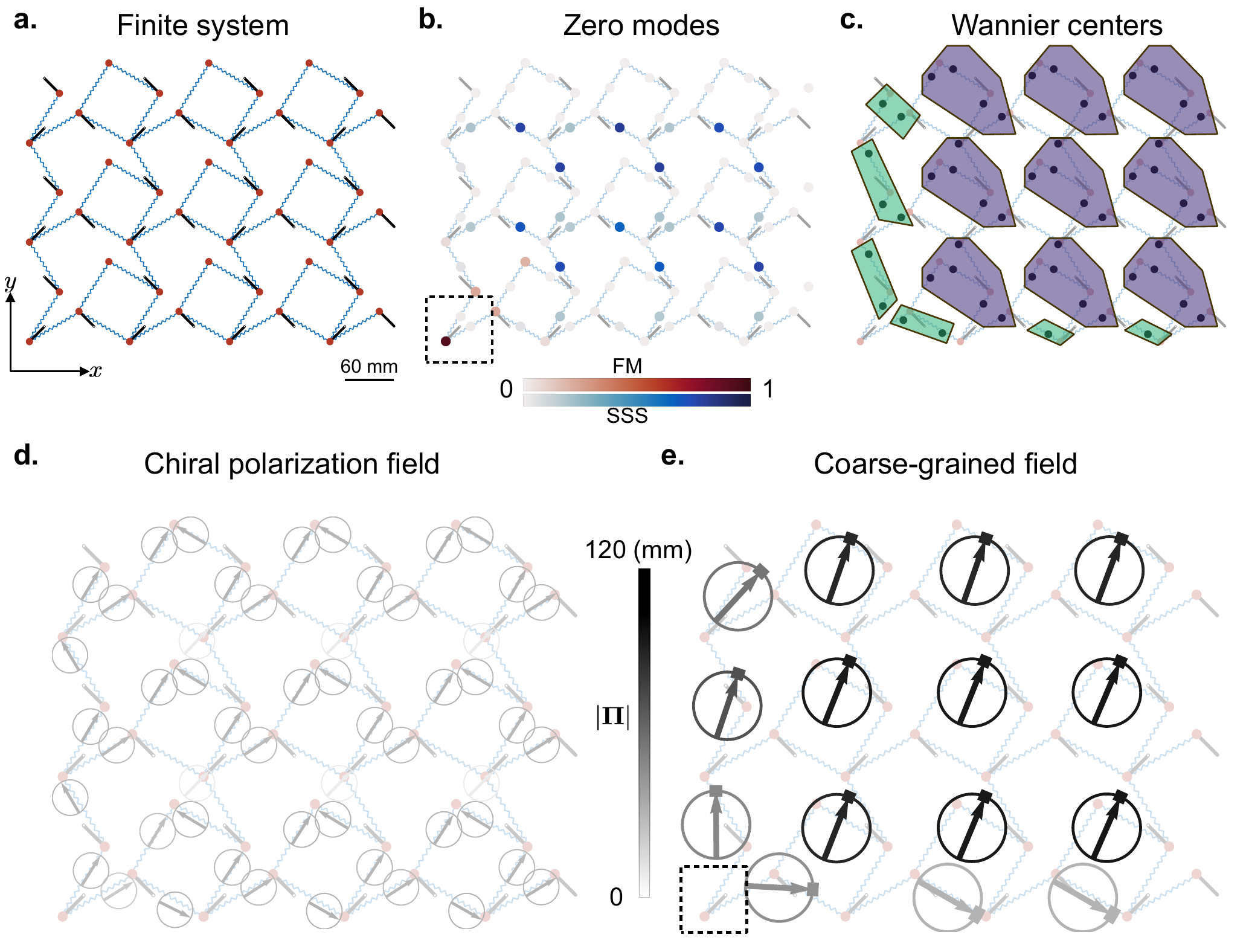}
        \caption{{\bf Model-based chiral polarization field of the 2D mechanical metamaterial.}
        {\bf a.} Finite system of $3\times 4$ unit cells, as used in the experimental setup.
        {\bf b.} Zero-energy mode distribution. The color indicates the weight on degrees of freedom (red) and constraints (blue). The bulk presents an extensive number of self-stress states, whereas only one floppy mode is localized in the bottom-left corner (dashed square).
        {\bf c.} Distribution of Wannier centers (black dots). Beginning from the bulk, we group the centers by 1) respecting the periodicity of the sample (i.e. groups of 5 Wannier centers) and 2) minimizing the spreading among them. These bulk molecules are highlighted in violet.
        The remaining centers are grouped into small edge molecules (green) according to their proximity.
        {\bf d.} Chiral polarization field issued from all the Wannier functions. For readability, we represent the vector field with arrows of fixed length enclosed by a circle. The magnitude of the vectors is depicted by the gray scale and the angular deviations are represented by the thicker portion of the circle.
        {\bf e.} Coarse-grained chiral polarization field. 
        In the bottom-left corner (dashed square) the discrete divergence is larger than the lattice spacing: $\Delta=207.9\text{mm}>120\text{mm}$. The discontinuity signals the presence of the localized floppy mode.
        }
        \label{fig.SI.theoreyCPolMech2D}
\end{figure}

In addition to the bulk states of self-stress, a finite system of $3\times4$ unit cells, Fig.~\ref{fig.SI.theoreyCPolMech2D}a, hosts a localized floppy mode in the bottom-left corner, Fig.~\ref{fig.SI.theoreyCPolMech2D}b.
In contrast to the one dimensional case, for which each unit-cell mapped exactly to one Wannier function, here the mapping is no longer bijective: each unit cell contains several Wannier functions.
This is explicitly shown by the distribution of Wannier centers, Fig.~\ref{fig.SI.theoreyCPolMech2D}c.
Each Wannier function defines its own polarization, Fig.~\ref{fig.SI.theoreyCPolMech2D}d,
leading to a discontinuous field at the sub unit-cell scale.
The spatial distribution of the centers hints towards a coarse-graining procedure: we add the chiral polarization from Wannier functions whose centers are separated by less that one lattice spacing.
The coarse grained field is shown in fig.~\ref{fig.SI.theoreyCPolMech2D}e, indicating a discontinuity on the bottom-left corner, at the scale of the unit-cell size, where the zero zero-energy mode is located.

We can quantify this discontinuity through the discrete divergence of the chiral polarization field, $\Delta$.
We define this quantity in the corner as $\Delta = |(\bm \Pi_{\text{top}}-\bm \Pi_{\text{corner}})\cdot \hat y+(\bm \Pi_{\text{right}}-\bm \Pi_{\text{corner}})\cdot\hat x|$, where top and right refer to the molecules next to the corner. Since no Wannier function is defined in the corner, we have $\bm \Pi_{\text{corner}}=\bm 0$.
From the coarse-grained field we obtain that $\Delta=207.9\text{mm}>120\text{mm}$.

\subsection{Numerical validation in 2D mechanical system}

\begin{figure}
        \centering
        \includegraphics[width=\columnwidth]{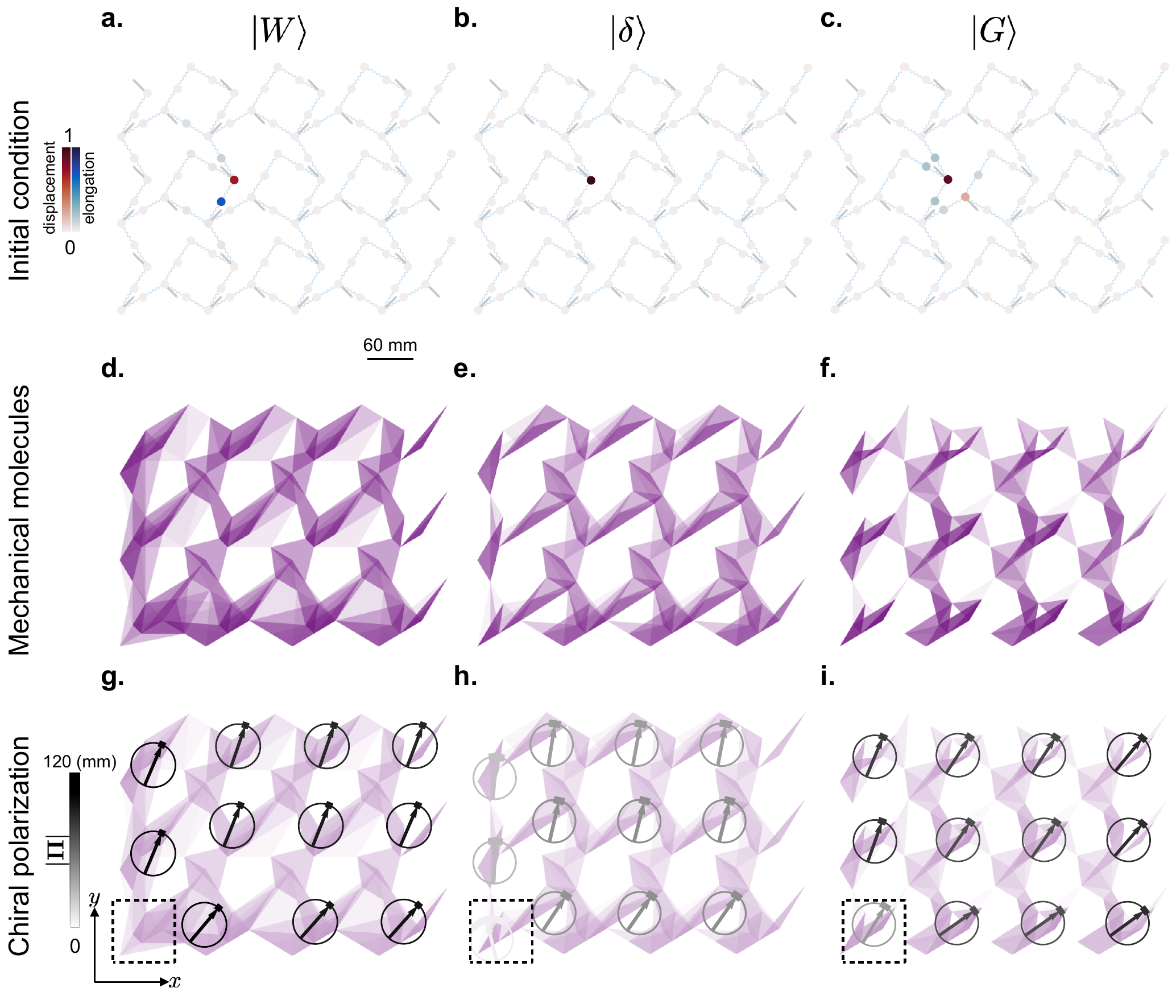}
        \caption{{\bf Local dynamics as proxies of the Wannier functions: Numerical validation in the two-dimensional model.} 
        Comparison of the chiral molecules and chiral polarization field issued from the dynamics of Wannier $\ket{W}$, fully localized $\ket{\delta}$, and Gaussian functions $\ket{G}$ over the same system.
        One example of each initial condition is depicted in {\bf a}, {\bf b}, and {\bf c}.
        The fully localized functions are taken only over the beads (degrees of freedom). The Gaussians have a standard deviation of $\sigma=20$mm.
        {\bf d, e, f.} Extent of the perturbations represented by the polygons for which the Mahalanobis distance $d$ is smaller than 1.8. This is done for every time step with polygons of light opacity.
        Thus, the darker regions are the most persistent.
        {\bf g, h, i.} Coarse-grained chiral polarization field for each case. The color of the arrow is linearly related to its magnitude.
        By definition, the Wannier functions do not span the regions of zero-energy mode, explaining the lack of polarization in the bottom-left corner.
        The discontinuity in the field matches the qualitative picture obtained from the spectral computation of fig.~\ref{fig.SI.theoreyCPolMech2D}e.
        }
        \label{fig.SI.validationHOTI}
\end{figure}

Following the same benchmark procedure as in our first 1D example, we  compare the full chiral polarization field obtained from three types of localized initial conditions: Wannier functions, perturbations localized on one bead, and Gaussian functions, Fig.~\ref{fig.SI.validationHOTI}a.
We compute the time evolution of the three initial conditions using the same Hamiltonian $\mathcal H$ for a finite system of $3\times 4$ unit cells, and choose units such that $k/m=1$.

Unlike the one-dimensional case, here the Wannier functions do not lead to time-independent chiral moments. Instead, they evolve in time.
The same observation holds for the other two initial conditions.
The centers explore confined areas, the beads and springs enclosed in these regions  define the mechanical molecules.
To identify the beads and springs within a mechanical molecule in 2D and 3D metamaterials, we use the Mahalanobis distance associated to the distribution $\Psi(t)$ and construct polygons spanning the strongly related sites. 
For details, see section~\ref{methods.DataReconstruction}\ref{methods.2DmechanicalSystem}.

While the exact shape of the polygons depends on the specifics of the initial perturbation, they all reveal a very similar underlying structure: the mechanical molecules.
We can now compute the associated  coarse-grained chiral polarization field averaged over time, Fig.~\ref{fig.SI.validationHOTI}g, h, and i.
In all three cases, we find chiral polarization fields with the same gross features and in particular sharing  the same discontinuity at the bottom-left corner: the signature of a topologically protected zero corner mode, Fig.~\ref{fig.SI.theoreyCPolMech2D}e.

As before, we can compute the discontinuity of the coarse-grained chiral polarization field, in the corner highlighted by the dashed square in the bottom row of Fig.~\ref{fig.SI.validationHOTI}. For the Wannier case, $\bm \Pi_{\text{corner}}=\bm 0$.
For the three cases explored, namely Wannier, fully localized, and Gaussian, we obtain respectively $\Delta=232.9\text{mm},\; 54.4\text{mm},\;93.3\text{mm}$, larger (or at least of the order) of the corner region/smallest molecule $\sim 60\text{mm}$.

\section{Definition of the wave functions and mechanical molecules   from raw experimental and numerical data}
\label{methods.DataReconstruction}

In this section we detail the data analysis carried out from the raw data of displacements and elongations in both the one- and two-dimensional mechanical systems discussed in the main text.

\subsection{Mechanical chain}

Both the simulations and experiments give us access to the planar displacements of the beads, $\bm u_i(t)=(u_{i,x}(t),u_{i,y}(t))$, with $i$ indexing each bead.
If two beads, say $i$ and $j$, are connected by a spring, its elongation is computed as
\begin{equation}
    e_{i,j}(t)=|(\bm r^{\text{eq}}_{j}+\bm u_{j}(t))-(\bm r^{\text{eq}}_{i}+\bm u_{i}(t))|-|\bm r^{\text{eq}}_{j}-\bm r^{\text{eq}}_{i}|,
\end{equation}
with $\bm r^{\text{eq}}_i$ the equilibrium position of the $i$-th bead.

\begin{figure}
        \centering
        \includegraphics[width=\columnwidth]{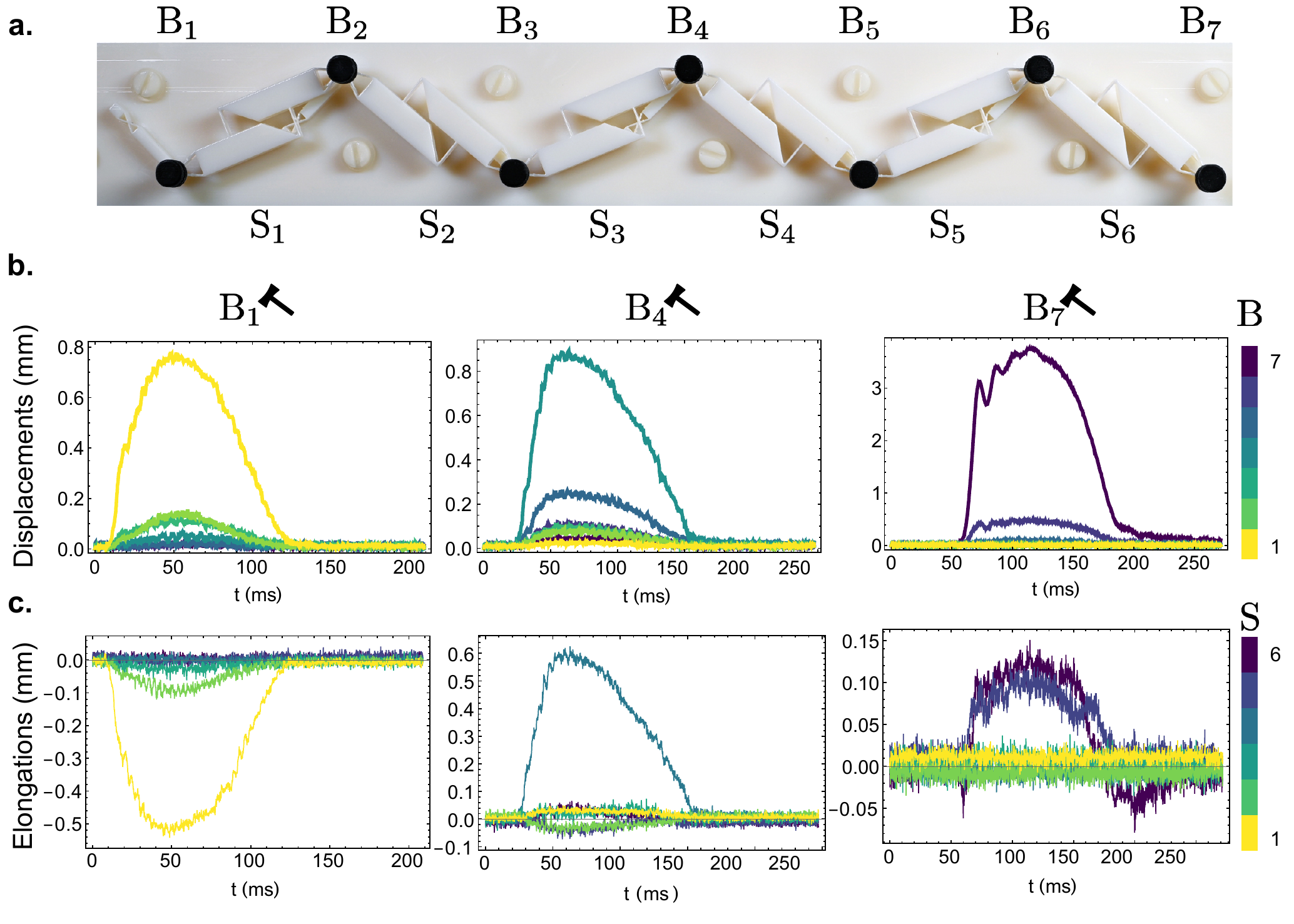}
        \caption{{\bf Raw displacements and elongations from local perturbations in the mechanical chain (Experiments).} 
        {\bf a.} Mechanical chain with labels for beads $\text{B}_i$ and springs $\text{S}_i$. The floppy mode is hosted on bead $\text{B}_7$.
        A local excitation leads to a displacement ({\bf b}) and an elongation ({\bf c}) response. Here we illustrate three distinct perturbations on beads $\text{B}_1$, $\text{B}_4$, and $\text{B}_7$ (from left to right).
        }
        \label{fig.SI.EXPSSH-DispElong}
\end{figure}

\begin{figure}
        \centering
        \includegraphics[width=\columnwidth]{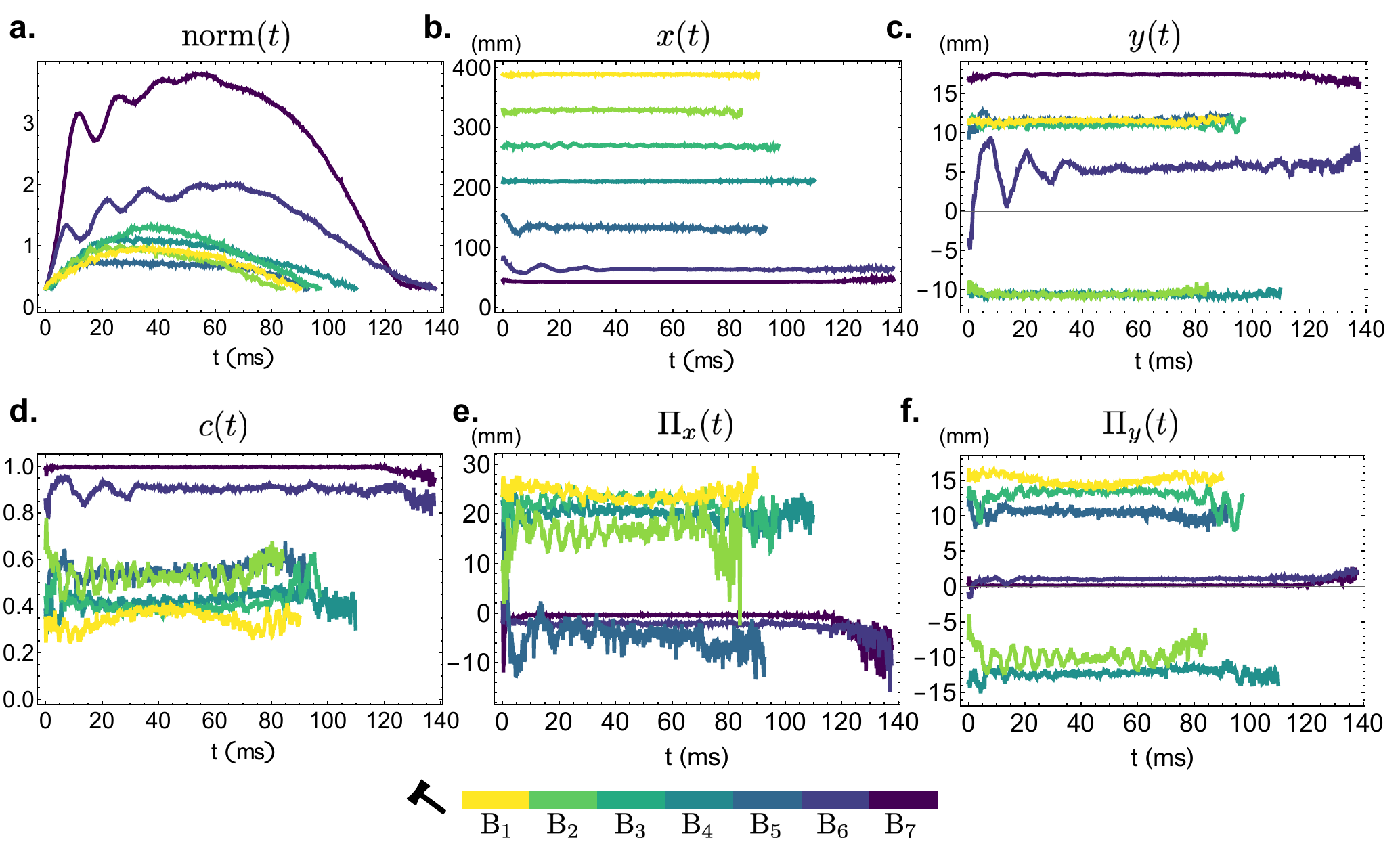}
        \caption{{\bf Moments from local perturbations in the mechanical chain (Experiments).} Norm ({\bf a}), positions ({\bf b} and {\bf c}), chiral charge ({\bf d}) and chiral polarization components ({\bf e} and {\bf f}), for all the different perturbations in the mechanical chain (color).
        }
        \label{fig.SI.EXPSSH-moments}
\end{figure}

Fig.~\ref{fig.SI.EXPSSH-DispElong} illustrates the raw  displacements and elongations of all the nodes for the one-dimensional mechanical chain of rotors.
From a perturbation applied to the bead $i$, we directly compute the wave functions $\Psi_i$, its norm, its center $\mathbf r_i(t)$, the chiral charge, and the chiral polarization, Fig.~\ref{fig.SI.EXPSSH-moments}.
In reality, the metamaterial dissipates mechanical energy. 
As a result the perturbations do not freely propagate but are damped in a finite time.
Beyond the damping time, the signal  corresponds to noise, leading to spurious measurements in terms of the chiral moments.
To filter out this unwanted noise we restrain our measurements to the time intervals in which the norm is higher than a threshold $\epsilon$.
For our experiments, we set $\epsilon_{\text{exp}}=0.3$.

Once the data are filtered, we compute the time average center and chiral polarization. We show in the SI that changing the extent of the time averaging window results in insignificant variations of our observables.

The mechanical molecules are directly extracted from the $N\%$ confidence interval of the distribution $\Psi(t)$. 
The value of $N$ depends on the nature of the experiment/simulation.
In our linear simulations there is no dissipation process. The perturbations hence unboundedly spreads across the whole system, as shown in  Fig.~\ref{fig.SI.validationSSH}. In this case, small values of $N$ (below 30\%) are well suited to capture the local asymmetric response. 
This educated choice is determined by the duration of the simulation.
We typically set the value of  $N$ so that the confidence interval of $\Psi$ does not exceed a distance much larger that the lattice spacing.  We would otherwise define aggregates of the elementary molecules

In the experiments, on the other hand, dissipation quickly dampens the response and high values of confidence (95\%) accurately captures the molecules, see Fig.2b.

The appropriate value of $N$ depends not only on the friction but also on the duration of the experiment, as well as the inherent speed of sound of the material.
In general, a good measurement strategy is to begin with small confidence values and slowly increase it until the first patterns of strongly correlated atoms separated by less than one lattice spacing emerge.

\subsection{2D mechanical system}
\label{methods.2DmechanicalSystem}

In the two dimensional system, the same data acquisition protocol applies (see SI for the raw measurements).

The detection of the molecules, however, relies on a generalization of the concept of confidence interval. 
In this work we chose the Mahalanobis distance~\cite{de2000mahalanobis} $d(\bm q,\Psi(t))$ corresponding to how many standard deviations away the point $\bm q$ is from the mean of the distribution $\Psi(t)$, $\bm r(t)$. 
It is mathematically defined as
\begin{equation}
    d(\bm q, \Psi(t)) = \sqrt{(\bm q-\bm r(t)) S^{-1}(\bm q-\bm r(t))},
\end{equation}
with $S$ being the covariance matrix.
We can then define the spread of the two-dimensional distribution $\Psi(t)$ as the region in space containing all the points $\bm q$ for which $d(\bm q,\Psi(t))$ is smaller than a threshold $\delta$. 
This threshold is akin to the confidence value in 1D. 
Similarly to the 1D case, a good general strategy is to begin with small values of $\delta$, see above.

\begin{figure}
        \centering
        \includegraphics[width=\columnwidth]{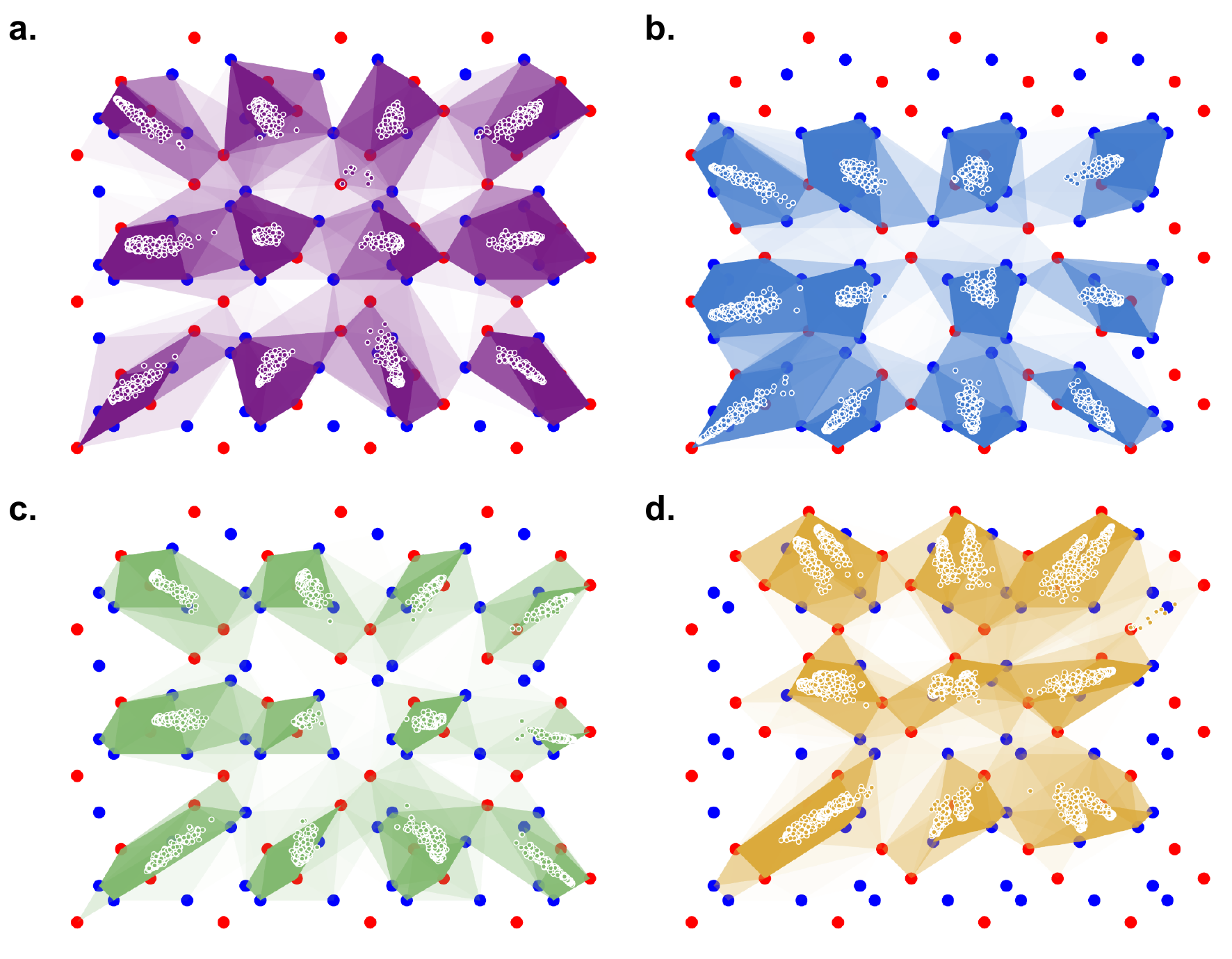}
        \caption{{\bf Mechanical molecules in the experimental poking of 2D mechanical system.}
        For each perturbation $\Psi(t)$ we draw the polygons spanning all the sites (red for beads, blue for springs) for which $d<\delta=0.8$. Each polygon, of light opacity, is superposed as time evolves. Opaque regions thus distinguish the strongly connected sites.
        Each panel ({\bf a} to {\bf d}) contains all the polygons associated to the perturbation of the same bead in different unit cells.
        We also highlight the centers $\bm r(t)$ of the perturbations as points of the same colors as the polygons.
        }
        \label{fig.SI.EXPHOTI-chiralMolecules}
\end{figure}

Fig.~\ref{fig.SI.EXPHOTI-chiralMolecules} shows the spreading of the perturbations extracted from the distance $d$ with a threshold $\delta =1.8$. 
The four panels corresponds to four ensembles of perturbations where only one type of bead is poked (all beads are related by a lattice translation). 
These measurements show that the structure is composed of  $3\times4$ molecules, all of them being revealed by the purple, blue, and green perturbations. 
However, the yellow perturbations only define $3\times 3$ polygonal regions. 
The chiral molecules are defined by the ensemble of beads and springs lying inside the superposition of all the polygonal regions. 
Correctly pairing the perturbations amounts to superimposing the colored polygons and  looking for the largest overlaps. 
In this case, the yellow polygons then contribute to the last three columns of molecules only.
The left edge of the sample is made of mechanical molecules that are distinct from the bulk molecules as clearly seen in Fig.1d.

Fig.1d reflects the final mechanical molecules. We then proceed as usual: we compute the average centers and the average chiral polarization field, see Fig.1e.

For the numerical study of the checkerboard lattice in Fig.4 we use $\delta=1.8$ over 20 time steps to avoid finite-size effects.

\section{Extended Figures}

In this section we provide the complementary plots and measurements obtained from experiments and FEM simulations.

\subsection{Experiments in the mechanical chain}
Fig.~\ref{fig.SI.EXPSSH-DispElong} illustrates the raw  displacements and elongations of all the nodes for the one-dimensional mechanical chain of rotors.

Fig.~\ref{fig.SI.EXPSSH-moments} shows the norm, the center, the chiral charge, and the chiral polarization for each perturbation in the mechanical chain. The data shown are those for which $\text{norm}$ is larger than $\epsilon_{exp}=0.3$.

\begin{figure}
        \centering
        \includegraphics[width=1\columnwidth]{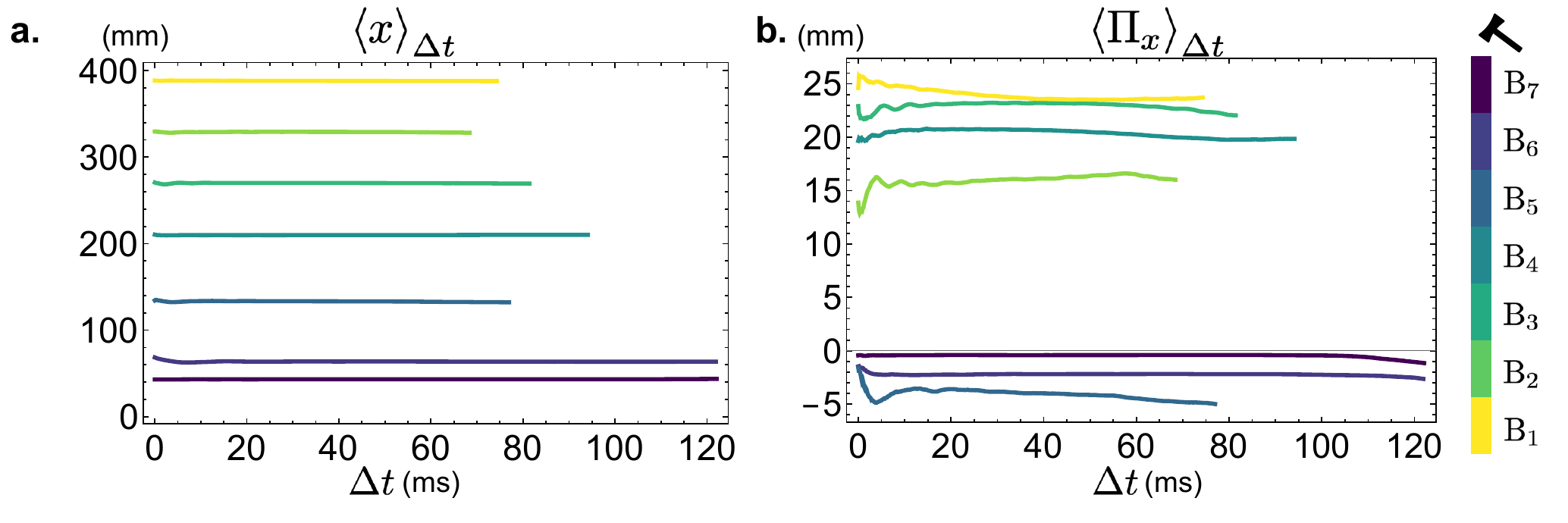}
        \caption{{\bf Time-averaged moments in the mechanical chain (Experiments).} 
        {\bf a.} Due to the constant-like signal of the horizontal position (Fig.~\ref{fig.SI.EXPSSH-moments}b), the time-averaged horizontal position, $\left<\bm x\right>_{\Delta t}$, is  independent of the period $\Delta t$.
        {\bf b.} The average chiral polarization, $\left<\Pi_x\right>_{\Delta t}$ is weakly dependent on $\Delta t$ yet the differences remain much smaller than the unit cell length $a=60$mm.
        }
        \label{fig.SI.EXPSSH-windowAverage}
\end{figure}

Fig.~\ref{fig.SI.EXPSSH-windowAverage} illustrates the dependence of the time-averaged values on the period $\Delta t$. 
In the main text we chose $\Delta t=80$ms.

\subsection{FEM simulations of the mechanical chain}
\begin{figure}
        \centering
        \includegraphics[width=1\columnwidth]{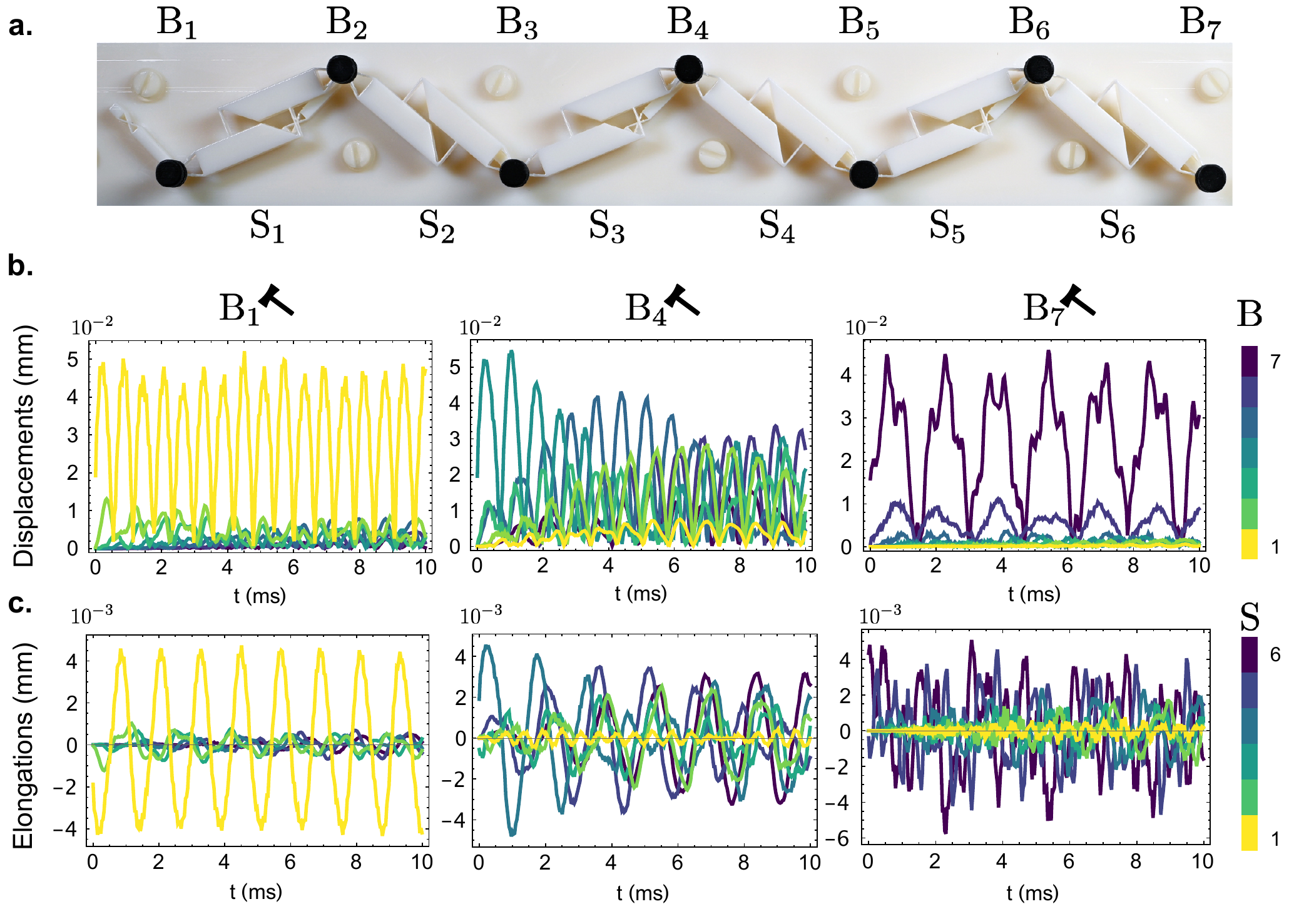}
        \caption{{\bf Raw displacements and elongations from local perturbations in the mechanical chain (FE simulations).} (top) Displacement of each node (color) when hammering the first (left), fourth (middle) and seventh (right) node. (bottom) Corresponding elongations for all the springs (color).}
        \label{fig.SI.FEMSSH-DispElong}
\end{figure}

\begin{figure}
        \centering
        \includegraphics[width=1\columnwidth]{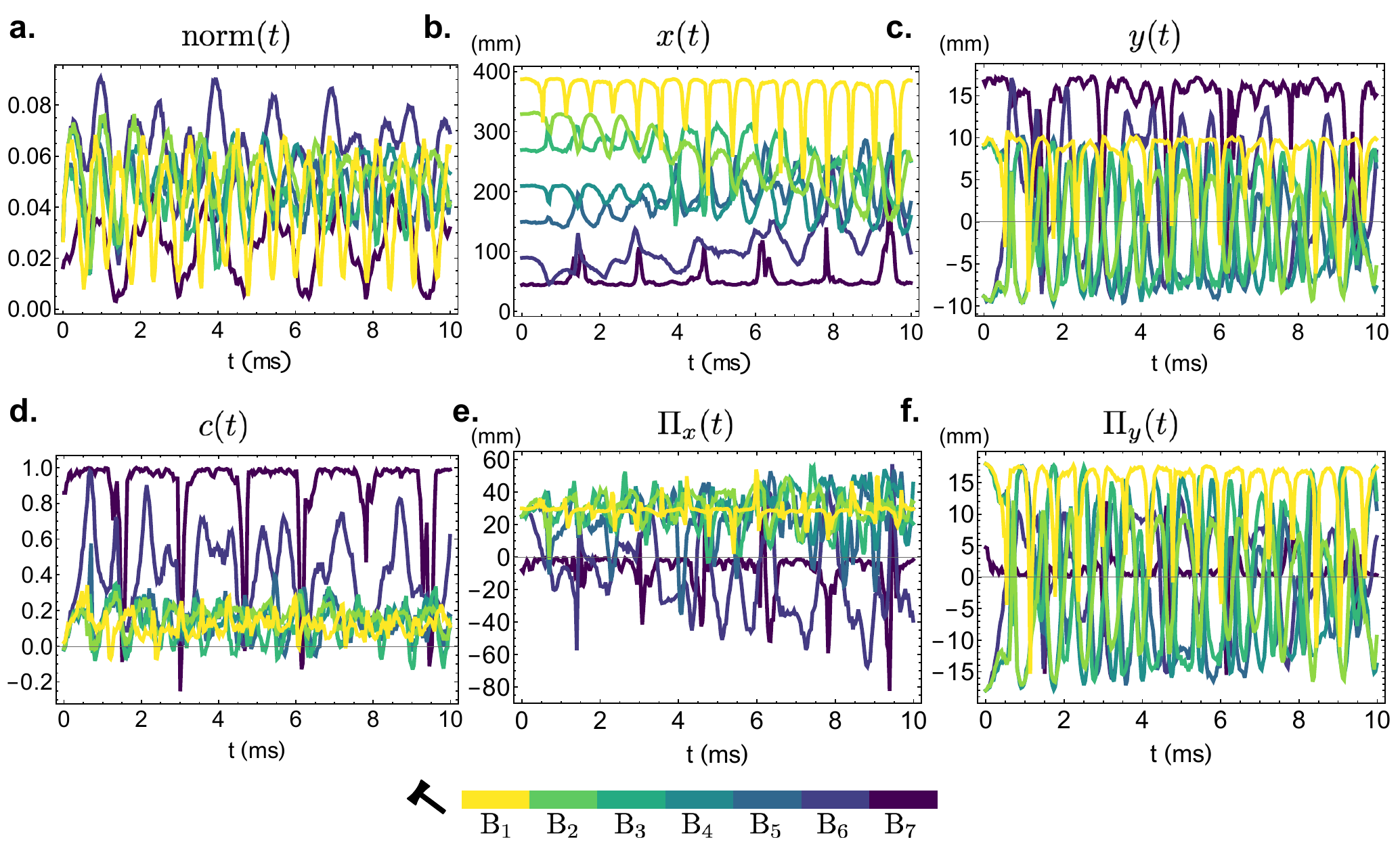}
        \caption{{\bf Moments from local perturbations in the mechanical chain (FE simulations).} Norm ({\bf a}), positions ({\bf b} and {\bf c}), chiral charge ({\bf d}) and chiral polarization components ({\bf e} and {\bf f}), for all the different perturbations in the mechanical chain (color). After ~100 ms from the impulsive perturbation, the response is completely dampen and the signal is noise-dominated.}
        \label{fig.SI.FEMSSH-moments}
\end{figure}

\begin{figure}
        \centering
        \includegraphics[width=1\columnwidth]{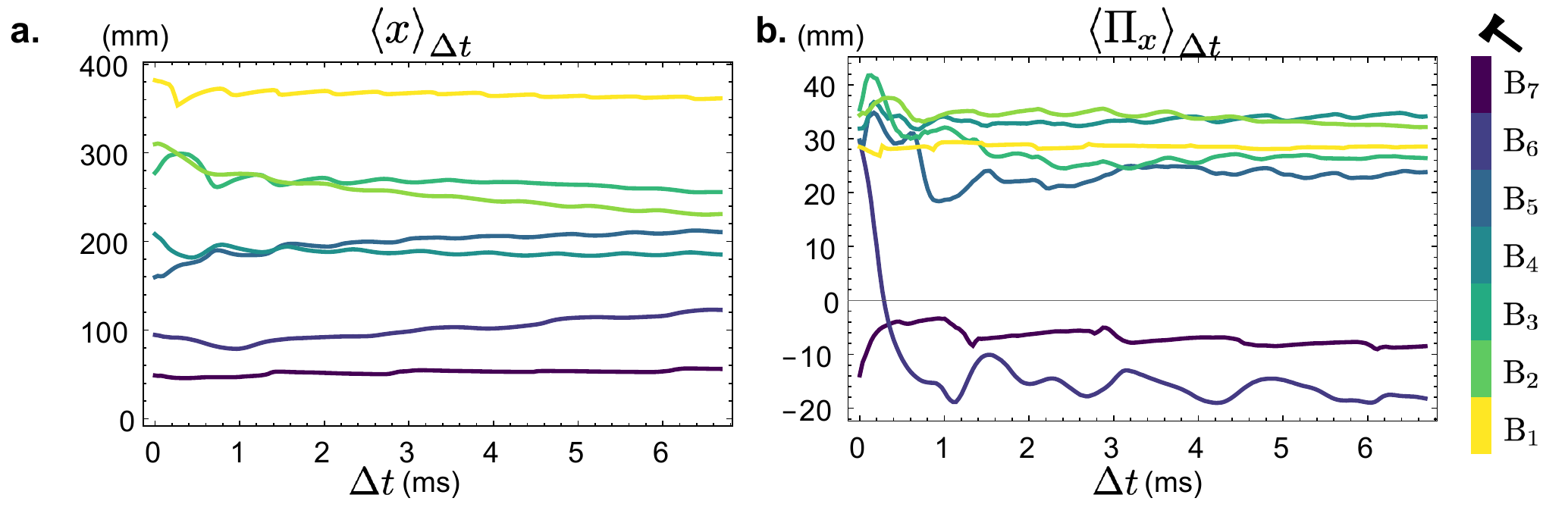}
        \caption{{\bf Time-averaged moments in the mechanical chain (FEM simulations).} 
        Time-averaged horizontal position ({\bf a}) and polarization ({\bf b}) for each perturbation.
        In both cases, due to the absence of friction, the averages are more dependent on the period $\Delta t$ than they are in the experiments.
        }
        \label{fig.SI.FEMSSH-windowAverage}
\end{figure}

Following the same structure, we show the raw data (Fig.~\ref{fig.SI.FEMSSH-DispElong}), the moments (Fig.~\ref{fig.SI.FEMSSH-moments}) filtered by $\epsilon_{text{FEM}}=0.01$, and the dependence of the time averages on the period $\Delta t$ (Fig.~\ref{fig.SI.FEMSSH-windowAverage}).
For the results shown in the main text, we use $\Delta t=2$ms.
As opposed to the experiments, the FEM simulations do not take into account any damping.

\subsection{Experiments in the two-dimensional metamaterial}

\begin{figure}
        \centering
        \includegraphics[width=1\columnwidth]{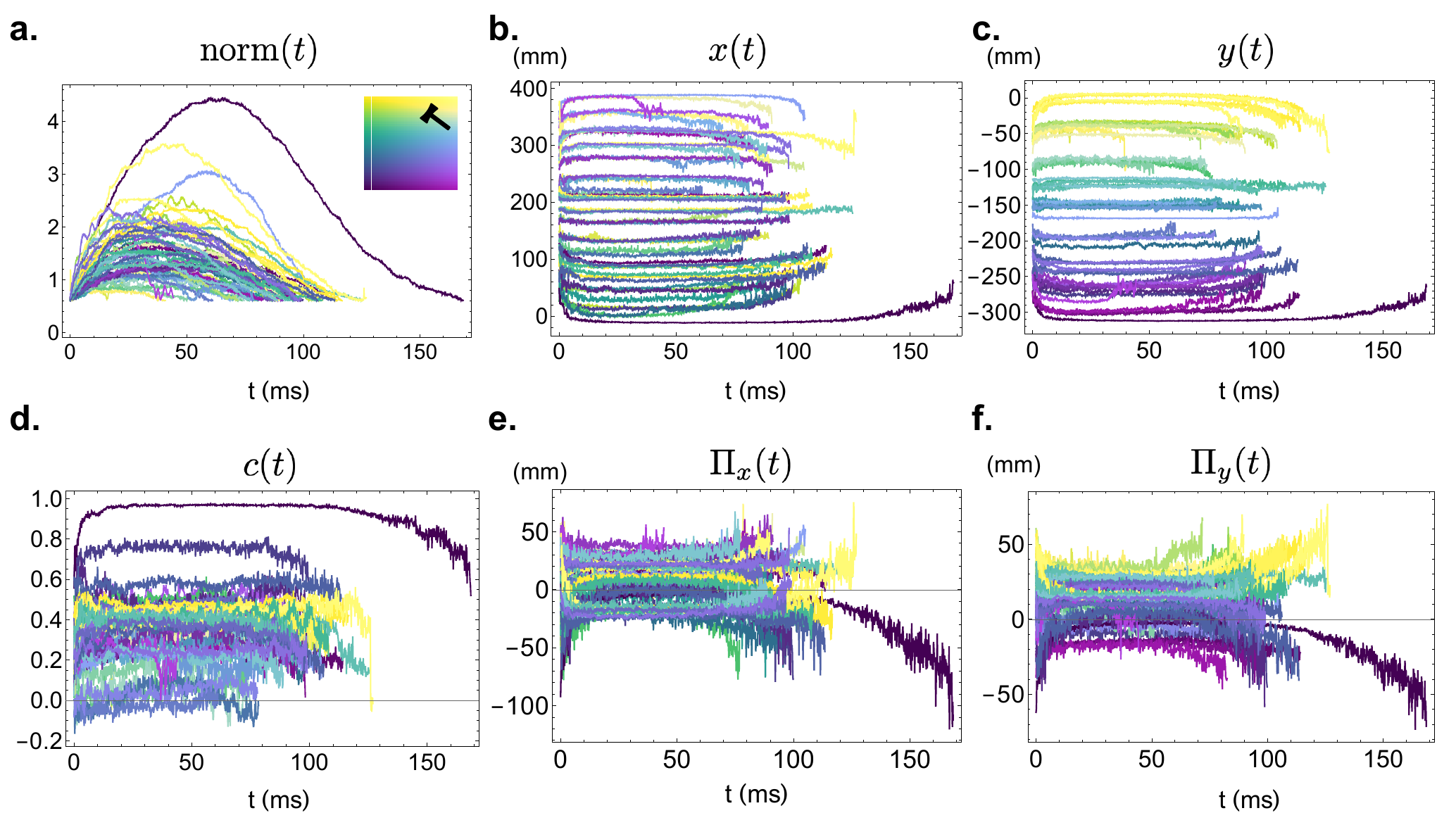}
        \caption{{\bf Moments from local perturbations in the mechanical two-dimensional metamaterial (Experiments).} Norm ({\bf a}), positions ({\bf b} and {\bf c}), chiral charge ({\bf d}) and chiral polarization components ({\bf e} and {\bf f}), for all the different perturbations in the mechanical metamaterial (colormap in the inset of {\bf a}). Only the data for which $\text{norm}(t)>\epsilon_{exp}$ is considered.
        }
        \label{fig.SI.EXPHOTI-moments}
\end{figure}

Fig.~\ref{fig.SI.EXPHOTI-moments} shows the norm, the center, the chiral charge, and the chiral polarization for each perturbation in the two-dimensional metamaterial.
We only show the data for which the norm is larger than $\epsilon_{\text{FEM}}=0.6$.

\begin{figure}
        \centering
        \includegraphics[width=1\columnwidth]{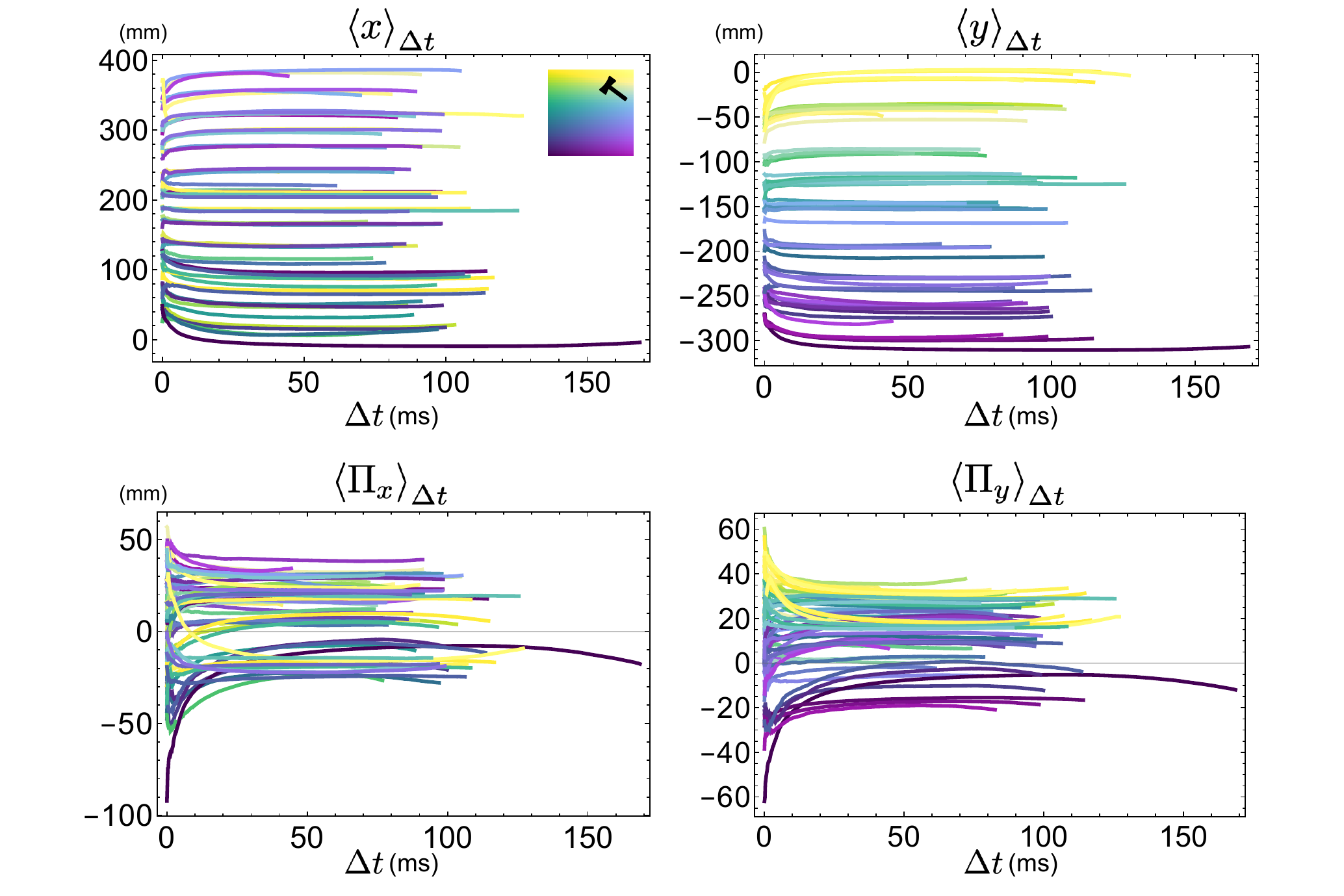}
        \caption{{\bf Time-averaged moments in the mechanical two-dimensional metamaterial (Experiments).} 
        Vertical and horizontal components of the average center and chiral polarization as a function of the period $\Delta_t$. Each color represents a different perturbation according to the colormap shown in the inset of the first panel.
        }
        \label{fig.SI.EXPHOTI-windowAverages}
\end{figure}

Fig.~\ref{fig.SI.EXPHOTI-windowAverages} illustrates the dependence of the time-averaged values on the period $\Delta t$. 
In the main text we chose $\Delta t=30$ms.

\subsection{FEM simulations of the two-dimensional metamaterial}

\begin{figure}
        \centering
        \includegraphics[width=1\columnwidth]{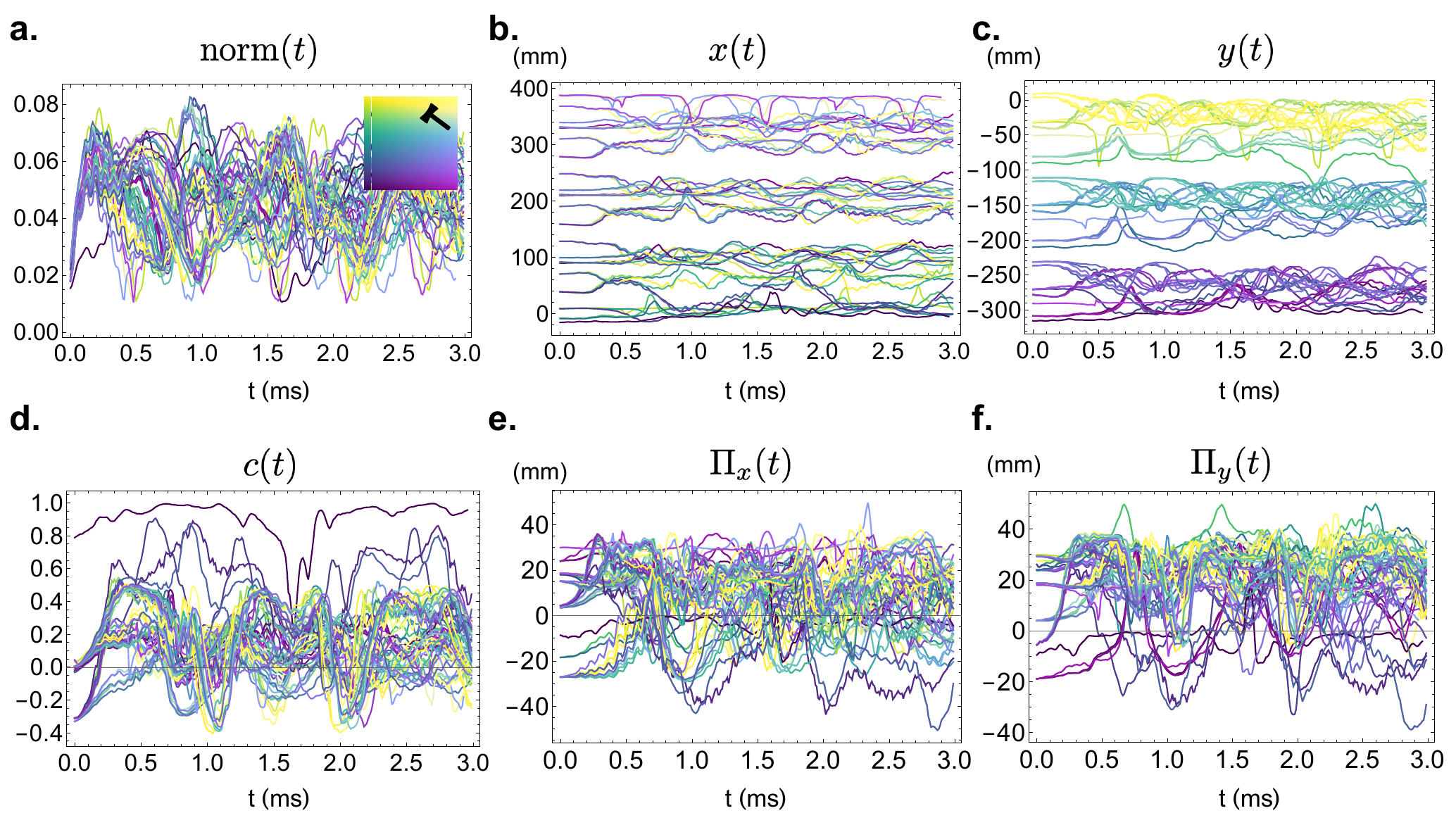}
        \caption{{\bf Moments from local perturbations in the mechanical two-dimensional metamaterial (FEM simulations).} Norm ({\bf a}), positions ({\bf b} and {\bf c}), chiral charge ({\bf d}) and chiral polarization components ({\bf e} and {\bf f}), for all the different perturbations in the mechanical metamaterial (colormap in the inset of {\bf a}). Only the data for which $\text{norm}(t)>\epsilon_{FEM}$ is considered.
        }
        \label{fig.SI.FEMHOTI-moments}
\end{figure}

Following the same structure, we show the moments (Fig.~\ref{fig.SI.FEMHOTI-moments}) filtered by $\epsilon_{text{FEM}}=0.01$, and the dependence of the time averages on the period $\Delta t$ (Fig.~\ref{fig.SI.FEMSSH-windowAverage}).
For the results shown in the main text, we use $\Delta t=0.5$ms.

Fig.~\ref{fig.SI.FEMHOTI-chiralMolecules}a to d. shows the extent of the perturbations for each type of node in a unit cell. Fig.~\ref{fig.SI.FEMHOTI-chiralMolecules}e and f reveal the final mechanical molecules and chiral polarization field of the simulated system.

\begin{figure}
        \centering
        \includegraphics[width=1\columnwidth]{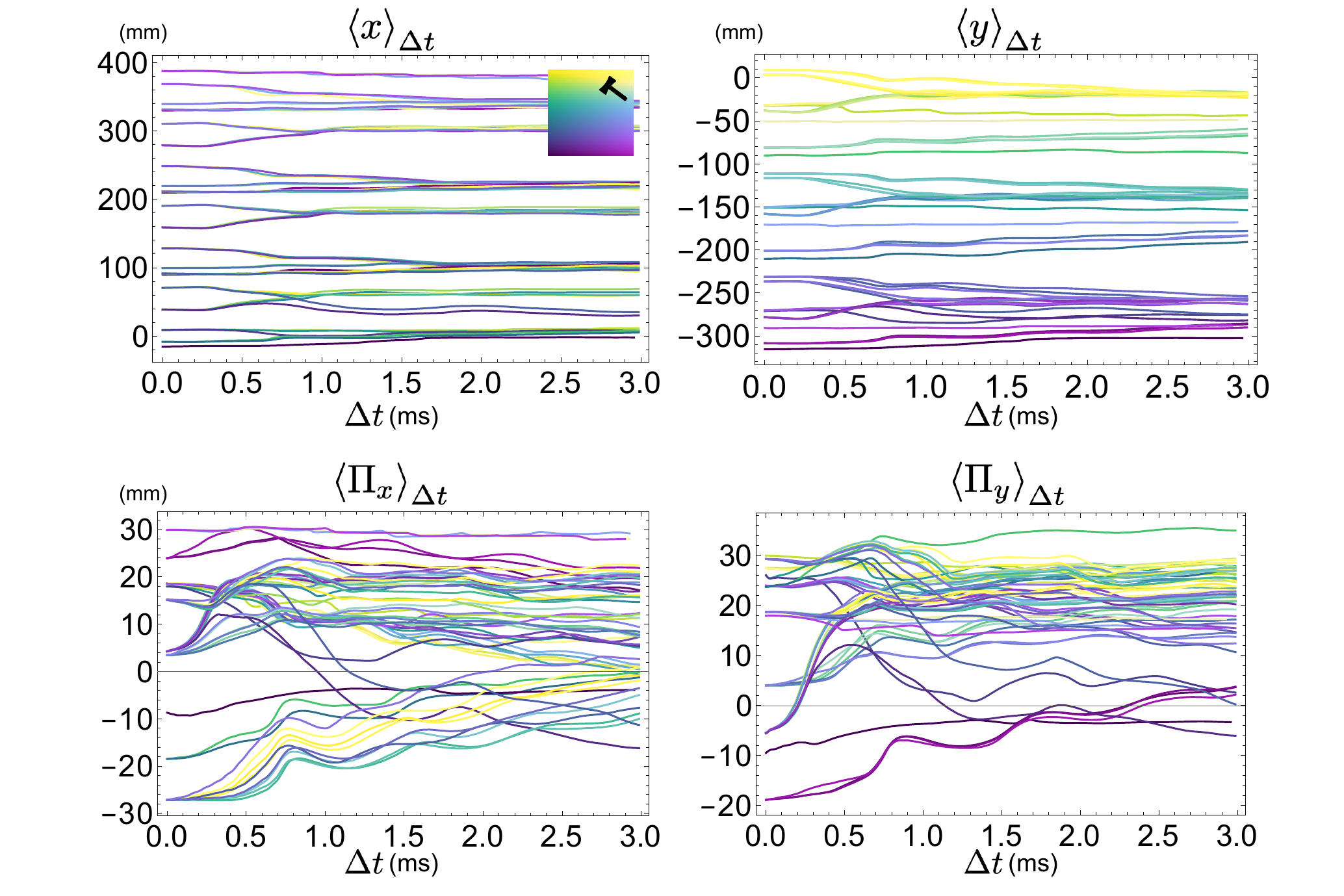}
        \caption{{\bf Time-averaged moments in the mechanical two-dimensional metamaterial (FEM simulations).} 
        Vertical and horizontal components of the average center and chiral polarization as a function of the period $\Delta_t$. Each color represents a different perturbation according to the colormap shown in the inset of the first panel.
        }
        \label{fig.SI.FEMHOTI-windowAverages}
\end{figure}

\begin{figure}
        \centering
        \includegraphics[width=\columnwidth]{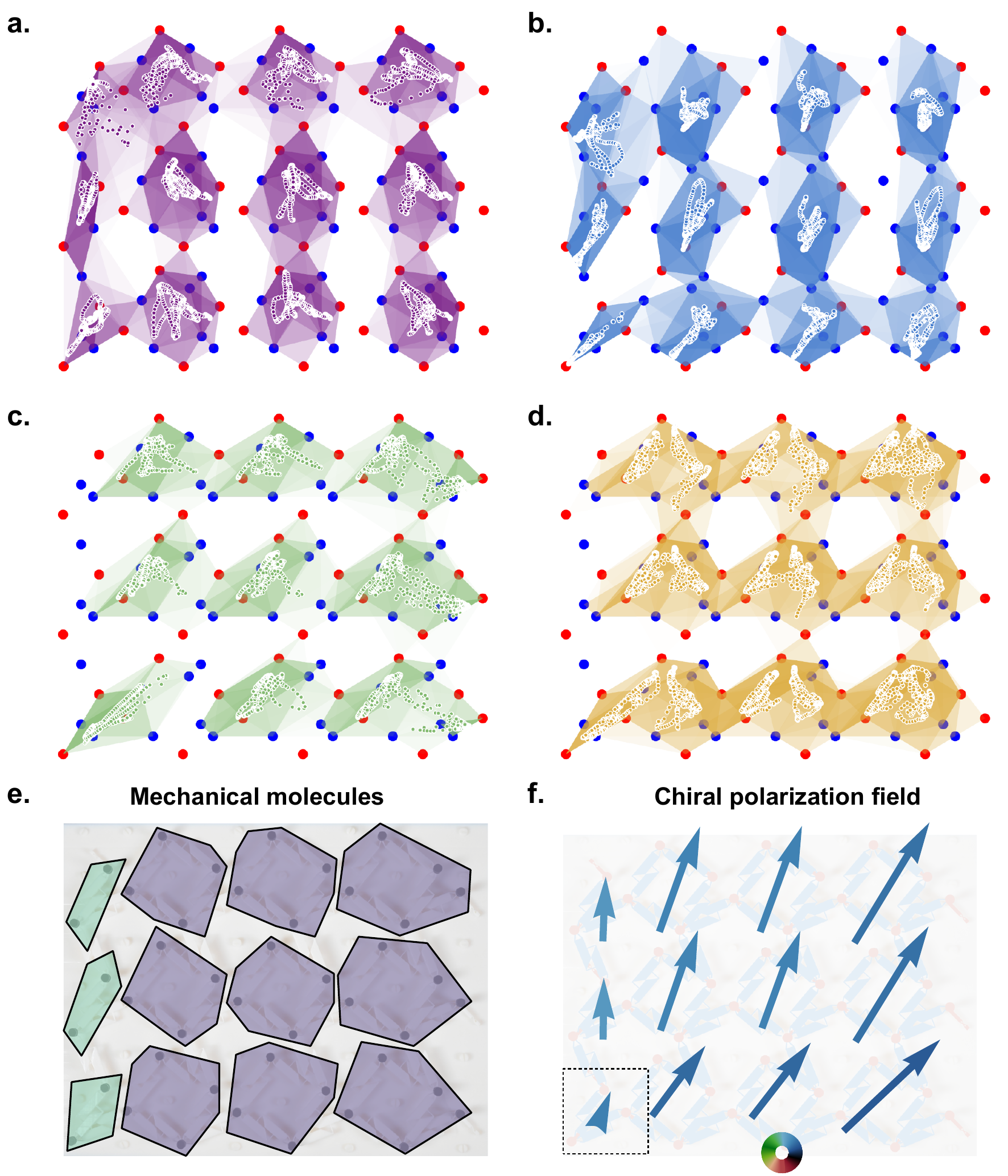}
        \caption{{\bf Mechanical molecules in the simulated poking of 2D mechanical system.}
        For each perturbation $\Psi(t)$ we draw the polygons spanning all the sites (red for beads, blue for springs) for which $d<\delta=1,2$. Each polygon, of light opacity, is superposed as time evolves. Opaque regions thus distinguish the strongly connected sites.
        Each panel ({\bf a} to {\bf d}) contains all the polygons associated to the perturbation of the same bead in different unit cells.
        We also highlight the centers $\bm r(t)$ of the perturbations as points of the same colors as the polygons.
        {\bf e.} As in the experimental poking (see Fig.1d) there are two types of mechanical molecules: large ones in the bulk (violet) and smaller ones in the left edge (green).
        {\bf f.} The chiral polarization exhibits the same discontinuity at the bottom left corner (dashed square).
        }
        \label{fig.SI.FEMHOTI-chiralMolecules}
\end{figure}

\section{Sample design}
\subsection{Basic units}
\begin{figure}
    \centering
    \includegraphics[width=0.75\columnwidth]{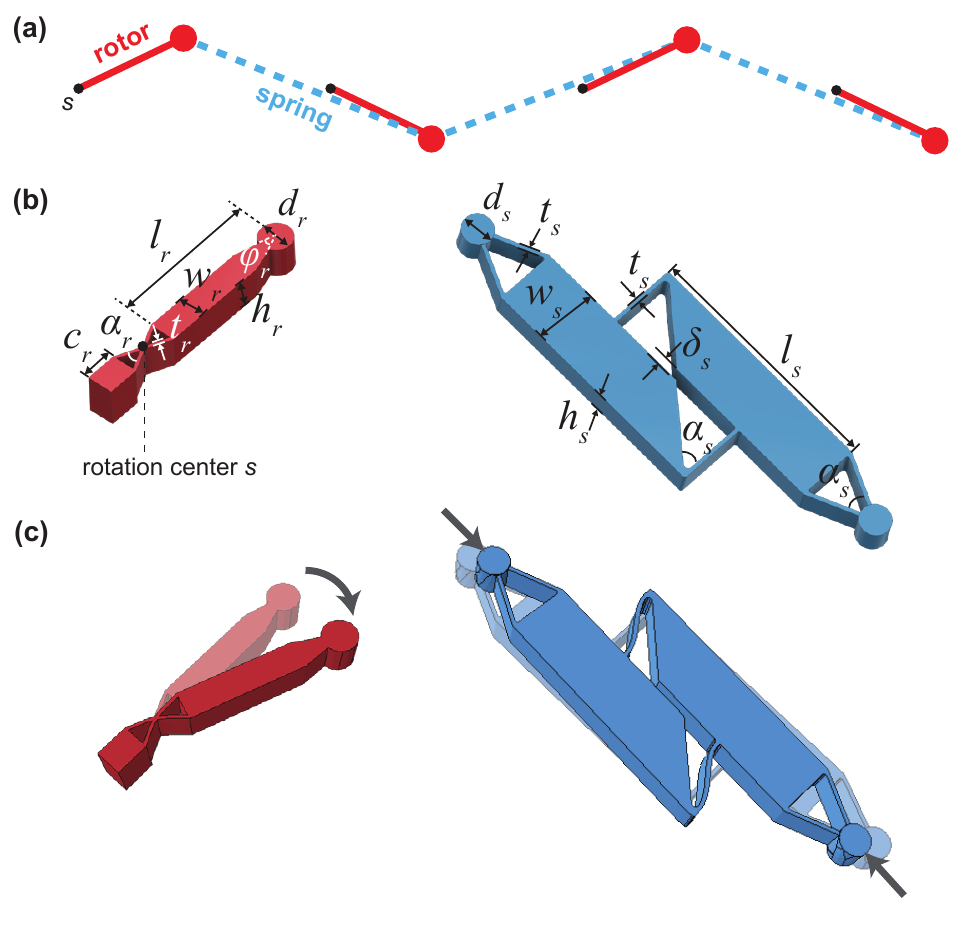}
    \caption{{\bf Basic units of the mechanical metamaterials.} 
    {\bf (a)} Ideal model of a 1D mechanical SSH chain. {\bf (b)} Design of two basic units: rotor (red) and spring (blue), where $l_r=21.9$ mm,  $w_r=3.9$ mm,  $h_r=4.8$ mm, $t_r=0.48$ mm, $d_r=4.8$ mm, $\alpha_r=50^ {\circ}$, $\varphi_r=31.5^ {\circ}$, $c_r=4.8$ mm, $l_s=37$ mm, $w_s=9.6$ mm, $h_s=4.8$ mm, $t_s=0.6$ mm, $d_s=4.8$ mm, $\alpha_s=50^{\circ}$ and $\delta_s=1.2$ mm.
    {\bf (c)} Illustration of the lowest energy eigenmode of both units. For a rotor (left), it corresponds to a rotation with respect to the pivot $s$, whereas for a spring it corresponds, primarily, to an axial compression (right).
    }
    \label{Fig.1}
\end{figure}

Inspired by the mechanical chain of ref.~\cite{Kane2014Topological}, the designs of the mechanical metamaterials presented in our paper are based on two basic units, respectively called rotors and springs (see Fig. \ref{Fig.1} (a) and (b)). 
Unlike the ideal model presented by Kane and Lukensky, the rotors and springs we present are not made from pure mechanisms, but from compliant mechanisms. Therefore their deformation does not cost zero elastic energy but a finite level of elastic energy~\cite{designbook1,designbook2}.
We optimize the geometrical parameters of our compliant mechanisms such that the first eigenmode of the rotor is a rotation with respect to the pivot $s$, and the first eigenmode of the spring is an axial compression/elongation, see Fig. \ref{Fig.1} (c). 
We further optimize the design by requiring that higher eigenmodes have much higher eigenvalues. Therefore, at low frequencies the target compliant mechanisms will dominate. These constraints typically require the use thin-walled hinges. However, there is a trade-off and to guarantee the printing quality, the minimum thickness of thin-wall structures is set to 0.48 mm. 

We conducted an analysis of the eigenmodes of mechanical metamaterials comprising $N$ rotors. We found that for the first $N$ eigenmodes of the mechanical metamaterial, all rotors deform as rotations and all springs deform as elongations or shortening. All deformations closely follow those prescribed by the compliant mechanism and no spurious mode pollutes the spectrum. 
In this article, we adopt the parameters shown in Fig. \ref{Fig.1} (b).

\subsection{Mechanical chain}
\begin{figure}
    \centering
    \includegraphics[width=\columnwidth]{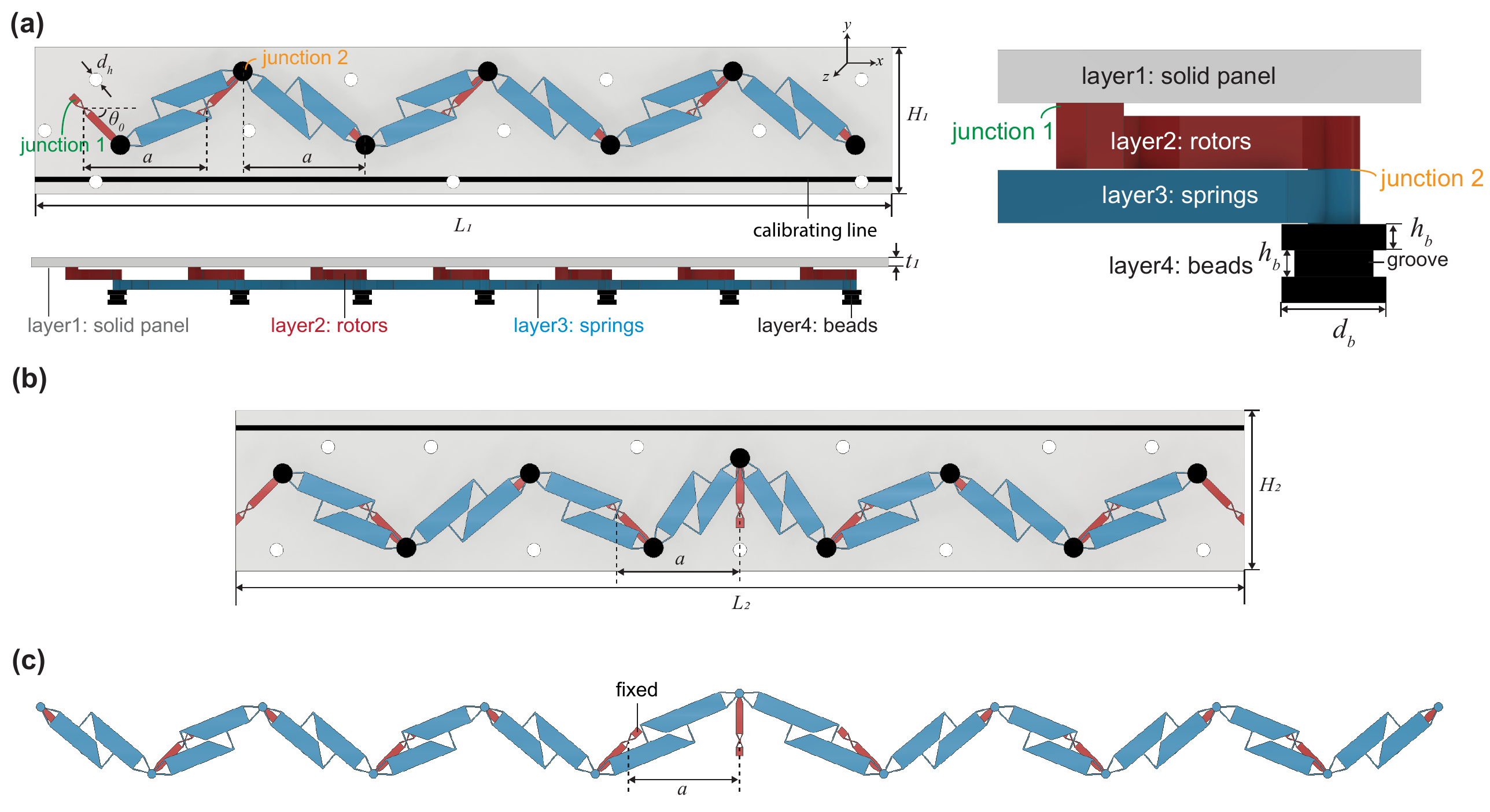}
    \caption{{\bf Design of the mechanical chains.} 
    {\bf (a)} Front view (top left) and top view (bottom left and right) of a homogeneous mechanical chain. 
    The size of the panel is $L_1=420$ mm, $H_1=72$ mm, and $t_1=4.8$ mm. The diameter of screw holes is $d_h=$ 6.5 mm. 
    The rotation centers of the rotors are fixed on the solid panel (junction 1). The rotor layer and spring layer are connected at junction 2. The dimensions of beads are $d_b=9.6$ mm and $h_b=2.4$ mm.
    {\bf (b)} Design of a heterogenous mechanical chain with a floppy mode in the middle. The panel size is $L_2=489.6$ mm and $H_2=78$ mm.
    {\bf (c)} Design of a heterogeneous mechanical chain with a self-stress state domain wall. 
    This model is designed for Finite Element Method (FEM) simulations. Here we prescind from both the solid panel and the bead layer.}
    \label{Fig.2}
\end{figure}

Here we detail the design of the mechanical chain used in the Main Text.
We begin with a solid optical panel on which we fix the 3D printed structure, see Fig.~\ref{Fig.2}(a). 
The dimensions of the panel are $L_1=420$ mm, $H_1=72$ mm and $t_1=4.8$ mm. 
The black strip on the panel helps to calibrate the camera angle. 
In order to improve the image analysis and data acquisition, we 3D print black disks on the rotors' tip, see Fig.~\ref{Fig.2}(a). 
The pivots' positions are horizontally aligned and equally spaced by $a=60$ mm. The angle between the rotors at equilibrium and the horizontal line is $\theta_0=\pm \pi /4$.
By design, rotors and springs will overlap with each other when assembled together. Therefore, we put rotors and springs in different planes. 
The 1D chain has four layers in total, see Fig.~\ref{Fig.2} (a). The bottom layer is the solid panel, the second layer hosts the rotors, the third one hosts the springs and the final one hosts the black disks. 
The pivots of the rotors are fixed on the solid panel (shown as junction 1) and springs connect the end of adjacent rotors (shown as junction 2).

The gap between the rotors and the solid panel along the z direction is 1.2 mm, the gap between the rotors and the springs is 0.12 mm and the gap between springs and beads is 0.12 mm. The gap between the layers are small to reduce influence of the out-of-plane deformation.

To implement a domain wall in the chain, we just need to change the tilt angle of some  rotors.
For a floppy mode domain wall, we connect a left a chain with a tilt-rotor angle $\theta_{\text{left}}=\pm \pi /4$ to a right chain with tilt angle $\theta_{\text{right}}=\pm 3\pi/4$, through a vertical middle rotor ($\theta_{\text{middle}}=\pi/2$), see Fig.~\ref{Fig.2}(b).
For a self-stress state domain wall, we interchange left and right  chains while keeping the same middle rotor, see Fig.~\ref{Fig.2}(c).

\subsection{Two-dimensional metamaterial}
\begin{figure}[t!]
    \centering
    \includegraphics[width=\columnwidth]{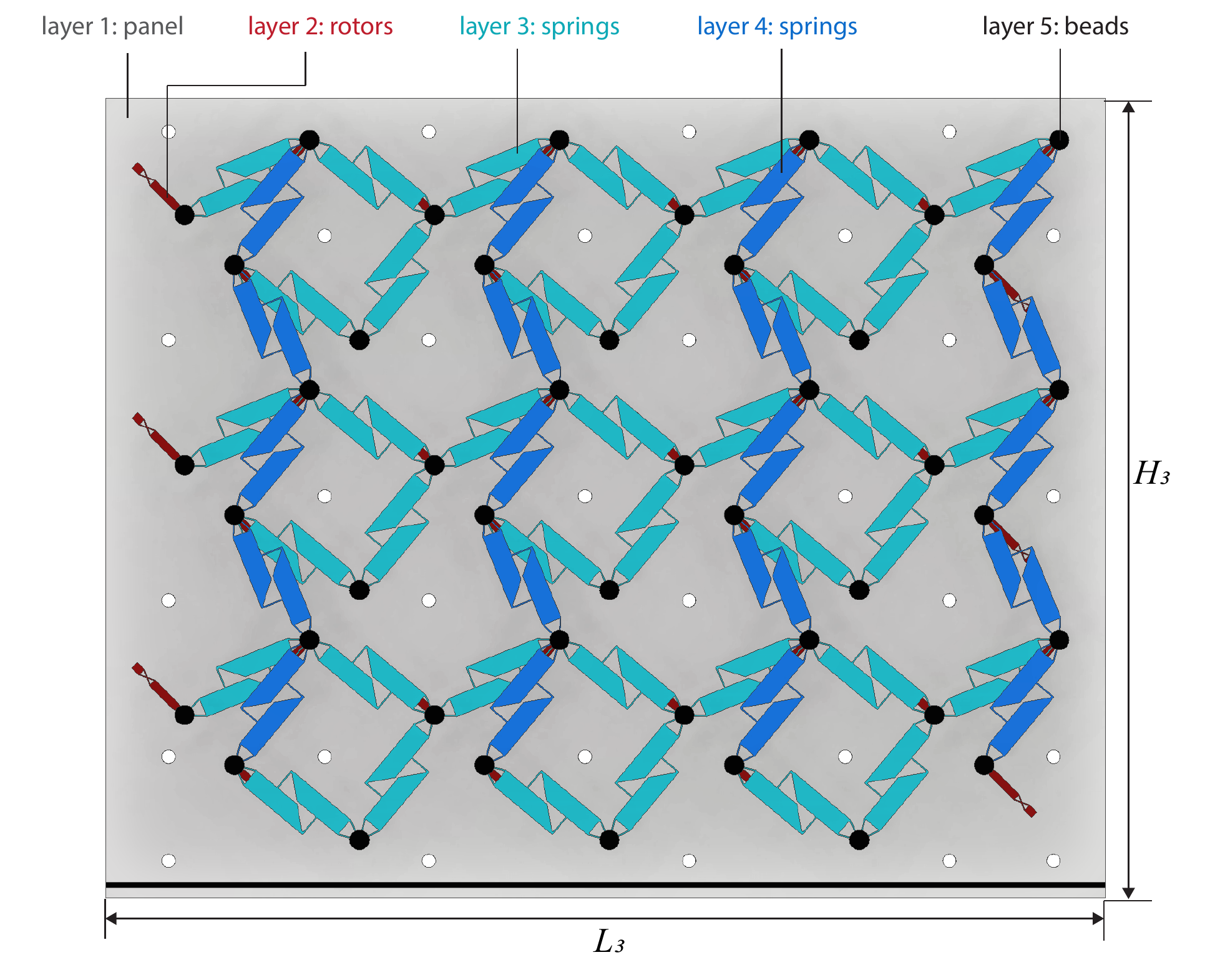}
    \caption{{\bf Design of the two-dimensional topological mechanical metamaterial.}
    }
    \label{Fig.3}
\end{figure}

The two-dimensional metamaterial assembles the 3D printed rotors  and springs of Fig.~\ref{Fig.1}, and in addition, single beads.
This corresponds to connecting two springs via a black bead with no rotor, see Fig.~\ref{Fig.3}. The unit cell of the metamaterial hosts 3 rotors, 4 springs, and 1 free bead, see Fig.~\ref{Fig.3}.
The whole metamaterial is made out of 3.5 units along the horizontal direction and 3 units along the vertical direction. 
The dimensions of the solid panel are $L_3=480$ mm and $H_3=384$ mm.
Due to the higher connectivity of the network, the springs are located in two different layers, see Fig.~\ref{Fig.3}.

\section{Sample fabrication}
All specimens are fabricated by additive manufacturing using a PolyJet 3D printer (Stratasys Object500 Connex3), whose build area is 490 mm $\times$ 390 mm  $\times$ 200 mm. The accuracy of the printer is 200 microns. 
The whole specimen is made of the same photopolymer, Stratasys Vero (Young's modulus $E\approx$ 2500 MPa). 
To improve the image analysis and data acquisition, we use a black material (Veroblack) for the beads and the calibration strip, and a white material (Verowhite) for the rest. 

\section{Experimental setup and data acquisition}
\subsection{Data acquisition}
To measure the chiral polarization, we need to perturb each bead, or rotor-tip, and measure the displacement of the black beads. 
To track the positions of beads, we record the experiments using a high-speed camera (Phantom VEO 640). We record all the tests with 6300 fps, and a resolution of 1792 px $\times$ 480 px for the mechanical chains and 1024 px $\times$ 768 px for the two-dimensional metamaterial.
We use the particle detection and tracking function in ImageJ to analyze the images and acquire the positions of the beads. 
We first threshold images (replace each pixel with black or white pixels) and then track the center of each pattern of black pixels. Each bead has around 18 to 35 pixels along the diameter. The accuracy of the measurement is around 0.006 mm (0.015 pixels).

\subsection{Perturbation experiments}
For the perturbations, we use a transparent plastic wire to pull the beads. 
We use the transparent wire for the following two reasons: i) the transparent wire is almost invisible on the photo and therefore improves the image analysis. ii) The direction of the perturbation is easier to control.
To simplify the pulling, we design a groove on beads, see Fig.~\ref{Fig.1}(a). The displacement applied is around 1-2 mm in a direction roughly perpendicular to the rotor.
Since the displacements are very small, the vibration of the whole system during the experiment can significantly affect the final data. To improve the steadiness, we screw the system to an optical panel attached to a steady table.

\subsection{Vibration experiments}
To observe the floppy corner mode of the two-dimensional metamaterial, we conducted vibration experiments by mounting the specimen onto an optical panel attached to a vibration machine (Tira Vibration Test System TV 5220-120). The vibration machine generated vertical vibrations that follow a sine wave function. We adjust the frequency using an Aim-TTi TG5011 function generator. The first mode appears at a vibration frequency of 88Hz.

\section{Numerical simulations}
We conducted finite element simulations using the commercial package ABAQUS. To capture the dynamic evolution, we employed the explicit solver, while the standard solver was used to acquire the eigenmodes.
We discretize the model with tetrahedrons (20-node quadratic brick with reduced integration, element type: C3D20R). Each pair of rotor and spring is divided in 2000 to 3000 tetrahedrons.
The simulated systems are composed of a linear elastic material with Young's modulus $E=$ 2500MPa, Poisson's ratio $\nu=$ 0.33, and density $\rho=$ 1.19 g/cm$^3$ (from Stratasys Vero product data sheet).
The rectangle panel is fixed. 
Out-of-plane deformations are also constrained in all simulations. 
The perturbation is given by displacement loads perpendicular to the rotor $|dx|=0.01$ mm and $|dy|=0.01$ mm within 0.0001s (smooth step). Then, the system is released, and we track the displacements of all nodes for a duration of 0.004s to 0.01s.


\section{Supplementary video}
Vertical vibration of the presented 2D metamaterial. a corner floppy mode is observed as the first mode at 88 Hz. The video was taken at 126 fps.

\bibliography{biblio}

\begin{thebibliography}{54}%
\makeatletter
\providecommand \@ifxundefined [1]{%
 \@ifx{#1\undefined}
}%
\providecommand \@ifnum [1]{%
 \ifnum #1\expandafter \@firstoftwo
 \else \expandafter \@secondoftwo
 \fi
}%
\providecommand \@ifx [1]{%
 \ifx #1\expandafter \@firstoftwo
 \else \expandafter \@secondoftwo
 \fi
}%
\providecommand \natexlab [1]{#1}%
\providecommand \enquote  [1]{``#1''}%
\providecommand \bibnamefont  [1]{#1}%
\providecommand \bibfnamefont [1]{#1}%
\providecommand \citenamefont [1]{#1}%
\providecommand \href@noop [0]{\@secondoftwo}%
\providecommand \href [0]{\begingroup \@sanitize@url \@href}%
\providecommand \@href[1]{\@@startlink{#1}\@@href}%
\providecommand \@@href[1]{\endgroup#1\@@endlink}%
\providecommand \@sanitize@url [0]{\catcode `\\12\catcode `\$12\catcode
  `\&12\catcode `\#12\catcode `\^12\catcode `\_12\catcode `\%12\relax}%
\providecommand \@@startlink[1]{}%
\providecommand \@@endlink[0]{}%
\providecommand \url  [0]{\begingroup\@sanitize@url \@url }%
\providecommand \@url [1]{\endgroup\@href {#1}{\urlprefix }}%
\providecommand \urlprefix  [0]{URL }%
\providecommand \Eprint [0]{\href }%
\providecommand \doibase [0]{https://doi.org/}%
\providecommand \selectlanguage [0]{\@gobble}%
\providecommand \bibinfo  [0]{\@secondoftwo}%
\providecommand \bibfield  [0]{\@secondoftwo}%
\providecommand \translation [1]{[#1]}%
\providecommand \BibitemOpen [0]{}%
\providecommand \bibitemStop [0]{}%
\providecommand \bibitemNoStop [0]{.\EOS\space}%
\providecommand \EOS [0]{\spacefactor3000\relax}%
\providecommand \BibitemShut  [1]{\csname bibitem#1\endcsname}%
\let\auto@bib@innerbib\@empty
\bibitem [{\citenamefont {Kelvin}(1891)}]{kelvin1891popular}%
  \BibitemOpen
  \bibfield  {author} {\bibinfo {author} {\bibfnamefont {W.~T.~B.}\
  \bibnamefont {Kelvin}},\ }\href@noop {} {\emph {\bibinfo {title} {Popular
  lectures and addresses}}},\ Vol.~\bibinfo {volume} {3}\ (\bibinfo
  {publisher} {Macmillan and Company},\ \bibinfo {year} {1891})\BibitemShut
  {NoStop}%
\bibitem [{\citenamefont {Hartmann}(1999)}]{Hartmann1999Magnetic}%
  \BibitemOpen
  \bibfield  {author} {\bibinfo {author} {\bibfnamefont {U.}~\bibnamefont
  {Hartmann}},\ }\bibfield  {title} {\bibinfo {title} {Magnetic force
  microscopy},\ }\href {https://doi.org/10.1146/annurev.matsci.29.1.53}
  {\bibfield  {journal} {\bibinfo  {journal} {Annual Review of Materials
  Science}\ }\textbf {\bibinfo {volume} {29}},\ \bibinfo {pages} {53} (\bibinfo
  {year} {1999})},\ \Eprint
  {https://arxiv.org/abs/https://doi.org/10.1146/annurev.matsci.29.1.53}
  {https://doi.org/10.1146/annurev.matsci.29.1.53} \BibitemShut {NoStop}%
\bibitem [{\citenamefont {Hooke}(1678)}]{Hooke}%
  \BibitemOpen
  \bibfield  {author} {\bibinfo {author} {\bibfnamefont {R.}~\bibnamefont
  {Hooke}},\ }\bibfield  {title} {\bibinfo {title} {De potentia restitutiva, or
  of spring explaining the power of springing bodies},\ }\href@noop {}
  {\bibfield  {journal} {\bibinfo  {journal} {London, UK: John Martyn}\
  }\textbf {\bibinfo {volume} {23}} (\bibinfo {year} {1678})}\BibitemShut
  {NoStop}%
\bibitem [{\citenamefont {Klitzing}\ \emph {et~al.}(1980)\citenamefont
  {Klitzing}, \citenamefont {Dorda},\ and\ \citenamefont {Pepper}}]{Klitzing}%
  \BibitemOpen
  \bibfield  {author} {\bibinfo {author} {\bibfnamefont {K.~v.}\ \bibnamefont
  {Klitzing}}, \bibinfo {author} {\bibfnamefont {G.}~\bibnamefont {Dorda}},\
  and\ \bibinfo {author} {\bibfnamefont {M.}~\bibnamefont {Pepper}},\
  }\bibfield  {title} {\bibinfo {title} {New method for high-accuracy
  determination of the fine-structure constant based on quantized hall
  resistance},\ }\href {https://doi.org/10.1103/PhysRevLett.45.494} {\bibfield
  {journal} {\bibinfo  {journal} {Phys. Rev. Lett.}\ }\textbf {\bibinfo
  {volume} {45}},\ \bibinfo {pages} {494} (\bibinfo {year} {1980})}\BibitemShut
  {NoStop}%
\bibitem [{\citenamefont {Thouless}(1998)}]{thouless1998topological}%
  \BibitemOpen
  \bibfield  {author} {\bibinfo {author} {\bibfnamefont {D.}~\bibnamefont
  {Thouless}},\ }\href@noop {} {\emph {\bibinfo {title} {Topological quantum
  numbers in nonrelativistic physics}}}\ (\bibinfo  {publisher} {World
  Scientific},\ \bibinfo {year} {1998})\BibitemShut {NoStop}%
\bibitem [{\citenamefont {Kadic}\ \emph {et~al.}(2013)\citenamefont {Kadic},
  \citenamefont {Bückmann}, \citenamefont {Schittny},\ and\ \citenamefont
  {Wegener}}]{Kadic_Review2013}%
  \BibitemOpen
  \bibfield  {author} {\bibinfo {author} {\bibfnamefont {M.}~\bibnamefont
  {Kadic}}, \bibinfo {author} {\bibfnamefont {T.}~\bibnamefont {Bückmann}},
  \bibinfo {author} {\bibfnamefont {R.}~\bibnamefont {Schittny}},\ and\
  \bibinfo {author} {\bibfnamefont {M.}~\bibnamefont {Wegener}},\ }\bibfield
  {title} {\bibinfo {title} {Metamaterials beyond electromagnetism},\ }\href
  {http://stacks.iop.org/0034-4885/76/i=12/a=126501} {\bibfield  {journal}
  {\bibinfo  {journal} {Rep. Prog. Phys.}\ }\textbf {\bibinfo {volume} {76}},\
  \bibinfo {pages} {126501} (\bibinfo {year} {2013})}\BibitemShut {NoStop}%
\bibitem [{\citenamefont {Bertoldi}\ \emph {et~al.}(2017)\citenamefont
  {Bertoldi}, \citenamefont {Vitelli}, \citenamefont {Christensen},\ and\
  \citenamefont {Van~Hecke}}]{Bertoldi2017Flexible}%
  \BibitemOpen
  \bibfield  {author} {\bibinfo {author} {\bibfnamefont {K.}~\bibnamefont
  {Bertoldi}}, \bibinfo {author} {\bibfnamefont {V.}~\bibnamefont {Vitelli}},
  \bibinfo {author} {\bibfnamefont {J.}~\bibnamefont {Christensen}},\ and\
  \bibinfo {author} {\bibfnamefont {M.}~\bibnamefont {Van~Hecke}},\ }\bibfield
  {title} {\bibinfo {title} {Flexible mechanical metamaterials},\ }\href@noop
  {} {\bibfield  {journal} {\bibinfo  {journal} {Nature Reviews Materials}\
  }\textbf {\bibinfo {volume} {2}},\ \bibinfo {pages} {1} (\bibinfo {year}
  {2017})}\BibitemShut {NoStop}%
\bibitem [{\citenamefont {Kane}\ and\ \citenamefont
  {Lubensky}(2014)}]{Kane2014Topological}%
  \BibitemOpen
  \bibfield  {author} {\bibinfo {author} {\bibfnamefont {C.~L.}\ \bibnamefont
  {Kane}}\ and\ \bibinfo {author} {\bibfnamefont {T.~C.}\ \bibnamefont
  {Lubensky}},\ }\bibfield  {title} {\bibinfo {title} {Topological boundary
  modes in isostatic lattices},\ }\href {https://doi.org/10.1038/nphys2835}
  {\bibfield  {journal} {\bibinfo  {journal} {Nature Physics}\ }\textbf
  {\bibinfo {volume} {10}},\ \bibinfo {pages} {39} (\bibinfo {year}
  {2014})}\BibitemShut {NoStop}%
\bibitem [{\citenamefont {Mao}\ and\ \citenamefont
  {Lubensky}(2018{\natexlab{a}})}]{Mao2018}%
  \BibitemOpen
  \bibfield  {author} {\bibinfo {author} {\bibfnamefont {X.}~\bibnamefont
  {Mao}}\ and\ \bibinfo {author} {\bibfnamefont {T.~C.}\ \bibnamefont
  {Lubensky}},\ }\bibfield  {title} {\bibinfo {title} {Maxwell lattices and
  topological mechanics},\ }\href@noop {} {\bibfield  {journal} {\bibinfo
  {journal} {Annual Review of Condensed Matter Physics}\ }\textbf {\bibinfo
  {volume} {9}},\ \bibinfo {pages} {413} (\bibinfo {year}
  {2018}{\natexlab{a}})}\BibitemShut {NoStop}%
\bibitem [{\citenamefont {Prodan}\ and\ \citenamefont
  {Prodan}(2009)}]{Prodan2009}%
  \BibitemOpen
  \bibfield  {author} {\bibinfo {author} {\bibfnamefont {E.}~\bibnamefont
  {Prodan}}\ and\ \bibinfo {author} {\bibfnamefont {C.}~\bibnamefont
  {Prodan}},\ }\bibfield  {title} {\bibinfo {title} {Topological phonon modes
  and their role in dynamic instability of microtubules},\ }\href@noop {}
  {\bibfield  {journal} {\bibinfo  {journal} {Physical review letters}\
  }\textbf {\bibinfo {volume} {103}},\ \bibinfo {pages} {248101} (\bibinfo
  {year} {2009})}\BibitemShut {NoStop}%
\bibitem [{\citenamefont {Rocklin}(2017)}]{Rocklin2017}%
  \BibitemOpen
  \bibfield  {author} {\bibinfo {author} {\bibfnamefont {D.~Z.}\ \bibnamefont
  {Rocklin}},\ }\bibfield  {title} {\bibinfo {title} {Directional mechanical
  response in the bulk of topological metamaterials},\ }\href@noop {}
  {\bibfield  {journal} {\bibinfo  {journal} {New Journal of Physics}\ }\textbf
  {\bibinfo {volume} {19}},\ \bibinfo {pages} {065004} (\bibinfo {year}
  {2017})}\BibitemShut {NoStop}%
\bibitem [{\citenamefont {Bilal}\ \emph {et~al.}(2017)\citenamefont {Bilal},
  \citenamefont {S{\"u}sstrunk}, \citenamefont {Daraio},\ and\ \citenamefont
  {Huber}}]{Bilal2017}%
  \BibitemOpen
  \bibfield  {author} {\bibinfo {author} {\bibfnamefont {O.~R.}\ \bibnamefont
  {Bilal}}, \bibinfo {author} {\bibfnamefont {R.}~\bibnamefont
  {S{\"u}sstrunk}}, \bibinfo {author} {\bibfnamefont {C.}~\bibnamefont
  {Daraio}},\ and\ \bibinfo {author} {\bibfnamefont {S.~D.}\ \bibnamefont
  {Huber}},\ }\bibfield  {title} {\bibinfo {title} {Intrinsically polar elastic
  metamaterials},\ }\href@noop {} {\bibfield  {journal} {\bibinfo  {journal}
  {Advanced Materials}\ }\textbf {\bibinfo {volume} {29}},\ \bibinfo {pages}
  {1700540} (\bibinfo {year} {2017})}\BibitemShut {NoStop}%
\bibitem [{\citenamefont {Coulais}\ \emph {et~al.}(2017)\citenamefont
  {Coulais}, \citenamefont {Sounas},\ and\ \citenamefont
  {Alù}}]{Coulais_Nature2017}%
  \BibitemOpen
  \bibfield  {author} {\bibinfo {author} {\bibfnamefont {C.}~\bibnamefont
  {Coulais}}, \bibinfo {author} {\bibfnamefont {D.}~\bibnamefont {Sounas}},\
  and\ \bibinfo {author} {\bibfnamefont {A.}~\bibnamefont {Alù}},\ }\bibfield
  {title} {\bibinfo {title} {Static non-reciprocity in mechanical
  metamaterials},\ }\href {https://doi.org/10.1038/nature21044} {\bibfield
  {journal} {\bibinfo  {journal} {Nature}\ }\textbf {\bibinfo {volume} {542}},\
  \bibinfo {pages} {461} (\bibinfo {year} {2017})}\BibitemShut {NoStop}%
\bibitem [{\citenamefont {Nash}\ \emph {et~al.}(2015)\citenamefont {Nash},
  \citenamefont {Kleckner}, \citenamefont {Read}, \citenamefont {Vitelli},
  \citenamefont {Turner},\ and\ \citenamefont {Irvine}}]{Nash2015}%
  \BibitemOpen
  \bibfield  {author} {\bibinfo {author} {\bibfnamefont {L.~M.}\ \bibnamefont
  {Nash}}, \bibinfo {author} {\bibfnamefont {D.}~\bibnamefont {Kleckner}},
  \bibinfo {author} {\bibfnamefont {A.}~\bibnamefont {Read}}, \bibinfo {author}
  {\bibfnamefont {V.}~\bibnamefont {Vitelli}}, \bibinfo {author} {\bibfnamefont
  {A.~M.}\ \bibnamefont {Turner}},\ and\ \bibinfo {author} {\bibfnamefont
  {W.~T.}\ \bibnamefont {Irvine}},\ }\bibfield  {title} {\bibinfo {title}
  {Topological mechanics of gyroscopic metamaterials},\ }\href@noop {}
  {\bibfield  {journal} {\bibinfo  {journal} {Proceedings of the National
  Academy of Sciences}\ }\textbf {\bibinfo {volume} {112}},\ \bibinfo {pages}
  {14495} (\bibinfo {year} {2015})}\BibitemShut {NoStop}%
\bibitem [{\citenamefont {Peano}\ \emph {et~al.}(2015)\citenamefont {Peano},
  \citenamefont {Brendel}, \citenamefont {Schmidt},\ and\ \citenamefont
  {Marquardt}}]{Peano2015}%
  \BibitemOpen
  \bibfield  {author} {\bibinfo {author} {\bibfnamefont {V.}~\bibnamefont
  {Peano}}, \bibinfo {author} {\bibfnamefont {C.}~\bibnamefont {Brendel}},
  \bibinfo {author} {\bibfnamefont {M.}~\bibnamefont {Schmidt}},\ and\ \bibinfo
  {author} {\bibfnamefont {F.}~\bibnamefont {Marquardt}},\ }\bibfield  {title}
  {\bibinfo {title} {Topological phases of sound and light},\ }\href@noop {}
  {\bibfield  {journal} {\bibinfo  {journal} {Physical Review X}\ }\textbf
  {\bibinfo {volume} {5}},\ \bibinfo {pages} {031011} (\bibinfo {year}
  {2015})}\BibitemShut {NoStop}%
\bibitem [{\citenamefont {S{\"u}sstrunk}\ and\ \citenamefont
  {Huber}(2016)}]{Susstrunk2016Classification}%
  \BibitemOpen
  \bibfield  {author} {\bibinfo {author} {\bibfnamefont {R.}~\bibnamefont
  {S{\"u}sstrunk}}\ and\ \bibinfo {author} {\bibfnamefont {S.~D.}\ \bibnamefont
  {Huber}},\ }\bibfield  {title} {\bibinfo {title} {Classification of
  topological phonons in linear mechanical metamaterials},\ }\href
  {https://doi.org/10.1073/pnas.1605462113} {\bibfield  {journal} {\bibinfo
  {journal} {Proceedings of the National Academy of Sciences}\ }\textbf
  {\bibinfo {volume} {113}},\ \bibinfo {pages} {E4767} (\bibinfo {year}
  {2016})},\ \Eprint
  {https://arxiv.org/abs/https://www.pnas.org/content/113/33/E4767.full.pdf}
  {https://www.pnas.org/content/113/33/E4767.full.pdf} \BibitemShut {NoStop}%
\bibitem [{\citenamefont {Ghatak}\ \emph {et~al.}(2020)\citenamefont {Ghatak},
  \citenamefont {Brandenbourger}, \citenamefont {Van~Wezel},\ and\
  \citenamefont {Coulais}}]{Ghatak2020}%
  \BibitemOpen
  \bibfield  {author} {\bibinfo {author} {\bibfnamefont {A.}~\bibnamefont
  {Ghatak}}, \bibinfo {author} {\bibfnamefont {M.}~\bibnamefont
  {Brandenbourger}}, \bibinfo {author} {\bibfnamefont {J.}~\bibnamefont
  {Van~Wezel}},\ and\ \bibinfo {author} {\bibfnamefont {C.}~\bibnamefont
  {Coulais}},\ }\bibfield  {title} {\bibinfo {title} {Observation of
  non-hermitian topology and its bulk--edge correspondence in an active
  mechanical metamaterial},\ }\href@noop {} {\bibfield  {journal} {\bibinfo
  {journal} {Proceedings of the National Academy of Sciences}\ }\textbf
  {\bibinfo {volume} {117}},\ \bibinfo {pages} {29561} (\bibinfo {year}
  {2020})}\BibitemShut {NoStop}%
\bibitem [{\citenamefont {Coulais}\ \emph {et~al.}(2021)\citenamefont
  {Coulais}, \citenamefont {Fleury},\ and\ \citenamefont {van
  Wezel}}]{Coulais_review2021}%
  \BibitemOpen
  \bibfield  {author} {\bibinfo {author} {\bibfnamefont {C.}~\bibnamefont
  {Coulais}}, \bibinfo {author} {\bibfnamefont {R.}~\bibnamefont {Fleury}},\
  and\ \bibinfo {author} {\bibfnamefont {J.}~\bibnamefont {van Wezel}},\
  }\bibfield  {title} {\bibinfo {title} {Topology and broken hermiticity},\
  }\href {https://doi.org/10.1038/s41567-020-01093-z} {\bibfield  {journal}
  {\bibinfo  {journal} {Nat. Phys.}\ }\textbf {\bibinfo {volume} {17}},\
  \bibinfo {pages} {9} (\bibinfo {year} {2021})}\BibitemShut {NoStop}%
\bibitem [{\citenamefont {S{\"u}sstrunk}\ and\ \citenamefont
  {Huber}(2015)}]{Susstrunk2015}%
  \BibitemOpen
  \bibfield  {author} {\bibinfo {author} {\bibfnamefont {R.}~\bibnamefont
  {S{\"u}sstrunk}}\ and\ \bibinfo {author} {\bibfnamefont {S.~D.}\ \bibnamefont
  {Huber}},\ }\bibfield  {title} {\bibinfo {title} {Observation of phononic
  helical edge states in a mechanical topological insulator},\ }\href@noop {}
  {\bibfield  {journal} {\bibinfo  {journal} {Science}\ }\textbf {\bibinfo
  {volume} {349}},\ \bibinfo {pages} {47} (\bibinfo {year} {2015})}\BibitemShut
  {NoStop}%
\bibitem [{\citenamefont {Serra-Garcia}\ \emph
  {et~al.}(2018{\natexlab{a}})\citenamefont {Serra-Garcia}, \citenamefont
  {Peri}, \citenamefont {S{\"u}sstrunk}, \citenamefont {Bilal}, \citenamefont
  {Larsen}, \citenamefont {Villanueva},\ and\ \citenamefont
  {Huber}}]{Serra2018}%
  \BibitemOpen
  \bibfield  {author} {\bibinfo {author} {\bibfnamefont {M.}~\bibnamefont
  {Serra-Garcia}}, \bibinfo {author} {\bibfnamefont {V.}~\bibnamefont {Peri}},
  \bibinfo {author} {\bibfnamefont {R.}~\bibnamefont {S{\"u}sstrunk}}, \bibinfo
  {author} {\bibfnamefont {O.~R.}\ \bibnamefont {Bilal}}, \bibinfo {author}
  {\bibfnamefont {T.}~\bibnamefont {Larsen}}, \bibinfo {author} {\bibfnamefont
  {L.~G.}\ \bibnamefont {Villanueva}},\ and\ \bibinfo {author} {\bibfnamefont
  {S.~D.}\ \bibnamefont {Huber}},\ }\bibfield  {title} {\bibinfo {title}
  {Observation of a phononic quadrupole topological insulator},\ }\href@noop {}
  {\bibfield  {journal} {\bibinfo  {journal} {Nature}\ }\textbf {\bibinfo
  {volume} {555}},\ \bibinfo {pages} {342} (\bibinfo {year}
  {2018}{\natexlab{a}})}\BibitemShut {NoStop}%
\bibitem [{\citenamefont {Saremi}\ and\ \citenamefont
  {Rocklin}(2018)}]{Saremi2018}%
  \BibitemOpen
  \bibfield  {author} {\bibinfo {author} {\bibfnamefont {A.}~\bibnamefont
  {Saremi}}\ and\ \bibinfo {author} {\bibfnamefont {Z.}~\bibnamefont
  {Rocklin}},\ }\bibfield  {title} {\bibinfo {title} {Controlling the
  deformation of metamaterials: Corner modes via topology},\ }\href
  {https://doi.org/10.1103/PhysRevB.98.180102} {\bibfield  {journal} {\bibinfo
  {journal} {Phys. Rev. B}\ }\textbf {\bibinfo {volume} {98}},\ \bibinfo
  {pages} {180102} (\bibinfo {year} {2018})}\BibitemShut {NoStop}%
\bibitem [{\citenamefont {Xue}\ \emph {et~al.}(2019)\citenamefont {Xue},
  \citenamefont {Yang}, \citenamefont {Gao}, \citenamefont {Chong},\ and\
  \citenamefont {Zhang}}]{Xue2019}%
  \BibitemOpen
  \bibfield  {author} {\bibinfo {author} {\bibfnamefont {H.}~\bibnamefont
  {Xue}}, \bibinfo {author} {\bibfnamefont {Y.}~\bibnamefont {Yang}}, \bibinfo
  {author} {\bibfnamefont {F.}~\bibnamefont {Gao}}, \bibinfo {author}
  {\bibfnamefont {Y.}~\bibnamefont {Chong}},\ and\ \bibinfo {author}
  {\bibfnamefont {B.}~\bibnamefont {Zhang}},\ }\bibfield  {title} {\bibinfo
  {title} {Acoustic higher-order topological insulator on a kagome lattice},\
  }\href@noop {} {\bibfield  {journal} {\bibinfo  {journal} {Nature materials}\
  }\textbf {\bibinfo {volume} {18}},\ \bibinfo {pages} {108} (\bibinfo {year}
  {2019})}\BibitemShut {NoStop}%
\bibitem [{\citenamefont {Ni}\ \emph {et~al.}(2019{\natexlab{a}})\citenamefont
  {Ni}, \citenamefont {Weiner}, \citenamefont {Alu},\ and\ \citenamefont
  {Khanikaev}}]{Ni2019}%
  \BibitemOpen
  \bibfield  {author} {\bibinfo {author} {\bibfnamefont {X.}~\bibnamefont
  {Ni}}, \bibinfo {author} {\bibfnamefont {M.}~\bibnamefont {Weiner}}, \bibinfo
  {author} {\bibfnamefont {A.}~\bibnamefont {Alu}},\ and\ \bibinfo {author}
  {\bibfnamefont {A.~B.}\ \bibnamefont {Khanikaev}},\ }\bibfield  {title}
  {\bibinfo {title} {Observation of higher-order topological acoustic states
  protected by generalized chiral symmetry},\ }\href@noop {} {\bibfield
  {journal} {\bibinfo  {journal} {Nature materials}\ }\textbf {\bibinfo
  {volume} {18}},\ \bibinfo {pages} {113} (\bibinfo {year}
  {2019}{\natexlab{a}})}\BibitemShut {NoStop}%
\bibitem [{\citenamefont {Hasan}\ and\ \citenamefont
  {Kane}(2010)}]{Hasan_Review}%
  \BibitemOpen
  \bibfield  {author} {\bibinfo {author} {\bibfnamefont {M.~Z.}\ \bibnamefont
  {Hasan}}\ and\ \bibinfo {author} {\bibfnamefont {C.~L.}\ \bibnamefont
  {Kane}},\ }\bibfield  {title} {\bibinfo {title} {Colloquium: Topological
  insulators},\ }\href {https://doi.org/10.1103/RevModPhys.82.3045} {\bibfield
  {journal} {\bibinfo  {journal} {Rev. Mod. Phys.}\ }\textbf {\bibinfo {volume}
  {82}},\ \bibinfo {pages} {3045} (\bibinfo {year} {2010})}\BibitemShut
  {NoStop}%
\bibitem [{\citenamefont {Chen}\ \emph {et~al.}(2014)\citenamefont {Chen},
  \citenamefont {Upadhyaya},\ and\ \citenamefont {Vitelli}}]{Chen2014}%
  \BibitemOpen
  \bibfield  {author} {\bibinfo {author} {\bibfnamefont {B.~G.-g.}\
  \bibnamefont {Chen}}, \bibinfo {author} {\bibfnamefont {N.}~\bibnamefont
  {Upadhyaya}},\ and\ \bibinfo {author} {\bibfnamefont {V.}~\bibnamefont
  {Vitelli}},\ }\bibfield  {title} {\bibinfo {title} {Nonlinear conduction via
  solitons in a topological mechanical insulator},\ }\href@noop {} {\bibfield
  {journal} {\bibinfo  {journal} {Proceedings of the National Academy of
  Sciences}\ }\textbf {\bibinfo {volume} {111}},\ \bibinfo {pages} {13004}
  (\bibinfo {year} {2014})}\BibitemShut {NoStop}%
\bibitem [{\citenamefont {Mao}\ and\ \citenamefont
  {Lubensky}(2018{\natexlab{b}})}]{Mao_Review}%
  \BibitemOpen
  \bibfield  {author} {\bibinfo {author} {\bibfnamefont {X.}~\bibnamefont
  {Mao}}\ and\ \bibinfo {author} {\bibfnamefont {T.~C.}\ \bibnamefont
  {Lubensky}},\ }\bibfield  {title} {\bibinfo {title} {Maxwell lattices and
  topological mechanics},\ }\href@noop {} {\bibfield  {journal} {\bibinfo
  {journal} {Annual Review of Condensed Matter Physics}\ }\textbf {\bibinfo
  {volume} {9}},\ \bibinfo {pages} {413} (\bibinfo {year}
  {2018}{\natexlab{b}})}\BibitemShut {NoStop}%
\bibitem [{\citenamefont {Paulose}\ \emph
  {et~al.}(2015{\natexlab{a}})\citenamefont {Paulose}, \citenamefont {Chen},\
  and\ \citenamefont {Vitelli}}]{Paulose2015}%
  \BibitemOpen
  \bibfield  {author} {\bibinfo {author} {\bibfnamefont {J.}~\bibnamefont
  {Paulose}}, \bibinfo {author} {\bibfnamefont {B.~G.-g.}\ \bibnamefont
  {Chen}},\ and\ \bibinfo {author} {\bibfnamefont {V.}~\bibnamefont
  {Vitelli}},\ }\bibfield  {title} {\bibinfo {title} {Topological modes bound
  to dislocations in mechanical metamaterials},\ }\href@noop {} {\bibfield
  {journal} {\bibinfo  {journal} {Nature Physics}\ }\textbf {\bibinfo {volume}
  {11}},\ \bibinfo {pages} {153} (\bibinfo {year}
  {2015}{\natexlab{a}})}\BibitemShut {NoStop}%
\bibitem [{\citenamefont {Ma}\ \emph {et~al.}(2018)\citenamefont {Ma},
  \citenamefont {Zhou}, \citenamefont {Sun}, \citenamefont {Mao},\ and\
  \citenamefont {Gonella}}]{Ma2018}%
  \BibitemOpen
  \bibfield  {author} {\bibinfo {author} {\bibfnamefont {J.}~\bibnamefont
  {Ma}}, \bibinfo {author} {\bibfnamefont {D.}~\bibnamefont {Zhou}}, \bibinfo
  {author} {\bibfnamefont {K.}~\bibnamefont {Sun}}, \bibinfo {author}
  {\bibfnamefont {X.}~\bibnamefont {Mao}},\ and\ \bibinfo {author}
  {\bibfnamefont {S.}~\bibnamefont {Gonella}},\ }\bibfield  {title} {\bibinfo
  {title} {Edge modes and asymmetric wave transport in topological lattices:
  Experimental characterization at finite frequencies},\ }\href
  {https://doi.org/10.1103/PhysRevLett.121.094301} {\bibfield  {journal}
  {\bibinfo  {journal} {Phys. Rev. Lett.}\ }\textbf {\bibinfo {volume} {121}},\
  \bibinfo {pages} {094301} (\bibinfo {year} {2018})}\BibitemShut {NoStop}%
\bibitem [{\citenamefont {Calladine}(1978)}]{Calladine1978}%
  \BibitemOpen
  \bibfield  {author} {\bibinfo {author} {\bibfnamefont {C.~R.}\ \bibnamefont
  {Calladine}},\ }\bibfield  {title} {\bibinfo {title} {Buckminster fuller's
  “tensegrity” structures and clerk maxwell's rules for the construction of
  stiff frames},\ }\href@noop {} {\bibfield  {journal} {\bibinfo  {journal}
  {International journal of solids and structures}\ }\textbf {\bibinfo {volume}
  {14}},\ \bibinfo {pages} {161} (\bibinfo {year} {1978})}\BibitemShut
  {NoStop}%
\bibitem [{\citenamefont {Zhang}\ and\ \citenamefont {Mao}(2018)}]{Zhang2018}%
  \BibitemOpen
  \bibfield  {author} {\bibinfo {author} {\bibfnamefont {L.}~\bibnamefont
  {Zhang}}\ and\ \bibinfo {author} {\bibfnamefont {X.}~\bibnamefont {Mao}},\
  }\bibfield  {title} {\bibinfo {title} {Fracturing of topological maxwell
  lattices},\ }\href@noop {} {\bibfield  {journal} {\bibinfo  {journal} {New
  Journal of Physics}\ }\textbf {\bibinfo {volume} {20}},\ \bibinfo {pages}
  {063034} (\bibinfo {year} {2018})}\BibitemShut {NoStop}%
\bibitem [{\citenamefont {Widstrand}\ \emph {et~al.}(2022)\citenamefont
  {Widstrand}, \citenamefont {Hu}, \citenamefont {Mao}, \citenamefont {Labuz},\
  and\ \citenamefont {Gonella}}]{Widstrand2022}%
  \BibitemOpen
  \bibfield  {author} {\bibinfo {author} {\bibfnamefont {C.}~\bibnamefont
  {Widstrand}}, \bibinfo {author} {\bibfnamefont {C.}~\bibnamefont {Hu}},
  \bibinfo {author} {\bibfnamefont {X.}~\bibnamefont {Mao}}, \bibinfo {author}
  {\bibfnamefont {J.}~\bibnamefont {Labuz}},\ and\ \bibinfo {author}
  {\bibfnamefont {S.}~\bibnamefont {Gonella}},\ }\bibfield  {title} {\bibinfo
  {title} {Stress focusing and damage protection in topological maxwell
  metamaterials},\ }\href@noop {} {\bibfield  {journal} {\bibinfo  {journal}
  {arXiv preprint arXiv:2209.14463}\ } (\bibinfo {year} {2022})}\BibitemShut
  {NoStop}%
\bibitem [{\citenamefont {Guzm{\'a}n}\ \emph {et~al.}(2022)\citenamefont
  {Guzm{\'a}n}, \citenamefont {Bartolo},\ and\ \citenamefont
  {Carpentier}}]{guzman2020geometry}%
  \BibitemOpen
  \bibfield  {author} {\bibinfo {author} {\bibfnamefont {M.}~\bibnamefont
  {Guzm{\'a}n}}, \bibinfo {author} {\bibfnamefont {D.}~\bibnamefont
  {Bartolo}},\ and\ \bibinfo {author} {\bibfnamefont {D.}~\bibnamefont
  {Carpentier}},\ }\bibfield  {title} {\bibinfo {title} {{Geometry and Topology
  Tango in Ordered and Amorphous Chiral Matter}},\ }\href
  {https://doi.org/10.21468/SciPostPhys.12.1.038} {\bibfield  {journal}
  {\bibinfo  {journal} {SciPost Phys.}\ }\textbf {\bibinfo {volume} {12}},\
  \bibinfo {pages} {38} (\bibinfo {year} {2022})}\BibitemShut {NoStop}%
\bibitem [{\citenamefont {Resta}\ and\ \citenamefont
  {Vanderbilt}(2007)}]{Resta2007}%
  \BibitemOpen
  \bibfield  {author} {\bibinfo {author} {\bibfnamefont {R.}~\bibnamefont
  {Resta}}\ and\ \bibinfo {author} {\bibfnamefont {D.}~\bibnamefont
  {Vanderbilt}},\ }\bibfield  {title} {\bibinfo {title} {Theory of
  polarization: a modern approach},\ }in\ \href@noop {} {\emph {\bibinfo
  {booktitle} {Physics of Ferroelectrics}}}\ (\bibinfo  {publisher}
  {Springer},\ \bibinfo {year} {2007})\ pp.\ \bibinfo {pages}
  {31--68}\BibitemShut {NoStop}%
\bibitem [{\citenamefont {Cardano}\ \emph
  {et~al.}(2017{\natexlab{a}})\citenamefont {Cardano}, \citenamefont
  {D’Errico}, \citenamefont {Dauphin}, \citenamefont {Maffei}, \citenamefont
  {Piccirillo}, \citenamefont {de~Lisio}, \citenamefont {De~Filippis},
  \citenamefont {Cataudella}, \citenamefont {Santamato}, \citenamefont
  {Marrucci} \emph {et~al.}}]{Cardano2017}%
  \BibitemOpen
  \bibfield  {author} {\bibinfo {author} {\bibfnamefont {F.}~\bibnamefont
  {Cardano}}, \bibinfo {author} {\bibfnamefont {A.}~\bibnamefont {D’Errico}},
  \bibinfo {author} {\bibfnamefont {A.}~\bibnamefont {Dauphin}}, \bibinfo
  {author} {\bibfnamefont {M.}~\bibnamefont {Maffei}}, \bibinfo {author}
  {\bibfnamefont {B.}~\bibnamefont {Piccirillo}}, \bibinfo {author}
  {\bibfnamefont {C.}~\bibnamefont {de~Lisio}}, \bibinfo {author}
  {\bibfnamefont {G.}~\bibnamefont {De~Filippis}}, \bibinfo {author}
  {\bibfnamefont {V.}~\bibnamefont {Cataudella}}, \bibinfo {author}
  {\bibfnamefont {E.}~\bibnamefont {Santamato}}, \bibinfo {author}
  {\bibfnamefont {L.}~\bibnamefont {Marrucci}}, \emph {et~al.},\ }\bibfield
  {title} {\bibinfo {title} {Detection of zak phases and topological invariants
  in a chiral quantum walk of twisted photons},\ }\href@noop {} {\bibfield
  {journal} {\bibinfo  {journal} {Nature communications}\ }\textbf {\bibinfo
  {volume} {8}},\ \bibinfo {pages} {1} (\bibinfo {year}
  {2017}{\natexlab{a}})}\BibitemShut {NoStop}%
\bibitem [{\citenamefont {Roberts}\ \emph {et~al.}(2022)\citenamefont
  {Roberts}, \citenamefont {Baardink}, \citenamefont {Nunn}, \citenamefont
  {Mosley},\ and\ \citenamefont {Souslov}}]{Roberts2022}%
  \BibitemOpen
  \bibfield  {author} {\bibinfo {author} {\bibfnamefont {N.}~\bibnamefont
  {Roberts}}, \bibinfo {author} {\bibfnamefont {G.}~\bibnamefont {Baardink}},
  \bibinfo {author} {\bibfnamefont {J.}~\bibnamefont {Nunn}}, \bibinfo {author}
  {\bibfnamefont {P.~J.}\ \bibnamefont {Mosley}},\ and\ \bibinfo {author}
  {\bibfnamefont {A.}~\bibnamefont {Souslov}},\ }\bibfield  {title} {\bibinfo
  {title} {Topological supermodes in photonic crystal fiber},\ }\href
  {https://doi.org/10.1126/sciadv.add3522} {\bibfield  {journal} {\bibinfo
  {journal} {Science Advances}\ }\textbf {\bibinfo {volume} {8}},\ \bibinfo
  {pages} {eadd3522} (\bibinfo {year} {2022})},\ \Eprint
  {https://arxiv.org/abs/https://www.science.org/doi/pdf/10.1126/sciadv.add3522}
  {https://www.science.org/doi/pdf/10.1126/sciadv.add3522} \BibitemShut
  {NoStop}%
\bibitem [{\citenamefont {Paulose}\ \emph
  {et~al.}(2015{\natexlab{b}})\citenamefont {Paulose}, \citenamefont {Chen},\
  and\ \citenamefont {Vitelli}}]{Paulose2015Topologica}%
  \BibitemOpen
  \bibfield  {author} {\bibinfo {author} {\bibfnamefont {J.}~\bibnamefont
  {Paulose}}, \bibinfo {author} {\bibfnamefont {B.~G.-g.}\ \bibnamefont
  {Chen}},\ and\ \bibinfo {author} {\bibfnamefont {V.}~\bibnamefont
  {Vitelli}},\ }\bibfield  {title} {\bibinfo {title} {Topological modes bound
  to dislocations in mechanical metamaterials},\ }\href
  {https://doi.org/10.1038/nphys3185} {\bibfield  {journal} {\bibinfo
  {journal} {Nature Physics}\ }\textbf {\bibinfo {volume} {11}},\ \bibinfo
  {pages} {153} (\bibinfo {year} {2015}{\natexlab{b}})}\BibitemShut {NoStop}%
\bibitem [{\citenamefont {Vanderbilt}(2018)}]{vanderbilt2018berry}%
  \BibitemOpen
  \bibfield  {author} {\bibinfo {author} {\bibfnamefont {D.}~\bibnamefont
  {Vanderbilt}},\ }\href@noop {} {\emph {\bibinfo {title} {Berry phases in
  electronic structure theory: electric polarization, orbital magnetization and
  topological insulators}}}\ (\bibinfo  {publisher} {Cambridge University
  Press},\ \bibinfo {year} {2018})\BibitemShut {NoStop}%
\bibitem [{\citenamefont {Jezequel}\ \emph {et~al.}(2022)\citenamefont
  {Jezequel}, \citenamefont {Tauber},\ and\ \citenamefont
  {Delplace}}]{Jezequel2022}%
  \BibitemOpen
  \bibfield  {author} {\bibinfo {author} {\bibfnamefont {L.}~\bibnamefont
  {Jezequel}}, \bibinfo {author} {\bibfnamefont {C.}~\bibnamefont {Tauber}},\
  and\ \bibinfo {author} {\bibfnamefont {P.}~\bibnamefont {Delplace}},\
  }\bibfield  {title} {\bibinfo {title} {Estimating bulk and edge topological
  indices in finite open chiral chains},\ }\href@noop {} {\bibfield  {journal}
  {\bibinfo  {journal} {Journal of Mathematical Physics}\ }\textbf {\bibinfo
  {volume} {63}},\ \bibinfo {pages} {121901} (\bibinfo {year}
  {2022})}\BibitemShut {NoStop}%
\bibitem [{\citenamefont {Benalcazar}\ \emph {et~al.}(2017)\citenamefont
  {Benalcazar}, \citenamefont {Bernevig},\ and\ \citenamefont
  {Hughes}}]{Benalcazar2017Quantized}%
  \BibitemOpen
  \bibfield  {author} {\bibinfo {author} {\bibfnamefont {W.~A.}\ \bibnamefont
  {Benalcazar}}, \bibinfo {author} {\bibfnamefont {B.~A.}\ \bibnamefont
  {Bernevig}},\ and\ \bibinfo {author} {\bibfnamefont {T.~L.}\ \bibnamefont
  {Hughes}},\ }\bibfield  {title} {\bibinfo {title} {Quantized electric
  multipole insulators},\ }\href {https://doi.org/10.1126/science.aah6442}
  {\bibfield  {journal} {\bibinfo  {journal} {Science}\ }\textbf {\bibinfo
  {volume} {357}},\ \bibinfo {pages} {61} (\bibinfo {year} {2017})},\ \Eprint
  {https://arxiv.org/abs/https://www.science.org/doi/pdf/10.1126/science.aah6442}
  {https://www.science.org/doi/pdf/10.1126/science.aah6442} \BibitemShut
  {NoStop}%
\bibitem [{\citenamefont {Serra-Garcia}\ \emph
  {et~al.}(2018{\natexlab{b}})\citenamefont {Serra-Garcia}, \citenamefont
  {Peri}, \citenamefont {S{\"u}sstrunk}, \citenamefont {Bilal}, \citenamefont
  {Larsen}, \citenamefont {Villanueva},\ and\ \citenamefont
  {Huber}}]{serra2018observation}%
  \BibitemOpen
  \bibfield  {author} {\bibinfo {author} {\bibfnamefont {M.}~\bibnamefont
  {Serra-Garcia}}, \bibinfo {author} {\bibfnamefont {V.}~\bibnamefont {Peri}},
  \bibinfo {author} {\bibfnamefont {R.}~\bibnamefont {S{\"u}sstrunk}}, \bibinfo
  {author} {\bibfnamefont {O.~R.}\ \bibnamefont {Bilal}}, \bibinfo {author}
  {\bibfnamefont {T.}~\bibnamefont {Larsen}}, \bibinfo {author} {\bibfnamefont
  {L.~G.}\ \bibnamefont {Villanueva}},\ and\ \bibinfo {author} {\bibfnamefont
  {S.~D.}\ \bibnamefont {Huber}},\ }\bibfield  {title} {\bibinfo {title}
  {Observation of a phononic quadrupole topological insulator},\ }\href
  {https://doi.org/10.1038/nature25156} {\bibfield  {journal} {\bibinfo
  {journal} {Nature}\ }\textbf {\bibinfo {volume} {555}},\ \bibinfo {pages}
  {342} (\bibinfo {year} {2018}{\natexlab{b}})}\BibitemShut {NoStop}%
\bibitem [{\citenamefont {Neupert}\ and\ \citenamefont
  {Schindler}(2018)}]{Neupert2018Topological}%
  \BibitemOpen
  \bibfield  {author} {\bibinfo {author} {\bibfnamefont {T.}~\bibnamefont
  {Neupert}}\ and\ \bibinfo {author} {\bibfnamefont {F.}~\bibnamefont
  {Schindler}},\ }\bibfield  {title} {\bibinfo {title} {Topological crystalline
  insulators},\ }in\ \href {https://doi.org/10.1007/978-3-319-76388-0_2} {\emph
  {\bibinfo {booktitle} {Topological Matter}}}\ (\bibinfo  {publisher}
  {Springer},\ \bibinfo {year} {2018})\ pp.\ \bibinfo {pages}
  {31--61}\BibitemShut {NoStop}%
\bibitem [{\citenamefont {Paulose}\ \emph
  {et~al.}(2015{\natexlab{c}})\citenamefont {Paulose}, \citenamefont
  {Meeussen},\ and\ \citenamefont {Vitelli}}]{Paulose2015Selectivebuckling}%
  \BibitemOpen
  \bibfield  {author} {\bibinfo {author} {\bibfnamefont {J.}~\bibnamefont
  {Paulose}}, \bibinfo {author} {\bibfnamefont {A.~S.}\ \bibnamefont
  {Meeussen}},\ and\ \bibinfo {author} {\bibfnamefont {V.}~\bibnamefont
  {Vitelli}},\ }\bibfield  {title} {\bibinfo {title} {Selective buckling via
  states of self-stress in topological metamaterials},\ }\href
  {https://doi.org/10.1073/pnas.1502939112} {\bibfield  {journal} {\bibinfo
  {journal} {Proceedings of the National Academy of Sciences}\ }\textbf
  {\bibinfo {volume} {112}},\ \bibinfo {pages} {7639} (\bibinfo {year}
  {2015}{\natexlab{c}})},\ \Eprint
  {https://arxiv.org/abs/https://www.pnas.org/doi/pdf/10.1073/pnas.1502939112}
  {https://www.pnas.org/doi/pdf/10.1073/pnas.1502939112} \BibitemShut {NoStop}%
\bibitem [{\citenamefont {Cardano}\ \emph
  {et~al.}(2017{\natexlab{b}})\citenamefont {Cardano}, \citenamefont
  {D'Errico}, \citenamefont {Dauphin}, \citenamefont {Maffei}, \citenamefont
  {Piccirillo}, \citenamefont {de~Lisio}, \citenamefont {De~Filippis},
  \citenamefont {Cataudella}, \citenamefont {Santamato}, \citenamefont
  {Marrucci}, \citenamefont {Lewenstein},\ and\ \citenamefont
  {Massignan}}]{cardano2017detection}%
  \BibitemOpen
  \bibfield  {author} {\bibinfo {author} {\bibfnamefont {F.}~\bibnamefont
  {Cardano}}, \bibinfo {author} {\bibfnamefont {A.}~\bibnamefont {D'Errico}},
  \bibinfo {author} {\bibfnamefont {A.}~\bibnamefont {Dauphin}}, \bibinfo
  {author} {\bibfnamefont {M.}~\bibnamefont {Maffei}}, \bibinfo {author}
  {\bibfnamefont {B.}~\bibnamefont {Piccirillo}}, \bibinfo {author}
  {\bibfnamefont {C.}~\bibnamefont {de~Lisio}}, \bibinfo {author}
  {\bibfnamefont {G.}~\bibnamefont {De~Filippis}}, \bibinfo {author}
  {\bibfnamefont {V.}~\bibnamefont {Cataudella}}, \bibinfo {author}
  {\bibfnamefont {E.}~\bibnamefont {Santamato}}, \bibinfo {author}
  {\bibfnamefont {L.}~\bibnamefont {Marrucci}}, \bibinfo {author}
  {\bibfnamefont {M.}~\bibnamefont {Lewenstein}},\ and\ \bibinfo {author}
  {\bibfnamefont {P.}~\bibnamefont {Massignan}},\ }\bibfield  {title} {\bibinfo
  {title} {Detection of zak phases and topological invariants in a chiral
  quantum walk of twisted photons},\ }\href
  {https://doi.org/10.1038/ncomms15516} {\bibfield  {journal} {\bibinfo
  {journal} {Nature Communications}\ }\textbf {\bibinfo {volume} {8}},\
  \bibinfo {pages} {15516} (\bibinfo {year} {2017}{\natexlab{b}})}\BibitemShut
  {NoStop}%
\bibitem [{\citenamefont {Maffei}\ \emph {et~al.}(2018)\citenamefont {Maffei},
  \citenamefont {Dauphin}, \citenamefont {Cardano}, \citenamefont
  {Lewenstein},\ and\ \citenamefont {Massignan}}]{maffei2018topological}%
  \BibitemOpen
  \bibfield  {author} {\bibinfo {author} {\bibfnamefont {M.}~\bibnamefont
  {Maffei}}, \bibinfo {author} {\bibfnamefont {A.}~\bibnamefont {Dauphin}},
  \bibinfo {author} {\bibfnamefont {F.}~\bibnamefont {Cardano}}, \bibinfo
  {author} {\bibfnamefont {M.}~\bibnamefont {Lewenstein}},\ and\ \bibinfo
  {author} {\bibfnamefont {P.}~\bibnamefont {Massignan}},\ }\bibfield  {title}
  {\bibinfo {title} {Topological characterization of chiral models through
  their long time dynamics},\ }\href {https://doi.org/10.1088/1367-2630/aa9d4c}
  {\bibfield  {journal} {\bibinfo  {journal} {New Journal of Physics}\ }\textbf
  {\bibinfo {volume} {20}},\ \bibinfo {pages} {013023} (\bibinfo {year}
  {2018})}\BibitemShut {NoStop}%
\bibitem [{\citenamefont {St-Jean}\ \emph {et~al.}(2017)\citenamefont
  {St-Jean}, \citenamefont {Goblot}, \citenamefont {Galopin}, \citenamefont
  {Lema{\^\i}tre}, \citenamefont {Ozawa}, \citenamefont {Le~Gratiet},
  \citenamefont {Sagnes}, \citenamefont {Bloch},\ and\ \citenamefont
  {Amo}}]{St-Jean2017Lasing-in-}%
  \BibitemOpen
  \bibfield  {author} {\bibinfo {author} {\bibfnamefont {P.}~\bibnamefont
  {St-Jean}}, \bibinfo {author} {\bibfnamefont {V.}~\bibnamefont {Goblot}},
  \bibinfo {author} {\bibfnamefont {E.}~\bibnamefont {Galopin}}, \bibinfo
  {author} {\bibfnamefont {A.}~\bibnamefont {Lema{\^\i}tre}}, \bibinfo {author}
  {\bibfnamefont {T.}~\bibnamefont {Ozawa}}, \bibinfo {author} {\bibfnamefont
  {L.}~\bibnamefont {Le~Gratiet}}, \bibinfo {author} {\bibfnamefont
  {I.}~\bibnamefont {Sagnes}}, \bibinfo {author} {\bibfnamefont
  {J.}~\bibnamefont {Bloch}},\ and\ \bibinfo {author} {\bibfnamefont
  {A.}~\bibnamefont {Amo}},\ }\bibfield  {title} {\bibinfo {title} {Lasing in
  topological edge states of a one-dimensionallattice},\ }\href
  {https://doi.org/10.1038/s41566-017-0006-2} {\bibfield  {journal} {\bibinfo
  {journal} {Nature Photonics}\ }\textbf {\bibinfo {volume} {11}},\ \bibinfo
  {pages} {651} (\bibinfo {year} {2017})}\BibitemShut {NoStop}%
\bibitem [{\citenamefont {Ni}\ \emph {et~al.}(2019{\natexlab{b}})\citenamefont
  {Ni}, \citenamefont {Weiner}, \citenamefont {Al{\`u}},\ and\ \citenamefont
  {Khanikaev}}]{Ni2019Observatio}%
  \BibitemOpen
  \bibfield  {author} {\bibinfo {author} {\bibfnamefont {X.}~\bibnamefont
  {Ni}}, \bibinfo {author} {\bibfnamefont {M.}~\bibnamefont {Weiner}}, \bibinfo
  {author} {\bibfnamefont {A.}~\bibnamefont {Al{\`u}}},\ and\ \bibinfo {author}
  {\bibfnamefont {A.~B.}\ \bibnamefont {Khanikaev}},\ }\bibfield  {title}
  {\bibinfo {title} {Observation of higher-order topological acoustic states
  protected by generalized chiral symmetry},\ }\href
  {https://doi.org/10.1038/s41563-018-0252-9} {\bibfield  {journal} {\bibinfo
  {journal} {Nature Materials}\ }\textbf {\bibinfo {volume} {18}},\ \bibinfo
  {pages} {113} (\bibinfo {year} {2019}{\natexlab{b}})}\BibitemShut {NoStop}%
\bibitem [{\citenamefont {Weiner}\ \emph {et~al.}(2020)\citenamefont {Weiner},
  \citenamefont {Ni}, \citenamefont {Li}, \citenamefont {Al{\`u}},\ and\
  \citenamefont {Khanikaev}}]{Weiner2020ThirdOrderAcoustic}%
  \BibitemOpen
  \bibfield  {author} {\bibinfo {author} {\bibfnamefont {M.}~\bibnamefont
  {Weiner}}, \bibinfo {author} {\bibfnamefont {X.}~\bibnamefont {Ni}}, \bibinfo
  {author} {\bibfnamefont {M.}~\bibnamefont {Li}}, \bibinfo {author}
  {\bibfnamefont {A.}~\bibnamefont {Al{\`u}}},\ and\ \bibinfo {author}
  {\bibfnamefont {A.~B.}\ \bibnamefont {Khanikaev}},\ }\bibfield  {title}
  {\bibinfo {title} {Demonstration of a third-order hierarchy of topological
  states in a three-dimensional acoustic metamaterial},\ }\href
  {https://doi.org/10.1126/sciadv.aay4166} {\bibfield  {journal} {\bibinfo
  {journal} {Science Advances}\ }\textbf {\bibinfo {volume} {6}},\ \bibinfo
  {pages} {eaay4166} (\bibinfo {year} {2020})},\ \Eprint
  {https://arxiv.org/abs/https://www.science.org/doi/pdf/10.1126/sciadv.aay4166}
  {https://www.science.org/doi/pdf/10.1126/sciadv.aay4166} \BibitemShut
  {NoStop}%
\bibitem [{\citenamefont {Bellec}\ \emph {et~al.}(2014)\citenamefont {Bellec},
  \citenamefont {Kuhl}, \citenamefont {Montambaux},\ and\ \citenamefont
  {Mortessagne}}]{Bellec2014Manipulation}%
  \BibitemOpen
  \bibfield  {author} {\bibinfo {author} {\bibfnamefont {M.}~\bibnamefont
  {Bellec}}, \bibinfo {author} {\bibfnamefont {U.}~\bibnamefont {Kuhl}},
  \bibinfo {author} {\bibfnamefont {G.}~\bibnamefont {Montambaux}},\ and\
  \bibinfo {author} {\bibfnamefont {F.}~\bibnamefont {Mortessagne}},\
  }\bibfield  {title} {\bibinfo {title} {Manipulation of edge states in
  microwave artificial graphene},\ }\href
  {https://doi.org/10.1088/1367-2630/16/11/113023} {\bibfield  {journal}
  {\bibinfo  {journal} {New Journal of Physics}\ }\textbf {\bibinfo {volume}
  {16}},\ \bibinfo {pages} {113023} (\bibinfo {year} {2014})}\BibitemShut
  {NoStop}%
\bibitem [{\citenamefont {Di~Ventra}\ \emph {et~al.}(2022)\citenamefont
  {Di~Ventra}, \citenamefont {Pershin},\ and\ \citenamefont
  {Chien}}]{DiVentra2022CustodialChiral}%
  \BibitemOpen
  \bibfield  {author} {\bibinfo {author} {\bibfnamefont {M.}~\bibnamefont
  {Di~Ventra}}, \bibinfo {author} {\bibfnamefont {Y.~V.}\ \bibnamefont
  {Pershin}},\ and\ \bibinfo {author} {\bibfnamefont {C.-C.}\ \bibnamefont
  {Chien}},\ }\bibfield  {title} {\bibinfo {title} {Custodial chiral symmetry
  in a su-schrieffer-heeger electrical circuit with memory},\ }\href
  {https://doi.org/10.1103/PhysRevLett.128.097701} {\bibfield  {journal}
  {\bibinfo  {journal} {Phys. Rev. Lett.}\ }\textbf {\bibinfo {volume} {128}},\
  \bibinfo {pages} {097701} (\bibinfo {year} {2022})}\BibitemShut {NoStop}%
\bibitem [{\citenamefont {Imhof}\ \emph {et~al.}(2018)\citenamefont {Imhof},
  \citenamefont {Berger}, \citenamefont {Bayer}, \citenamefont {Brehm},
  \citenamefont {Molenkamp}, \citenamefont {Kiessling}, \citenamefont
  {Schindler}, \citenamefont {Lee}, \citenamefont {Greiter}, \citenamefont
  {Neupert},\ and\ \citenamefont {Thomale}}]{Imhof2018Topolectri}%
  \BibitemOpen
  \bibfield  {author} {\bibinfo {author} {\bibfnamefont {S.}~\bibnamefont
  {Imhof}}, \bibinfo {author} {\bibfnamefont {C.}~\bibnamefont {Berger}},
  \bibinfo {author} {\bibfnamefont {F.}~\bibnamefont {Bayer}}, \bibinfo
  {author} {\bibfnamefont {J.}~\bibnamefont {Brehm}}, \bibinfo {author}
  {\bibfnamefont {L.~W.}\ \bibnamefont {Molenkamp}}, \bibinfo {author}
  {\bibfnamefont {T.}~\bibnamefont {Kiessling}}, \bibinfo {author}
  {\bibfnamefont {F.}~\bibnamefont {Schindler}}, \bibinfo {author}
  {\bibfnamefont {C.~H.}\ \bibnamefont {Lee}}, \bibinfo {author} {\bibfnamefont
  {M.}~\bibnamefont {Greiter}}, \bibinfo {author} {\bibfnamefont
  {T.}~\bibnamefont {Neupert}},\ and\ \bibinfo {author} {\bibfnamefont
  {R.}~\bibnamefont {Thomale}},\ }\bibfield  {title} {\bibinfo {title}
  {Topolectrical-circuit realization of topological corner modes},\ }\href
  {https://doi.org/10.1038/s41567-018-0246-1} {\bibfield  {journal} {\bibinfo
  {journal} {Nature Physics}\ }\textbf {\bibinfo {volume} {14}},\ \bibinfo
  {pages} {925} (\bibinfo {year} {2018})}\BibitemShut {NoStop}%
\bibitem [{\citenamefont {Kotwal}\ \emph {et~al.}(2021)\citenamefont {Kotwal},
  \citenamefont {Moseley}, \citenamefont {Stegmaier}, \citenamefont {Imhof},
  \citenamefont {Brand}, \citenamefont {Kie{\ss}ling}, \citenamefont {Thomale},
  \citenamefont {Ronellenfitsch},\ and\ \citenamefont
  {Dunkel}}]{kotwal2021active}%
  \BibitemOpen
  \bibfield  {author} {\bibinfo {author} {\bibfnamefont {T.}~\bibnamefont
  {Kotwal}}, \bibinfo {author} {\bibfnamefont {F.}~\bibnamefont {Moseley}},
  \bibinfo {author} {\bibfnamefont {A.}~\bibnamefont {Stegmaier}}, \bibinfo
  {author} {\bibfnamefont {S.}~\bibnamefont {Imhof}}, \bibinfo {author}
  {\bibfnamefont {H.}~\bibnamefont {Brand}}, \bibinfo {author} {\bibfnamefont
  {T.}~\bibnamefont {Kie{\ss}ling}}, \bibinfo {author} {\bibfnamefont
  {R.}~\bibnamefont {Thomale}}, \bibinfo {author} {\bibfnamefont
  {H.}~\bibnamefont {Ronellenfitsch}},\ and\ \bibinfo {author} {\bibfnamefont
  {J.}~\bibnamefont {Dunkel}},\ }\bibfield  {title} {\bibinfo {title} {Active
  topolectrical circuits},\ }\href@noop {} {\bibfield  {journal} {\bibinfo
  {journal} {Proceedings of the National Academy of Sciences}\ }\textbf
  {\bibinfo {volume} {118}},\ \bibinfo {pages} {e2106411118} (\bibinfo {year}
  {2021})}\BibitemShut {NoStop}%
\bibitem [{\citenamefont {De~Maesschalck}\ \emph {et~al.}(2000)\citenamefont
  {De~Maesschalck}, \citenamefont {Jouan-Rimbaud},\ and\ \citenamefont
  {Massart}}]{de2000mahalanobis}%
  \BibitemOpen
  \bibfield  {author} {\bibinfo {author} {\bibfnamefont {R.}~\bibnamefont
  {De~Maesschalck}}, \bibinfo {author} {\bibfnamefont {D.}~\bibnamefont
  {Jouan-Rimbaud}},\ and\ \bibinfo {author} {\bibfnamefont {D.~L.}\
  \bibnamefont {Massart}},\ }\bibfield  {title} {\bibinfo {title} {The
  mahalanobis distance},\ }\href@noop {} {\bibfield  {journal} {\bibinfo
  {journal} {Chemometrics and intelligent laboratory systems}\ }\textbf
  {\bibinfo {volume} {50}},\ \bibinfo {pages} {1} (\bibinfo {year}
  {2000})}\BibitemShut {NoStop}%
\bibitem [{\citenamefont {van Beek}(2019)}]{designbook1}%
  \BibitemOpen
  \bibfield  {author} {\bibinfo {author} {\bibfnamefont {A.}~\bibnamefont {van
  Beek}},\ }\href@noop {} {\emph {\bibinfo {title} {Advanced engineering
  design: lifetime performance and reliability}}}\ (\bibinfo {year}
  {2019})\BibitemShut {NoStop}%
\bibitem [{\citenamefont {Koster}(1996)}]{designbook2}%
  \BibitemOpen
  \bibfield  {author} {\bibinfo {author} {\bibfnamefont {M.}~\bibnamefont
  {Koster}},\ }\href@noop {} {\emph {\bibinfo {title} {Constructieprincipes:
  voor het nauwkeurig bewegen en positioneren}}}\ (\bibinfo {year}
  {1996})\BibitemShut {NoStop}%
\end{thebibliography}%


%

\end{document}